\documentclass{aa}  

\usepackage{graphicx}
\usepackage{txfonts}
\usepackage[colorlinks=true, linkcolor=blue, citecolor=blue, urlcolor=blue]{hyperref}
\usepackage{cleveref}
\usepackage{xcolor}
\usepackage{caption}
\usepackage{longtable}
\usepackage[switch]{lineno}
\newcommand{\kms}{\,\mathrm{km\,s^{-1}}}
\newcommand{\Ni}{\mbox{$^{56}$Ni}}
\newcommand{\Co}{\mbox{$^{56}$Co}}

\crefname{figure}{Fig.}{Figs.}
\Crefname{figure}{Fig.}{Figs.}
\crefname{table}{Table}{Tables}
\Crefname{table}{Table}{Tables}
\crefname{appendix}{Appendix}{Appendices}
\Crefname{appendix}{Appendix}{Appendices}
\crefname{section}{Sect.}{Sects.}
\Crefname{section}{Sect.}{Sects.}

\begin{document} 

   \title{Probing circumstellar interaction with spectroscopy of ZTF hydrogen-rich superluminous supernovae}

\author{
Yang Hu \inst{1} \href{https://orcid.org/0000-0002-9744-3910}{\includegraphics[scale=0.5]{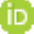}}
\and
Ragnhild Lunnan \inst{1} \href{https://orcid.org/0000-0001-9454-4639}{\includegraphics[scale=0.5]{plots/icon.pdf}}
\and
Priscila J. Pessi \inst{2} \href{https://orcid.org/0000-0002-8041-8559}{\includegraphics[scale=0.5]{plots/icon.pdf}}
\and
Lin Yan \inst{3} \href{https://orcid.org/0000-0003-1710-9339}{\includegraphics[scale=0.5]{plots/icon.pdf}}
\and
Anders Jerkstrand \inst{1} \href{https://orcid.org/0000-0001-8005-4030}{\includegraphics[scale=0.5]{plots/icon.pdf}}
\and
Jesper Sollerman \inst{1} \href{https://orcid.org/0000-0003-1546-6615}{\includegraphics[scale=0.5]{plots/icon.pdf}}
\and
Claes Fransson \inst{1} \href{https://orcid.org/0000-0001-8532-3594}{\includegraphics[scale=0.5]{plots/icon.pdf}}
\and
Steve Schulze \inst{4} \href{https://orcid.org/0000-0001-6797-1889}{\includegraphics[scale=0.5]{plots/icon.pdf}}
\and
Igor Andreoni \inst{5} \href{https://orcid.org/0000-0002-8977-1498}{\includegraphics[scale=0.5]{plots/icon.pdf}}
\and
Bart F. A. van Baal \inst{1,6,7} \href{https://orcid.org/0009-0001-3767-942X}{\includegraphics[scale=0.5]{plots/icon.pdf}}
\and
Se\'an J. Brennan \inst{8} \href{https://orcid.org/0000-0003-1325-6235}{\includegraphics[scale=0.5]{plots/icon.pdf}}
\and
Panos Charalampopoulos \inst{9} \href{https://orcid.org/0000-0002-0326-6715}{\includegraphics[scale=0.5]{plots/icon.pdf}}
\and
Ting-Wan Chen \inst{10} \href{https://orcid.org/0000-0002-1066-6098}{\includegraphics[scale=0.5]{plots/icon.pdf}}
\and
Eduardo Concepci\'on \inst{11} \href{https://orcid.org/0009-0001-6354-2494}{\includegraphics[scale=0.5]{plots/icon.pdf}}
\and
Morgan Fraser \inst{12} \href{https://orcid.org/0000-0003-2191-1674}{\includegraphics[scale=0.5]{plots/icon.pdf}}
\and
Christoffer Fremling \inst{3} \href{https://orcid.org/0000-0002-4223-103X}{\includegraphics[scale=0.5]{plots/icon.pdf}}
\and
Anamaria Gkini \inst{1,13} \href{https://orcid.org/0009-0000-9383-2305}{\includegraphics[scale=0.5]{plots/icon.pdf}}
\and
Mariusz Gromadzki \inst{14} \href{https://orcid.org/0000-0002-1650-1518}{\includegraphics[scale=0.5]{plots/icon.pdf}}
\and
Cosimo Inserra \inst{15} \href{https://orcid.org/0000-0002-3968-4409}{\includegraphics[scale=0.5]{plots/icon.pdf}}
\and
Giorgos Leloudas \inst{16} \href{https://orcid.org/0000-0002-8597-0756}{\includegraphics[scale=0.5]{plots/icon.pdf}}
\and
Tom\'as E. M\"uller-Bravo \inst{17,18} \href{https://orcid.org/0000-0003-3939-7167}{\includegraphics[scale=0.5]{plots/icon.pdf}}
\and
Matt Nicholl \inst{19} \href{https://orcid.org/0000-0002-2555-3192}{\includegraphics[scale=0.5]{plots/icon.pdf}}
\and
Etienne Russeil \inst{1} \href{https://orcid.org/0000-0001-9923-2407}{\includegraphics[scale=0.5]{plots/icon.pdf}}
\and
Yashvi Sharma \inst{3} \href{https://orcid.org/0000-0003-4531-1745}{\includegraphics[scale=0.5]{plots/icon.pdf}}
\and
Claudia Skoglund \inst{1} \href{https://orcid.org/0009-0006-7528-0168}{\includegraphics[scale=0.5]{plots/icon.pdf}}
\and
Yuhan Yao \inst{20} \href{https://orcid.org/0000-0001-6747-8509}{\includegraphics[scale=0.5]{plots/icon.pdf}}
\and
David R. Young \inst{19} \href{https://orcid.org/0000-0002-1229-2499}{\includegraphics[scale=0.5]{plots/icon.pdf}}
\and
Matthew J. Graham \inst{13} \href{https://orcid.org/0000-0002-3168-0139}{\includegraphics[scale=0.5]{plots/icon.pdf}}
\and
Steven L. Groom \inst{21} \href{https://orcid.org/0000-0001-5668-3507}{\includegraphics[scale=0.5]{plots/icon.pdf}}
\and
Mansi M. Kasliwal \inst{13} \href{https://orcid.org/0000-0002-5619-4938}{\includegraphics[scale=0.5]{plots/icon.pdf}}
\and
Frank J. Masci \inst{21} \href{https://orcid.org/0000-0002-8532-9395}{\includegraphics[scale=0.5]{plots/icon.pdf}}
\and
Josiah Purdum \inst{3} \href{https://orcid.org/0000-0003-1227-3738}{\includegraphics[scale=0.5]{plots/icon.pdf}}
\and
Roger Smith \inst{3} \href{https://orcid.org/0000-0001-7062-9726}{\includegraphics[scale=0.5]{plots/icon.pdf}}
}

\institute{
The Oskar Klein Centre, Department of Astronomy, Stockholm University, AlbaNova, SE-10691 Stockholm, Sweden 
\and
Astrophysics Division, National Centre for Nuclear Research, Pasteura 7, 02-093 Warsaw, Poland 
\and
Caltech Optical Observatories, California Institute of Technology, Pasadena, CA 91125, USA 
\and
Department of Particle Physics and Astrophysics, Weizmann Institute of Science, 234 Herzl St, 76100 Rehovot, Israel 
\and
University of North Carolina at Chapel Hill, 120 E. Cameron Ave., Chapel Hill, NC 27514, USA 
\and
Finnish Centre for Astronomy with ESO (FINCA), University of Turku, FI-20014 Turku, Finland 
\and
Department of Physics and Astronomy, University of Turku, FI-20014 Turku, Finland 
\and
Max-Planck-Institut f\"ur extraterrestrische Physik, Giessenbachstrasse 1, 85748 Garching, Germany 
\and
Institute of Space Sciences (ICE-CSIC), Campus UAB, Carrer de Can Magrans, s/n, E-08193 Barcelona, Spain 
\and
Graduate Institute of Astronomy, National Central University, 300 Jhongda Road, 32001 Jhongli, Taiwan 
\and
Center for Astrophysics and Cosmology, University of Nova Gorica, Vipavska 11c, 5270 Ajdovščina, Slovenia 
\and
School of Physics, University College Dublin, O'Brien Center for Science North, Belfield, Dublin 4, Ireland 
\and
Division of Physics, Mathematics, and Astronomy, California Institute of Technology, Pasadena, CA 91125, USA 
\and
Astronomical Observatory, University of Warsaw, Al. Ujazdowskie 4, 00-478 Warszawa, Poland 
\and
Cardiff Hub for Astrophysics Research and Technology, Department of Physics, School of Physical, Chemical and Environmental Sciences, Cardiff University, Queens Buildings, The Parade, Cardiff, CF24 3AA, UK 
\and
DTU Space, Department of Space Research and Space Technology, Technical University of Denmark, Elektrovej 327, 2800 Kgs. Lyngby, Denmark 
\and
School of Physics, Trinity College Dublin, The University of Dublin, Dublin 2, Ireland 
\and
Instituto de Ciencias Exactas y Naturales (ICEN), Universidad Arturo Prat, Chile 
\and
Astrophysics Research Centre, School of Mathematics and Physics, Queen's University Belfast, Belfast BT7 1NN, UK 
\and
The Kavli Institute for Astronomy and Astrophysics, Peking University, Beijing 100871, China 
\and
IPAC, California Institute of Technology, 1200 E. California Blvd, Pasadena, CA 91125, USA 
\\ \email{yang.hu@astro.su.se}
}

 
  \abstract
   {Hydrogen-rich superluminous supernovae (H-rich SLSNe) are rare, exceptionally luminous transients. Most exhibit narrow Balmer emission and can be further classified as SLSNe-IIn, indicating interaction with dense circumstellar material (CSM). Their spectroscopic evolution and the diversity of their circumstellar environments remain poorly understood.}
   {We characterise the spectroscopic diversity of H-rich SLSNe, focusing on the dominant SLSN-IIn population. We investigate how this diversity relates to CSM properties and whether SLSNe-IIn are spectroscopically distinct from lower-luminosity SNe IIn.}
   {We present and analyse 304 optical spectra of 106 H-rich SLSNe observed by the Zwicky Transient Facility (ZTF). We examine their spectral evolution and Balmer-line properties, and model the H$\alpha$ profiles using Monte Carlo electron-scattering calculations. We compare the sample with other types of H-rich SNe.}
   {Of the 106 objects, 105 exhibit narrow Balmer emission consistent with SLSN-IIn classifications, including six with additional Ia-CSM-like characteristics and three with possible flash-ionisation features; the remaining one object exhibits only broad Balmer emission and is classified as a SLSN-II. Post-peak Balmer profiles range from approximately symmetric, exponential profiles to asymmetric profiles with prominent blue excesses or shoulders. Objects with Balmer line blue excess favour higher fitted photon-emitting shell velocities and tend to be fainter than those with persistently symmetric profiles. The H$\alpha$ pseudo-equivalent width (pEW) and FWHM of SLSNe-IIn overlap with those of lower-luminosity SNe IIn, while the pEW evolution is more diverse than in the broad-lined SLSN-II comparison sample. Our SLSNe-IIn sample's median Balmer decrement increases from 2.85 before $+100$\,d to 4.40 afterwards, remaining consistently below that of the comparison Ia-CSM sample.}
  {We identify no Balmer-line morphology unique to SLSNe-IIn in the available comparison with lower-luminosity SNe IIn, although some SLSNe-IIn exhibit prolonged high-ionisation emission. Our results support a close spectroscopic similarity between these populations and interaction with dense CSM formed through enhanced pre-explosion mass loss. We also highlight a possible hidden Ia-CSM population among luminous H-rich interacting SNe.}

   \maketitle
%
\section{Introduction}
Hydrogen-rich superluminous supernovae (H-rich SLSNe) are an observationally defined class of supernovae characterised by the presence of hydrogen in their spectra and peak optical absolute magnitudes of $\lesssim-20$\,mag \citep{Smith_2007_2006gyDiscovery, Ofek_2007_2006gy, Gal-Yam_2012_SLSNe, Perley_2016_SLSNe, Pessi_2025_SLSNeIISample}, reaching as bright as $-22.6$\,mag in the most extreme cases \citep{Kangas_2022_SLSNeIISample}. These luminosities exceed those of ordinary H-rich core-collapse SNe by factors of 10--100, requiring an additional power source beyond the canonical radioactive decay expected in normal core-collapse explosions \citep[e.g.][]{Gal-Yam_2012_SLSNe}.

The observationally defined class of H-rich SLSNe may encompass events with luminosities powered by different mechanisms. One proposed mechanism is the interaction between the SN ejecta and dense circumstellar material (CSM; \citealt{Chevalier_1994_CSM, Chugai_1997_CSM, Smith_2007_2006gy}) expelled by the progenitor system prior to the explosion. In this case, radiative shocks efficiently convert the kinetic energy of the ejecta into radiation. The presence of such dense CSM typically requires extreme pre-explosion mass loss, potentially through winds or eruptions associated with luminous blue variable stars (LBVs; \citealt{Humphreys_1994_LBV, Smith_2008_2006tf}), eruptions with pulsational pair-instability origins \citep{Woosley_2007_PPI, Woosley_2017_PPI, Lunnan_2018_iPTF16eh, Gkini_2025_EruptiveMassloss, Gkini_2026_EruptiveMassloss}, or mass transfer in binary systems \citep{Petrovic_2005_Binary, Langer_2012_Binary, Ercolino_2026_Binary}. Alternative models have also been proposed, primarily in the context of H-poor SLSNe (SLSNe-I; e.g. \citealt{Quimby_2011_SLSNeI}), but efforts have been made to extend their applicability to the H-rich population. One class of models invokes a central engine that continuously injects energy into the expanding SN ejecta, such as a rapidly rotating magnetar \citep{Ostriker_1971_Magnetar, Kasen_2010_Magnetar, Woosley_2010_Magnetar} or fallback accretion onto a newly formed black hole \citep{Dexter_2013_Fallback}. Another model attributes the luminosity to the radioactive decay of large amounts of \Ni\ \citep[e.g.][]{Smith_2007_2006gyDiscovery, Gal-Yam_2009_2007bi}, with the required \Ni\ masses generally expected only in pair-instability supernovae (PISNe) of very massive stars \citep{Barkat_1967_PISNe, Woosley_2002_StellarEvolution, Heger_2003_StellarEvolution, Schulze_2024_2018ibb}. 

While photometric observations alone are often insufficient to distinguish among the proposed powering mechanisms for H-rich SLSNe, spectroscopy provides a crucial additional diagnostic. H-rich SLSNe exhibit considerable diversity in their hydrogen Balmer line profiles. Approximately 20\% of H-rich SLSNe exhibit only broad Balmer emission lines ($\mathrm{FWHM}\geq5000\,\kms$) after peak \citep{Kangas_2022_SLSNeIISample}, and \citet{Inserra_2018_SLSNeII} presented the first study of a small sample of such events. Following these studies, we refer specifically to these broad-lined objects as SLSNe-II throughout this work. The 80\% majority exhibit narrow Balmer emission lines resembling those seen in Type IIn SNe (SNe IIn; \citealt{Schlegel_1990_SNeIIn, Smith_2017_IIn, Nyholm_2020_SNeIIn}), characterised by a narrow emission core ($\mathrm{FWHM}\sim$ a few hundred $\kms$) together with exponential wings\footnote{These wings are also commonly fitted phenomenologically with Lorentzian profiles.} (generally with $\mathrm{FWHM}\leq3000\,\kms$). Because of this spectral similarity, these superluminous events are sometimes referred to as SLSNe-IIn \citep[but see also][]{Hiramatsu_2024_IIn}; we use this term for the narrow-lined objects throughout this work. 

For the dominant narrow-lined population, there is broad consensus that CSM interaction supplies a substantial fraction of the observed luminosity. The narrow Balmer components are generally interpreted as direct evidence for CSM interaction \citep{Chevalier_1994_CSM, Chugai_1997_CSM, Smith_2017_IIn}, as they originate from slowly moving, H-rich CSM that is ionised by shock breakout and subsequent ejecta--CSM interaction. Nevertheless, several key questions remain. One of the central issues, highlighted by both \citet{Perley_2016_SLSNe} and \citet{Hiramatsu_2024_IIn}, is whether the objects referred to as SLSNe-IIn simply represent the luminous extreme of the SN IIn population. This question can be approached from both observational and physical perspectives. Observationally, one may ask whether there is a continuous distribution in the observed properties, such as peak luminosity, evolution timescale, and the duration and strength of narrow lines, across the full population of H-rich interacting SNe. Physically, one may instead ask whether the underlying progenitor and CSM properties form a continuum, including parameters such as CSM mass and extent, pre-explosion mass-loss rate, and even progenitor type. 

A second complication is that narrow Balmer emission and high luminosity do not uniquely identify a core-collapse SN. SNe Ia-CSM are thermonuclear explosions interacting with dense H-rich CSM and can produce luminous light curves and narrow Balmer-dominated spectra resembling those of SNe IIn \citep{Silverman_2013_IaCSM, Leloudas_2015_IaCSM, Sharma_2023_IaCSM}. Early well-studied examples include SNe\,2002ic \citep[e.g.][]{Hamuy_2003_2002ic} and 2005gj \citep{Aldering_2006_2005gj, Prieto_2007_2005gj}, whose spectra were dominated by narrow Balmer emission with diluted SN Ia-like features, and PTF11kx \citep[SN\,2011km;][]{Dilday_2012_PTF11kx, Silverman_2013_PTF11kx}, which exhibited clearer SN Ia features with increasingly strong Balmer emission. However, in some cases \citep[e.g.][]{Benetti_2006_2002ic}, the thermonuclear nature of these Ia-CSM/IIn objects was ambiguous. \citet{Leloudas_2015_IaCSM}, using simulated composite spectra, argued that an underlying SN Ia can be hidden in interacting events as faint as $M_V=-20.1$\,mag, and that SNe Ia-CSM can span $-21.6<M_V<-19.5$\,mag. This range substantially overlaps with H-rich SLSNe. A notable example of this complication is the H-rich SLSN 2006gy, for which a thermonuclear origin has also been proposed \citep{Jerkstrand_2020_2006gy}. Further classification ambiguities arise for transients projected near galaxy centres, where luminous interacting SNe, tidal disruption events (TDEs), and active galactic nuclei (AGNs) can share similar properties. For example, photometrically selected TDE candidate samples can include both SNe and AGN flares \citep{Yao_2023_TDE}. Identifying such events is therefore essential before drawing population-level conclusions on H-rich SLSNe.

Addressing these questions requires statistically meaningful samples. To date, most spectral investigations of H-rich SLSNe have focused on individual objects, such as SN\,2006gy \citep{Ofek_2007_2006gy, Smith_2007_2006gyDiscovery, Smith_2007_2006gy}, SN\,2006tf \citep{Smith_2008_2006tf}, SN\,2008am \citep{Chatzopoulos_2011_2008am}, SN\,2010jl \citep{Stoll_2011_2010jl, Zhang_2012_2010jl, Fransson_2014_2010jl}, SN\,2016aps \citep{Suzuki_2021_2016aps}, SN\,2017hcc \citep{Moran_2023_2017hcc}, SN\,2021aaev \citep{Hu_2026_2021aaev} and SN\,2021adxl \citep{Brennan_2024_2021adxl}, with a few small sample studies \citep[e.g.][]{Inserra_2018_SLSNeII, Salmaso_2025_SNeIIn}. Consequently, our understanding of the population as a whole remains incomplete. In recent years, the Zwicky Transient Facility (ZTF; \citealt{Bellm_2019_ZTF, Graham_2019_ZTF, Masci_2019_ZTF, Dekany_2020_ZTF}) survey, together with extensive spectroscopic follow-up programmes, has dramatically increased the number of observed H-rich SLSNe \citep{Kangas_2022_SLSNeIISample, Pessi_2025_SLSNeIISample}. The photometric sample of more than one hundred H-rich SLSNe presented by \citet{Pessi_2025_SLSNeIISample} now enables population-level studies on their spectra at this scale for the first time. 

In this paper, we present 304 optical spectra of 106 H-rich SLSNe drawn from the ZTF sample identified by \citet{Pessi_2025_SLSNeIISample}. We reassess their spectroscopic classifications and distinguish between SLSNe-IIn and SLSNe-II, and identify objects with possible Ia-CSM characteristics. We then investigate the spectral evolution, Balmer-line properties, and diversity, with the population analysis focusing primarily on the dominant SLSN-IIn subset. This paper is organised as follows. In \Cref{sec:spectroscopy}, we introduce the spectroscopic sample, the instruments used, and the data reduction and calibration procedures. In \cref{sec:analysis}, we present our analysis of the spectroscopic sample. In \cref{sec:modelling}, we derive constraints on the physical properties of the sample using spectroscopic line-profile modelling. In \cref{sec:discussion}, we discuss the implications of our findings for the distinction between SNe IIn and SLSNe-IIn, the CSM properties, the progenitor systems, and the Ia-CSM contamination in the H-rich SLSNe population. We conclude the paper in \cref{sec:conclusion}.


\section{Spectroscopy} \label{sec:spectroscopy}

\subsection{Sample selection and statistics}
Our spectroscopic sample is derived from the parent sample of H-rich SLSNe presented by \citet{Pessi_2025_SLSNeIISample}. That study presented and analysed the photometric properties of 107 events observed by the ZTF survey between 17 March 2018 (MJD 58194.0) and 12 December 2022 (MJD 59925.0)\footnote{The time range was chosen by \citet{Pessi_2025_SLSNeIISample} as extending from the beginning of the ZTF survey to the beginning of their study.} with spectroscopic classifications as being H-rich on the GROWTH marshal \citep{Kasliwal_2019_GROWTH} and Fritz platform \citep{vanderWalt_2019_Fritz, Coughlin_2023_Fritz} regardless of their classification label on the Transient Name Server\footnote{\url{https://www.wis-tns.org/}} (TNS) at the time. The superluminous criterion adopted by \citet{Pessi_2025_SLSNeIISample} requires a peak rest-frame absolute magnitude of $M\leq-19.9$~mag in at least one of the ZTF $gri$ bands. The parent sample was constructed to include all H-rich SLSNe observed by ZTF during this time range, except for 13 SNe with only broad ($\mathrm{FWHM}\geq5000\,\kms$) Balmer emission lines presented by \citet{Kangas_2022_SLSNeIISample}\footnote{The paper presented 14 events, but one object, AT\,2020yue, was later re-classified as a TDE \citep{Yao_2023_2023-08-17}.}. We use the objects in \citet{Kangas_2022_SLSNeIISample} as a comparison sample, hereafter referred to as K22. Together, K22 and the parent sample comprise the full ZTF H-rich SLSN sample observed through 12 December 2022. As a result, the parent sample is strongly weighted towards SLSNe-IIn. However, because the time ranges of the two studies do not fully overlap, the parent sample does not fully exclude all SLSNe-II (see \Cref{sec:SLSN-II}). Moreover, because the parent sample was constructed based primarily on photometry, further spectroscopic inspection in this work can provide evidence for alternative classifications (e.g. SNe Ia-CSM; see \Cref{sec:Ia-CSM}). From this parent sample, we exclude AT\,2022rze, which has since been reclassified as a changing-look active galactic nucleus (CLAGN; \citealt{Pessi_2025_2022rze}). This leaves a final sample of 106 events in this work. 

\begin{figure}
    \centering
    \includegraphics[width=\linewidth]{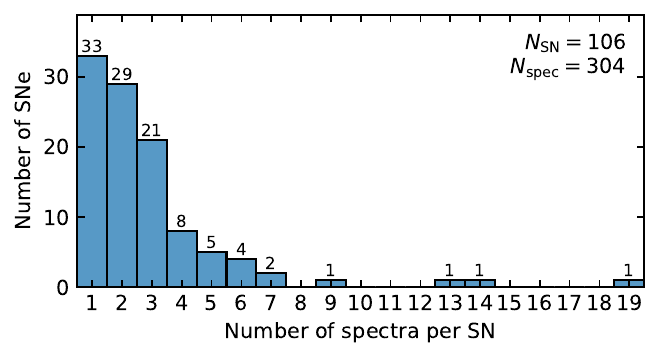}
    \caption{Distribution of the number of spectra per SN in the sample.}
    \label{fig:count}
\end{figure}

This final sample spans a redshift range of $z=0.018$--$0.369$, the nearest being SN\,2021adxl ($z=0.018$) and the next nearest being SN\,2022mma ($z=0.038$). The distribution of the number of spectra per SN is shown in \Cref{fig:count}. A total of 33 objects have only one spectrum, typically the classification spectrum, and the majority of the sample has fewer than five spectra. Only a small subset of objects have more extensive spectroscopic coverage. The general photometric characteristics of the 106 objects have already been summarised in the appendices of \citet{Pessi_2025_SLSNeIISample}. Here, we summarise their spectroscopic characteristics in Appendix~\ref{app:A} and ~\ref{app:B} and in the supplementary material.

\subsection{Instruments and reduction}
Our sample contains spectra from various programmes. Most spectra were obtained through the ZTF collaboration, and a substantial fraction comes from the extended Public ESO Spectroscopic Survey of Transient Objects (ePESSTO, the extension of PESSTO; \citealt{Smartt_2015_PESSTO}) and its successor ePESSTO+. The rest comes from publicly available spectra on Weizmann Interactive Supernova data REPository (\textsc{WISeREP}; \citealt{Yaron_2012_WISeREP}). The instruments, telescopes and associated reduction software are summarised in \Cref{tab:instruments}.

\begin{table}[!htbp]
\centering
\caption{Spectrographs used in this work and their corresponding
reduction software. Most spectra come from the ZTF collaboration,
with additional contributions from ePESSTO/ePESSTO+.
Additional spectra were obtained from publicly available spectra on WISeREP.}
\label{tab:instruments}
\footnotesize
\setlength{\tabcolsep}{3pt}
\renewcommand{\arraystretch}{1.1}
\begin{tabular}{
@{}p{0.21\linewidth}
p{0.45\linewidth}
p{0.29\linewidth}@{}
}
\hline
\hline
Instrument & Telescope & Reduction software \\
\hline
SEDM$^{1}$     & Palomar 60-inch (P60)
              & \texttt{pysedm}$^{2}$ \\
DBSP$^{3}$     & Palomar 200-inch (P200)
              & \texttt{DBSP\_DRP}$^{4}$;
                \texttt{DBSP\_quicklook}$^{5}$ \\
SPRAT$^{6}$    & Liverpool Telescope (LT)
              & LT pipeline$^{7}$ \\
ALFOSC$^{8}$   & NOT$^{9}$
              & IRAF$^{10}$; \texttt{PyNOT}$^{11}$ \\
LRIS$^{12}$    & Keck I
              & \texttt{LPipe}$^{13}$ \\
DIS$^{14}$     & APO$^{15}$
              & IRAF$^{10}$ \\
ISIS$^{16}$    & WHT$^{17}$
              & IRAF$^{10}$ \\
Kast$^{18}$    & Shane, 3 m
              & IRAF$^{10}$; archival \\
EFOSC2$^{19}$  & NTT$^{20}$, 3.58 m
              & PESSTO pipeline$^{21}$ \\
SNIFS$^{22}$   & UH88$^{23}$
              & Archival \\
Binospec$^{24}$ & MMT$^{25}$
               & Archival \\
Blue Channel  & MMT
              & Archival \\
FLOYDS        & LCO$^{26}$, 2 m
              & Archival \\
AFOSC$^{27}$   & Ekar, 1.82 m
              & Archival \\
DOLORES$^{28}$ & TNG$^{29}$, 3.58 m
              & Archival \\
FORS2$^{30}$   & Very Large Telescope (VLT)
              & Archival \\
X-shooter$^{31}$ & VLT
                & Archival \\
\hline
\end{tabular}
\par\smallskip
\begin{minipage}{\linewidth}
\footnotesize
\raggedright
$^{1}$ Spectral Energy Distribution Machine
\citep{Blagorodnova_2018_SEDM, Kim_2022_SEDM}\par
$^{2}$ \citet{Rigault_2019_pysedm}\par
$^{3}$ Double Beam Spectrograph
\citep{Oke_1983_DBSP}\par
$^{4}$ \texttt{DBSP\_DRP}
\citep{Roberson_2021_DBSPDRP}, based on
\texttt{PypeIt} \citep{Prochaska_2020_PypeIt}\par
$^{5}$ MATLAB-based pipeline
(Fremling et al., in prep.)\par
$^{6}$ SPectrograph for the Rapid Acquisition of Transients
\citep{Piascik_2014_SPRAT},
on the Liverpool Telescope \citep{Steele_2004_LT}\par
$^{7}$ Adapted from the FRODOSpec pipeline
\citep{Barnsley_2012_LT}\par
$^{8}$ Alhambra Faint Object Spectrograph and Camera\par
$^{9}$ Nordic Optical Telescope\par
$^{10}$ \citet{Tody_1986_IRAF, Tody_1993_IRAF}\par
$^{11}$ \url{https://github.com/jkrogager/PyNOT}\par
$^{12}$ Low Resolution Imaging Spectrometer
\citep{Oke_1995_Keck}\par
$^{13}$ \citet{Perley_2019_LPipe}\par
$^{14}$ Dual Imaging Spectrograph\par
$^{15}$ Astrophysical Research Consortium 3.5-m telescope
at Apache Point Observatory\par
$^{16}$ Intermediate-dispersion Spectrograph and Imaging System\par
$^{17}$ William Herschel Telescope, 4.2 m\par
$^{18}$ Kast Double Spectrograph\par
$^{19}$ ESO Faint Object Spectrograph and Camera v2
\citep{Wiersema_2018_EFOSC2}\par
$^{20}$ New Technology Telescope\par
$^{21}$ \url{https://github.com/svalenti/pessto}\par
$^{22}$ SuperNova Integral Field Spectrograph
\citep{Lantz_2004_SNIFS}\par
$^{23}$ University of Hawaii 88-inch telescope\par
$^{24}$ \citet{Fabricant_2019_Binospec}\par
$^{25}$ Multiple Mirror Telescope, 6.5 m\par
$^{26}$ Las Cumbres Observatory\par
$^{27}$ Asiago Faint Object Spectrograph and Camera\par
$^{28}$ Device Optimized for the LOw RESolution\par
$^{29}$ Telescopio Nazionale Galileo\par
$^{30}$ FOcal Reducer and low dispersion Spectrograph 2\par
$^{31}$ \citet{Vernet_2011_XShooter}\par
\end{minipage}
\end{table}

\subsection{Calibration} \label{sec:calibration}
We use ZTF photometry to place the spectra in our sample on an absolute flux scale. Spectra obtained before the first ZTF photometric detection or after the last detection of a given object cannot be reliably calibrated in this way. In total, we are able to absolute-calibrate 287 of the total 304 spectra. The calibration procedure is described below. 

We first collect the ZTF light curves of the sample from \citet{Pessi_2025_SLSNeIISample}. For each SN, we jointly interpolate the observed ZTF $gri$ light curves using a two-dimensional Gaussian process (GP)\footnote{\citet{Pessi_2025_SLSNeIISample} have already performed a 1-D GP to extract the light curve characteristics. Here, we use 2-D GP because the calibration requires a more accurate interpolation across both time and wavelength.} implemented with \textsc{george} \citep{Ambikasaran_2015_George}. If a band has fewer than four epochs of observations, we exclude it to avoid poor constraints and biases for other bands. We fit the apparent magnitudes as a function of observer-frame MJD and filters' effective wavelength. Our default covariance function is the product of a Mat\'ern-$5/2$ kernel in time and a squared-exponential kernel in wavelength. We interpolate the magnitude on a common observer-frame grid with one-day spacing. We define the phase of each object relative to the observed peak of the ZTF $r$-band light curve, as tabulated in \citet{Pessi_2025_SLSNeIISample}. If the $r$-band peak cannot be determined, we use the $g$-band peak, or the $i$-band peak if necessary. For objects with multi-peaked light curves, we adopt the first peak (e.g. SN\,2018dfa in our sample, \cref{sec:Ia-CSM}) for phase calculation. Unless otherwise stated, all phases quoted in this work refer to rest-frame days relative to the photometric peak, calculated as $(t-t_{\rm peak})/(1+z)$.

Now we scale each spectrum relative to these GP-interpolated light curves. For each spectrum, we compute synthetic AB magnitudes $m_{\rm syn}$ in the ZTF $gri$ bands using the filter transmission curves from the SVO Filter Profile Service\footnote{\url{https://svo2.cab.inta-csic.es/theory/fps/}}. We compare each synthetic magnitude, $m_{\rm syn}$, with the GP-interpolated apparent magnitude, $m_{\rm GP}$, at the nearest GP grid point within one day of the observation. The flux scale factor for that band is $10^{-0.4(m_{\rm GP}-m_{\rm syn})}$. If several bands are available, we combine their scale factors using an inverse-variance-weighted mean. We then multiply the entire spectrum by the resulting factor.

For the DBSP spectra, we find that the blackbody colours inferred from the spectra are inconsistent with those derived from spectra obtained with other instruments at neighbouring epochs, despite our best reduction attempt using existing pipelines. We attribute these discrepancies to systematic effects associated with the instrument or reduction pipelines. To mitigate this, we recalibrate the affected DBSP spectra using a wavelength-dependent scaling. The scaling is chosen such that the resulting spectral colour matches the weighted mean blackbody temperature inferred from the nearest spectra obtained before and after the DBSP observation.

Finally, all spectra were corrected for Milky Way extinction using the $A_V$ values reported for each SN by \citet{Pessi_2025_SLSNeIISample}, assuming $R_V = 3.1$ and the extinction law of \citet{Fitzpatrick_1999_MWExtinction}. We do not correct for host-galaxy extinction, as we do not have a homogeneous way to estimate it for the full sample. Therefore, the intrinsic luminosities of individual objects may be underestimated, but this does not affect the main analysis presented in this work.


\section{Analysis} \label{sec:analysis}
The 106 H-rich SLSNe in the parent sample do not form a homogeneous spectroscopic class. Of these, 105 exhibit narrow SN-associated Balmer emission lines and are therefore spectroscopically consistent with SLSNe-IIn. The bias towards narrow lines in our sample is not surprising, since the parent sample excluded broad-lined objects in K22. Six of these 105 objects show additional spectral characteristics resembling those of SNe Ia-CSM, although the available data neither establish a thermonuclear origin nor rule out a core-collapse origin (see \Cref{sec:Ia-CSM}). Because of this ambiguity, we keep the six objects in our SLSN-IIn sample, but we flag them as a subgroup having Ia-CSM spectral characteristics. The remaining one object, SN\,2022lvm, exhibits only broad SN-associated Balmer emission in the available spectra and is consistent with the broad-lined SLSN-II class defined by K22 (see \Cref{sec:SLSN-II}). The measured properties of all 304 spectra are tabulated in Appendix~\ref{app:B} and the supplementary material.

\subsection{Spectroscopic evolution of the SLSN-IIn sample}

\subsubsection{Pre-peak spectra} \label{sec:early}
\begin{figure*}
    \centering
    \includegraphics[width=\linewidth]{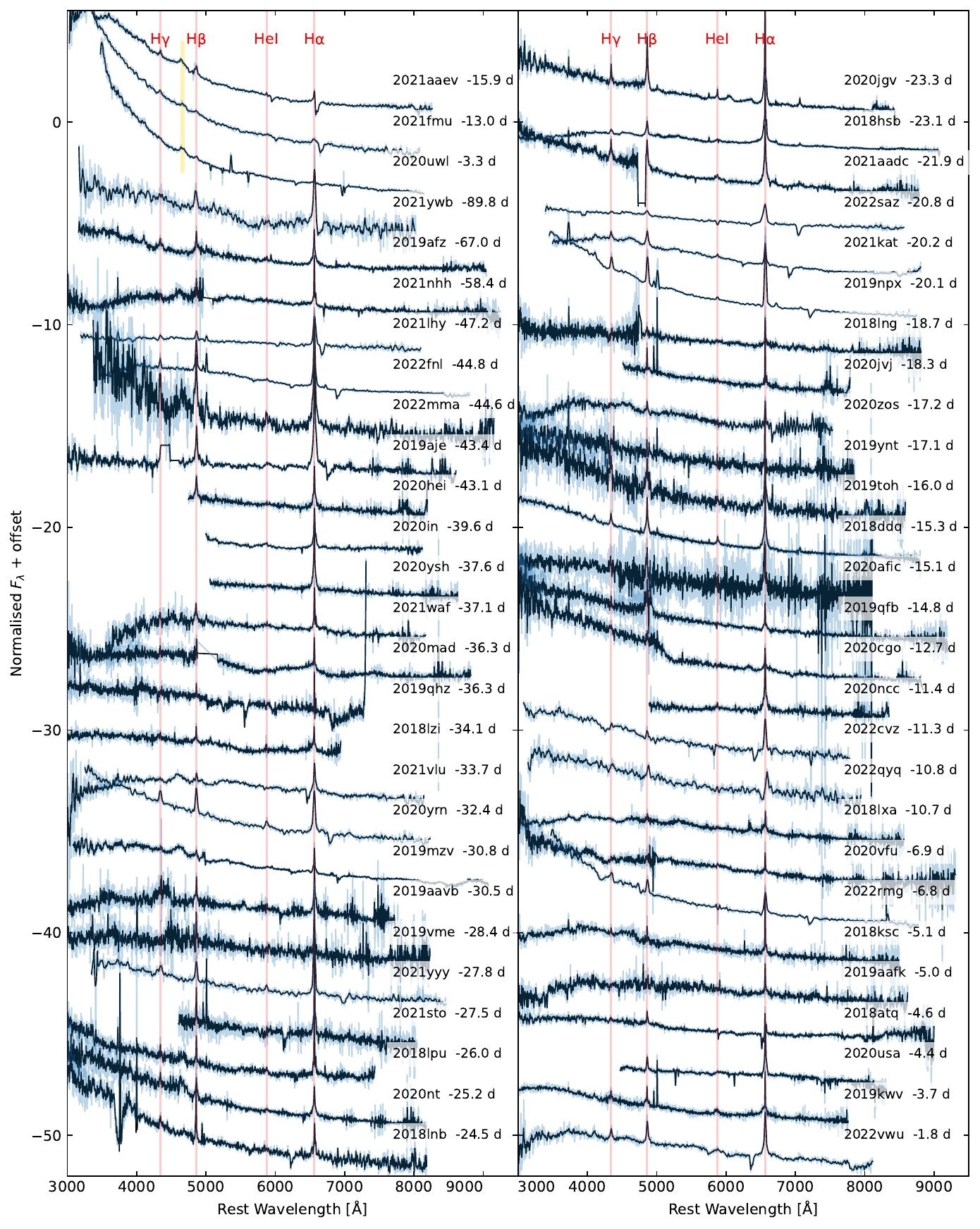}
    \caption{First non-SEDM/SPRAT spectra of all objects in the SLSN-IIn subset with pre-peak spectra (54 spectra for 54 objects). Spectra are dereddened according to the $A_V$ values reported for each SN by \citet{Pessi_2025_SLSNeIISample}. The first three spectra are significantly bluer than the others at comparable phases and also show high-ionisation features marked in gold; they are placed at the top. The remaining spectra are ordered by increasing rest-frame phase from top to bottom. Narrow emission lines are marked in red.}
    \label{fig:sample_early_spectra}
\end{figure*}

We first describe the general spectroscopic evolution of the 105 SLSNe-IIn, starting from their pre-peak spectra. In total, 73 out of the 105 objects in this subset have pre-peak spectra; 54 have non-SEDM/SPRAT pre-peak spectra. For visual clarity, \Cref{fig:sample_early_spectra} shows the first spectrum obtained before rest-frame $0$\,d with respect to $r$-band peak for each object. We exclude SEDM and SPRAT spectra from this figure because their resolution is typically insufficient to resolve narrow lines seen in our sample, but they are available in the supplementary material.

In the pre-peak spectra, most objects show a blue continuum with strong, narrow (FWHM $\lesssim1000\,\kms$) and exponential Balmer lines. Such features are common in H-rich interacting SNe and are generally attributed to interaction between the SN ejecta and dense H-rich CSM. Radiation from the shock interaction ionises the slowly moving unshocked CSM, producing narrow recombination lines, while electron scattering in the optically thick CSM broadens the line wings. This is similar to the case of ordinary Type IIn SNe, but contrasts with SLSNe-II, which generally show blue, nearly featureless spectra \citep{Kangas_2022_SLSNeIISample}. A fraction of objects, including e.g. SNe\,2021aadc, 2022fnl, 2019aje, 2020in, 2018hsb, 2021kat, 2018ddq and 2020usa in \Cref{fig:sample_early_spectra}, also show He\,I $\lambda5876$ before peak with comparable widths, indicating that He is present and sufficiently excited in the line-forming region.

\begin{figure}
    \centering
    \includegraphics[width=\linewidth]{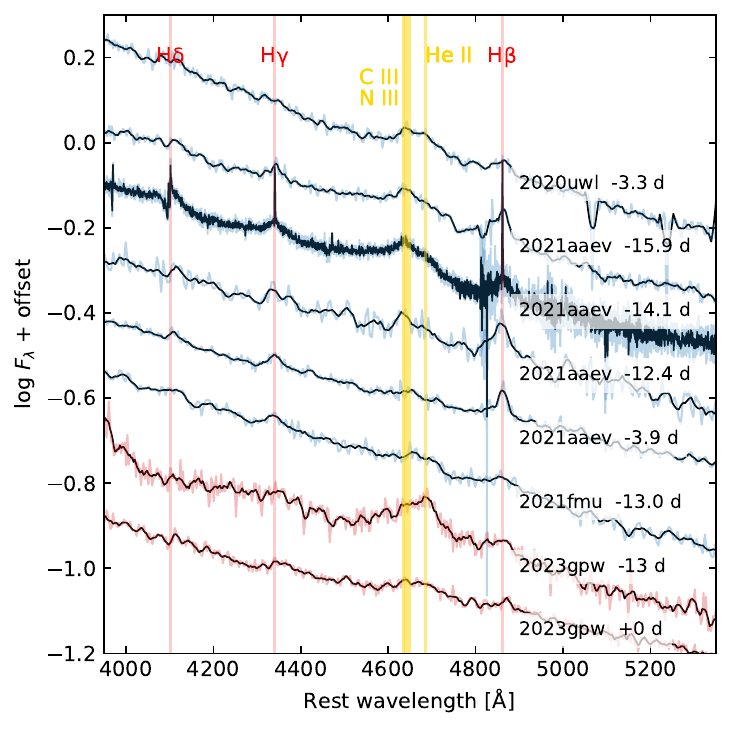}
    \caption{Objects in the sample showing high-ionisation lines of He\,II $\lambda4686$, C\,III $\lambda4649$ and N\,III $\lambda\lambda4634,4641$, plotted in blue. The spectra for SN\,2023gpw are from \citet{Kangas_2026_2023gpw} and are plotted in red.}
    \label{fig:sample_flash_spectra}
\end{figure}

Among the 73 objects with pre-peak spectra, we identify three objects with high-ionisation features of He\,II $\lambda4686$, N\,III $\lambda\lambda4634,4641$, C\,III $\lambda4649$, or blends of these transitions: SNe\,2021aaev, 2020uwl, and 2021fmu. These spectra are shown at the top of \Cref{fig:sample_flash_spectra} and the features are marked in gold. They are also significantly bluer than the other spectra at comparable phases, with blackbody temperatures in the range of 15,000--30,000\,K. High-ionisation features can also appear as narrow emission lines in very early spectra of ordinarily bright SNe and typically only last for a few hours to a few days after explosion \citep{Gal-Yam_2014_Flash, Bruch_2023_Flash}. Because of this, these features are called `flash' features in ordinary SNe and they trace dense, confined CSM close to the progenitor \citep[e.g.][]{Gal-Yam_2014_Flash, Khazov_2016_Flash, Bruch_2023_Flash}. The observed fraction of objects showing such high-ionisation features in our sample is lower than the $30$--$40\%$ estimated for ordinary H-rich SNe in \citet{Bruch_2023_Flash} or the $23.5\%$ estimated in \citet{Ertini_2026_TypeII}. This difference is likely driven by observational selection effects. Detecting these features requires very early spectroscopy, whereas SLSN candidates are often identified only after their light curves show a slow rise and high luminosity \citep{Kangas_2022_SLSNeIISample, Chen_2023_SLSNeI, Russeil_2026_NOMAI}. By that time, these short-lived features may already have disappeared.

SN\,2021aaev shows a two-component emission blend between H$\beta$ and H$\gamma$ in the $-14.1$\,d spectrum\footnote{The phases of SN\,2021aaev reported in this work differ slightly from \citet{Hu_2026_2021aaev}, as they adopted the ATLAS $o$-band maximum rather than the ZTF $r$-band maximum used here.} in \Cref{fig:sample_flash_spectra}, with a stronger component near 4640\,\AA\ and a weaker component near He\,II $\lambda4686$. The stronger component is attributed to C\,III/N\,III, producing a ledge-like profile \citep[e.g.][]{Andrews_2019_2017gmr} that may represent a later stage of the high-ionisation spectral evolution \citep{Hu_2026_2021aaev}. A detailed analysis by \citet{Hu_2026_2021aaev} estimates an emission duration of approximately 24\,d using the midpoint prescription described in the paper\footnote{Flash duration is measured between the midpoint of first light and the first flash spectrum, and the midpoint of the last flash spectrum and the first non-flash spectrum.}. SN\,2020uwl shows C\,III/N\,III and He\,II emission of comparable strength in its $-3.3$\,d spectrum, obtained 3.3 rest-frame days after first detection. SN\,2021fmu shows weaker, tentative high-ionisation features in its $-13.0$\,d spectrum, obtained 18.2 rest-frame days after first detection. For comparison, \Cref{fig:sample_flash_spectra} also includes the SLSN-II SN\,2023gpw \citep{Kangas_2026_2023gpw}. Its high-ionisation features are detected from approximately $-13$\,d to maximum light. This gives an observed detection span of approximately 13\,d, or 24\,d using the mid-point prescription \citep{Hu_2026_2021aaev}. Unlike SN\,2021aaev, its broad emission blend is dominated by He\,II rather than C\,III/N\,III. Similar high-ionisation spectral evolution has recently been reported for the SLSN-IIn SN\,2019vxm \citep{Rameshan_2026_2019vxm, Cai_2026_2019vxm}.

Although the sample of SLSNe showing high-ionisation features remains small, comparisons with ordinary H-rich SNe suggest a connection between peak brightness and the persistence of these features. \citet{Bruch_2023_Flash} and \citet{Hu_2026_2021aaev} find that events with longer flash timescales tend to reach brighter peaks. Furthermore, our study provides SN\,2021fmu as another example of high-ionisation emission at a relatively late epoch. One possible interpretation is that more luminous events sustain an ionising radiation field for longer. More extended dense CSM could also contribute to prolonged emission, allowing high-ionisation features to emerge beyond the first few hours to days.

Other features appearing at this phase for a minority of objects are broad features around 4500--4800\,\AA\ and 5000--5500\,\AA. Examples include SNe\,2021nhh ($-58.4$\,d), 2021vlu ($-33.7$\,d), 2019mzv ($-30.8$\,d), 2021kat ($-20.2$\,d), as shown in \Cref{fig:sample_early_spectra}. Although the 4500--4800\,\AA\ region overlaps with the high-ionisation feature discussed in the previous paragraphs, we identify them as Fe\,II-dominated line complex because their profiles are broad and non-exponential, and they are always accompanied by the second complex at 5000--5500\,\AA. This means a cooler or lower-ionisation photospheric component is already contributing to the spectrum before their optical peak. This may indicate that in some objects, the optically thick CSM interaction is less dominant, so that features from the underlying ejecta/photosphere are starting to emerge.

\subsubsection{Post-peak spectra}

\begin{figure*}
    \centering
    \includegraphics[width=\linewidth]{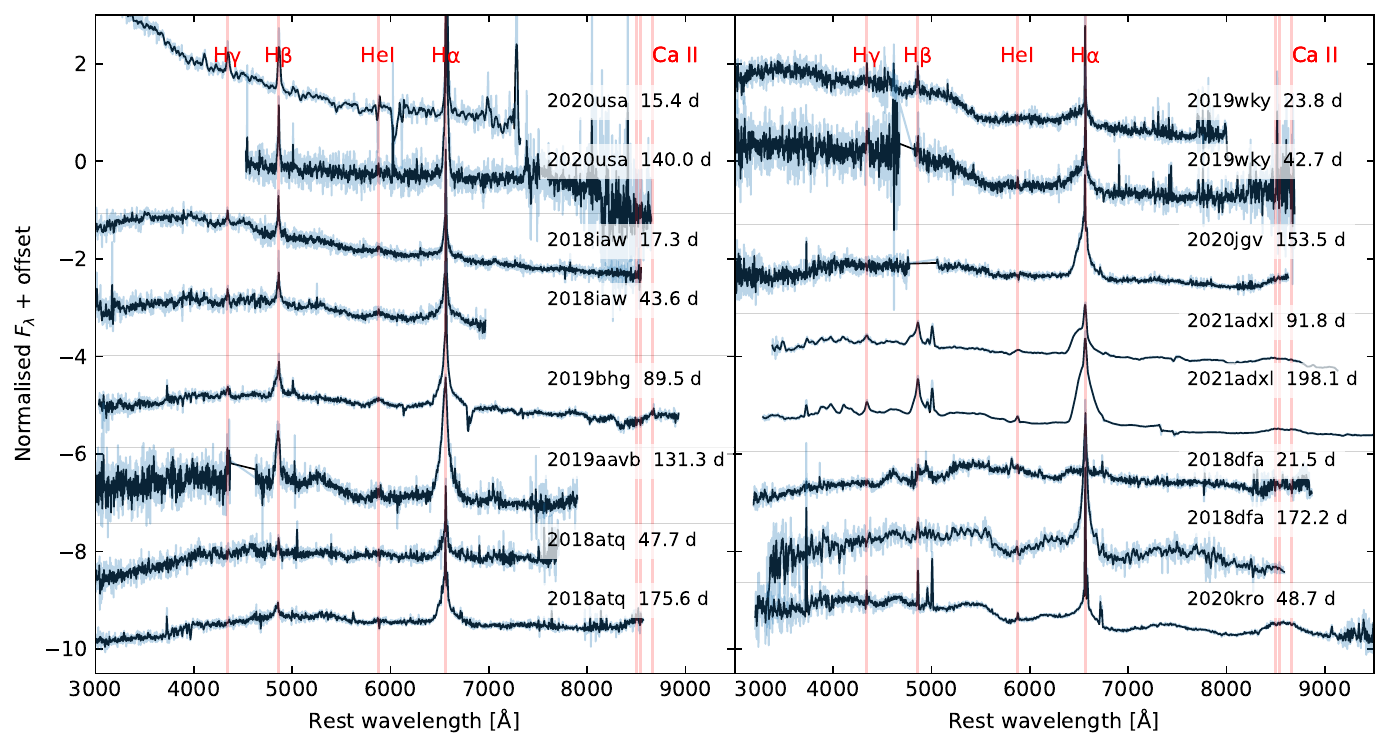}
    \caption{Selected rest-frame spectra of ten objects obtained between peak and $+200$\,d, chosen to illustrate the diversity of post-peak spectral evolution. The  spectra are dereddened using the $A_V$ values reported by \citet{Pessi_2025_SLSNeIISample}, normalised, and vertically offset. Light blue curves show the observed spectra and dark curves show their smoothed versions. Narrow emission lines are marked in red, and horizontal grey lines separate different objects.}
    \label{fig:sample_post_peak_spectra}
\end{figure*}

\begin{figure}[h!]
    \centering
    \includegraphics[width=\linewidth]{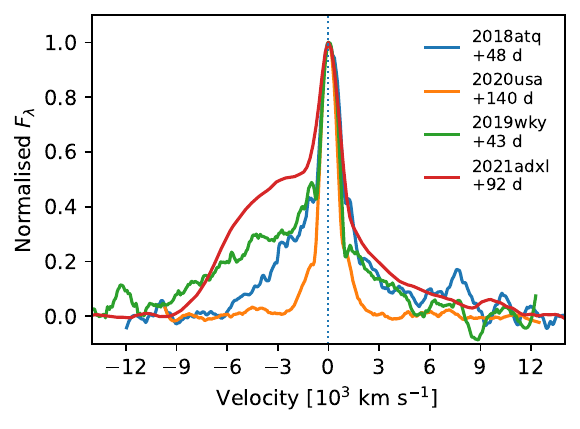}
    \caption{Selected post-peak H$\alpha$ profiles illustrating the diversity of Balmer-line morphologies in the sample. SN\,2020usa shows a narrow profile; SN\,2018atq shows mild blue excess; SN\,2019wky has more extended slanted blue excess; SN\,2021adxl has a narrow core and a broad, relatively flat-topped blue shoulder.}
    \label{fig:halpha_post_peak}
\end{figure}

As the SNe evolve past peak, the Balmer lines generally persist, but the spectra become less uniformly blue and increasingly diverse. We identify 62 objects with spectra obtained between peak and 200\,d, where spectra beyond 200\,d are defined as late-time in this work. For visual clarity, \Cref{fig:sample_post_peak_spectra} shows 16 selected spectra of ten objects between peak and $+200$\,d, chosen to illustrate the range of observed morphologies. We show one or two spectra per object depending on the available coverage. 

Broad, Fe-dominated complexes at 4500--4800\,\AA\ and 5000--5500\,\AA, which appear only in a minority of the pre-peak spectra, become more common and prominent after peak. These features are particularly pronounced in SNe\,2018dfa and 2020kro, and are also accompanied by Fe-dominated complexes between 7000--7700\,\AA. We argue in \Cref{sec:Ia-CSM} that these objects with strong Fe complexes and spectral morphology resembling SNe\,2018dfa and 2020kro exhibit Ia-CSM-like characteristics. However, Fe-dominated complexes are not ubiquitous. For example, they are not clearly detected in SN\,2020usa in \Cref{fig:sample_post_peak_spectra}. When the wavelength coverage is sufficient, most objects also show the Ca\,II $\lambda\lambda8498,8542,8662$ triplet. The increasing prominence of these metal-line features is consistent with a growing contribution from cooler, lower-ionisation material and possibly the underlying SN ejecta as the optically thick CSM interaction evolves.

The post-peak H$\alpha$ profiles also show substantial diversity. Based on a visual assessment of the spectra between peak and $+200$\,d, we can assess their morphology for 45 of the 62 objects, while the remaining 17 are ambiguous because of low SNR or insufficient spectral resolution. Of these 45 objects, 29 show narrow H$\alpha$ emission without a clear blue excess ($64\%$). For example, SNe\,2020usa, 2018iaw and 2019bhg in \Cref{fig:sample_post_peak_spectra} retain a narrow, approximately symmetric profile, with wings limited to velocities of $-3000\,\kms$ (see \Cref{fig:halpha_post_peak} for SN\,2020usa). A fraction of objects with sufficient spectral quality develop multi-component H$\alpha$ profiles, consisting of a narrow core plus a blue excess of varying strength. About 10 of the 45 objects show a very mild blue excess ($22\%$), such as SN\,2018atq at 48\,d in \Cref{fig:halpha_post_peak} where the blue wing extends to $-6000\,\kms$ and the red wing extends to between 3000--6000\,$\kms$. Five objects ($11\%$), exemplified by SN\,2019wky at $43$\,d, develop a more extended, slanted blue excess reaching $\sim-10000\,\kms$. One object, SN\,2021adxl, shows a broad, relatively flat-topped structure extending beyond $-9000\,\kms$, often referred to as a blue shoulder \citep[e.g.][]{Taddia_2020_2013L, Brennan_2024_2021adxl}. These morphologies appear to form a continuous sequence, ranging from approximately symmetric profiles to increasingly prominent emission on the blue side. A blue excess can trace emission from fast-moving shocked material, with the corresponding redshifted emission suppressed by occultation or optical-depth effects \citep{Taddia_2020_2013L, Brennan_2024_2021adxl}, although an intrinsically asymmetric CSM distribution may produce a similar profile \citep[e.g.][]{Andrews_2017_2013L, Nagao_2025_2021adxl, Reynolds_2025_2021irp}.

\subsubsection{Late-time spectra}
\begin{figure*}
    \centering
    \includegraphics[width=\linewidth]{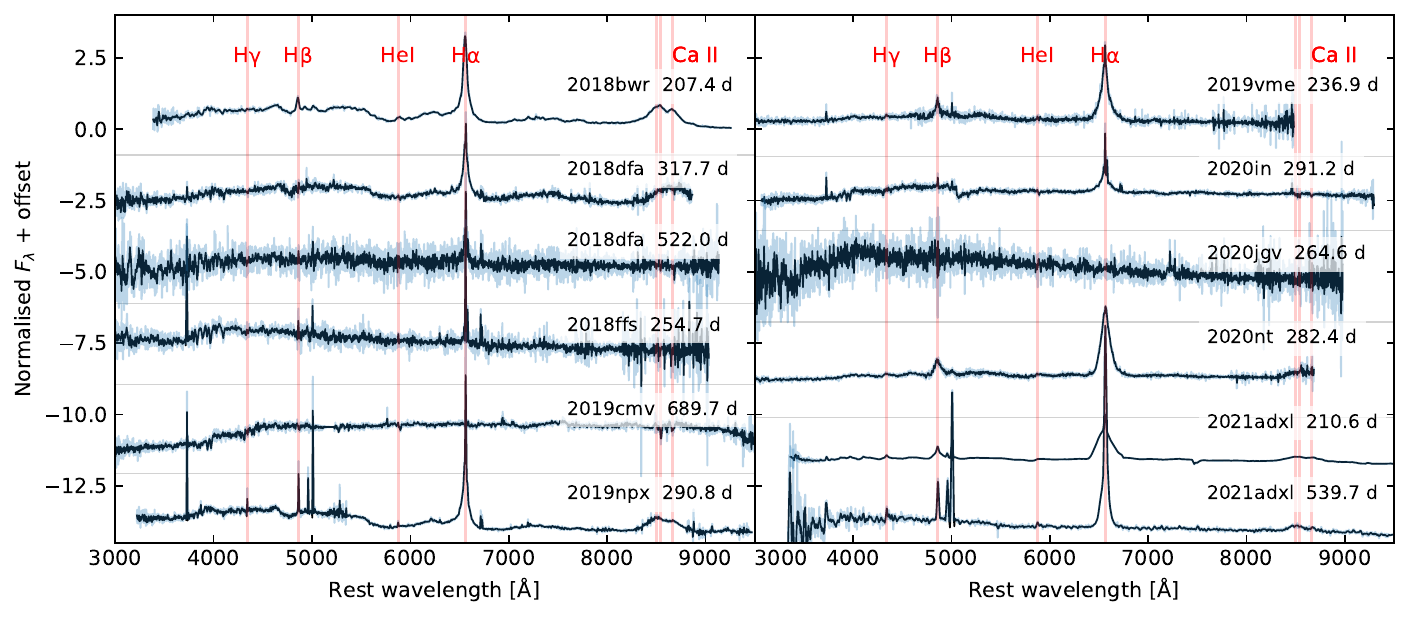}
    \caption{Late-time non-SEDM/SPRAT spectra of 10 objects in the sample, one or two from each object selected for visual clarity.}
    \label{fig:sample_late_spectra}
\end{figure*}

We define spectra beyond 200\,d as late-time spectra in this work, and 10 objects in the sample have spectral coverage at these epochs. The late-time spectra of SNe\,2018ffs, 2019cmv, and 2020jgv show no clear SN-associated emission lines, and in addition SNe\,2018ffs and 2019cmv show narrow emission lines from their host galaxies. The spectra of SN\,2018ffs will also be presented in the forthcoming ePESSTO+ data-release paper by Aryan et al. (in prep.). Among the objects with detectable SN emission, the spectra are typically dominated by prominent H$\alpha$ emission, while H$\beta$ is much weaker or, in some cases, barely detected. The H$\alpha$ profiles remain diverse. Unlike in typical nebular-phase Type II SNe, or SLSNe-II in K22, the line is not always dominated by broad emission components. SN\,2020in, for example, retains a relatively narrow H$\alpha$ profile with exponential wings. The persistence of narrow Balmer emission and interaction-related line wings indicates that CSM interaction continues to dominate the optical spectra of at least some objects beyond $+200$\,d. 

When the wavelength coverage is sufficient, we commonly detect the Ca\,II near-infrared triplet. In contrast, we do not find any object with significant [O\,I] $\lambda\lambda6300,6364$ emission, a common nebular-phase diagnostic in core-collapse SNe \citep{Jerkstrand_2014_NebularSpectra}. This resembles the findings for SLSNe-II in K22. However, the absence of obvious [O\,I] does not by itself distinguish between thermonuclear and core-collapse origins, as continued CSM interaction can veil emission from the inner ejecta. It should be noted that some luminous SNe IIn, such as SN\,2013L, do develop detectable [O\,I] at comparable phases \citep{Taddia_2020_2013L}.

Several objects show a broad feature spanning approximately 6000--6300\,\AA\ which lies blueward of [O\,I] $\lambda6300$ (see \Cref{fig:sample_late_spectra}). The clearest examples are SNe\,2018bwr, 2018dfa, and 2019npx, which we argue exhibit Ia-CSM-like characteristics (see \Cref{sec:Ia-CSM}). These spectra also contain strong Fe-dominated complexes at 4500--4800\,\AA, 5000--5500\,\AA\ and approximately 7000--7700\,\AA. The wavelength and morphology of the 6000--6300\,\AA\ feature, together with its occurrence alongside other Fe-rich blends, lead us to favour an Fe-dominated blend over [O\,I] emission. Fe emission near 6248\,\AA\ has also been identified in several SNe Ia-CSM, such as SNe\,2012ca \citep{Inserra_2016_2012ca} and 2020uem \citep{Sharma_2023_IaCSM}. A specific Fe\,II identification was adopted for this feature in SN\,Ia-CSM 2020aeuh by \citet{Tsalapatas_2025_2020aeuh}. 

Overall, H-rich SLSNe with narrow Balmer lines show substantial spectroscopic diversity despite sharing an interaction-powered appearance. Their common feature is the presence of narrow Balmer emission at some phase of evolution, but their line-profile evolution differs significantly from object to object. This diversity may reflect differences in the density, geometry, and radial extent of the CSM and potentially different underlying explosion mechanisms. Understanding their physical origin therefore requires long-term spectroscopic follow-up, especially into the nebular phase when the declining optical depth of the interaction region reveals the underlying ejecta and deeper layers.

\subsection{H-rich SLSN with broad Balmer lines} \label{sec:SLSN-II}

\begin{figure*}
    \centering
    \includegraphics[width=\linewidth]{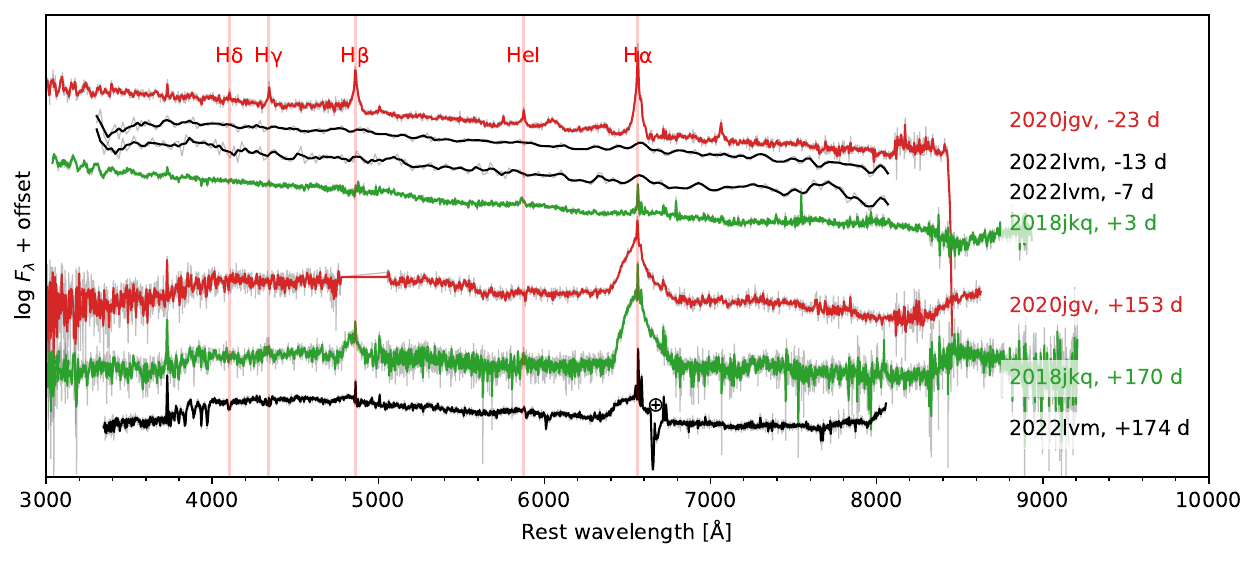}
    \caption{Spectral sequence of SN\,2022lvm shown black in the rest frame. For comparison, SN\,2020jgv is plotted in red and 2018jkq \citep{Kangas_2022_SLSNeIISample} in green. Balmer lines and telluric bands are labelled.}
    \label{fig:2022lvm_spec}
\end{figure*}

\begin{figure}
    \centering
    \includegraphics[width=\linewidth]{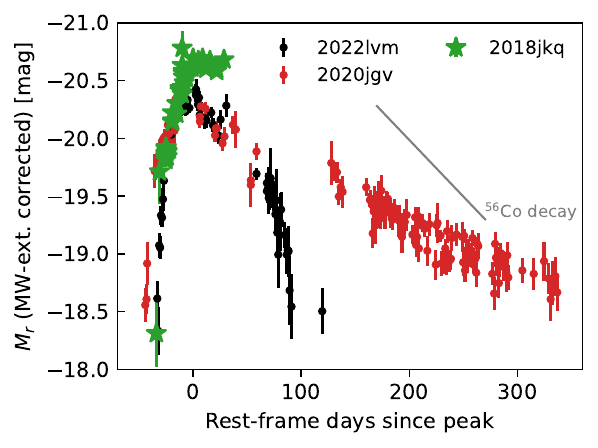}
    \caption{Milky Way-extinction-corrected absolute $r$-band light curves of SN\,2022lvm, the broad-lined SLSN-II SN\,2018jkq, and the SLSN-IIn SN\,2020jgv, shown in rest-frame phase relative to their $r$-band peak. The grey line indicates the decline rate expected from $^{56}$Co decay.}
    \label{fig:2022lvm_phot}
\end{figure}

\begin{figure}
    \centering
    \includegraphics[width=\linewidth]{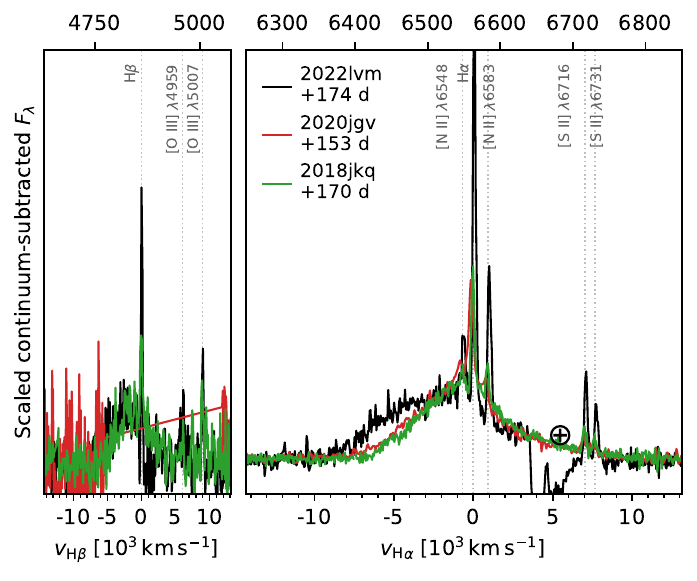}
    \caption{Continuum-subtracted H$\beta$ (left) and H$\alpha$ (right) profiles of SN\,2022lvm, compared with those of the broad-lined SLSN-II 2018jkq and the SLSN-IIn 2020jgv. Each profile is independently scaled by its blue shoulder. Dotted lines mark common host-galaxy emission lines. Telluric absorptions are also marked.}
    \label{fig:2022lvm_Halpha}
\end{figure}

Among the 106 objects in our full sample, SN\,2022lvm is the only event without a securely detected narrow SN component. Its available spectra show only broad (defined as FWHM$\geq5000\kms$) SN-associated Balmer emission. It is therefore most consistent with the broad-lined SLSN-II class described by K22. SN\,2022lvm was initially classified on TNS as a Type II SN \citep{Gomez_2023_2023-01-26}, and its peak $r$-band magnitude of $-20.3$~mag places it in the superluminous regime. We show its spectral sequence in \Cref{fig:2022lvm_spec}, its $r$-band light curve in \Cref{fig:2022lvm_phot}, and zoomed-in views of the Balmer-line profiles in \Cref{fig:2022lvm_Halpha}. For comparison, we include the broad-lined SLSN-II 2018jkq from K22 and SN\,2020jgv from our narrow-lined sample, which develops similarly broad Balmer profiles at late phases despite exhibiting clear IIn-like spectra before peak. 

The three objects occupy a similar peak luminosity range, although SN\,2018jkq is somewhat brighter, reaching $M_r\simeq-20.7$\,mag and remaining near this luminosity for the first $\sim30$\,d after maximum. SNe\,2022lvm and 2020jgv reach peak luminosities of $M_r\simeq-20.4$, but show different post-peak evolution. SN\,2022lvm fades rapidly after $\sim+50$\,d, whereas SN\,2020jgv remains detectable beyond $+300$\,d and declines more slowly than the $^{56}$Co decay rate at late phases, consistent with continued CSM interaction. The overlapping peak luminosities and early light-curve evolution show that photometry alone does not clearly distinguish broad-lined SLSNe-II from SLSNe-IIn. This is consistent with the findings of \citet{Pessi_2025_SLSNeIISample}.

In terms of spectroscopy, for SN\,2022lvm, the two SEDM spectra obtained before peak do not have sufficient spectral resolution to determine whether the SN-associated Balmer features are narrow or broad. The spectrum obtained at $+174.0$\,d exhibits a broad H$\alpha$ profile extending to velocities of approximately $-8000$ to $-9000\,\kms$ (\Cref{fig:2022lvm_Halpha}). H$\beta$ shows a comparably broad profile in velocity space. The blue excess is greater than that observed in SN\,2018jkq, whose H$\alpha$ profile reaches approximately $-7000\,\kms$. Although the red side of H$\alpha$ is affected by strong telluric absorption, its red wing appears less extended than the blue wing. We identify the narrow emission lines superposed on the spectrum as host-galaxy H\,II-region emission, including H$\alpha$, H$\beta$, [N\,II] $\lambda\lambda6548,6583$, and [S\,II] $\lambda\lambda6716,6731$. The only securely identified SN-associated Balmer components are broad with FWHM$>5000\,\kms$. Following the classification scheme of K22, we therefore classify SN\,2022lvm as a SLSN-II. A another useful literature comparison is PS15br  \citep{Inserra_2018_SLSNeII}, which exhibited a broad, asymmetric H$\alpha$ profile at approximately $+102$\,d. SN\,2022lvm shares these qualitative characteristics. At approximately $+337$\,d, PS15br exhibited a multicomponent H$\alpha$ profile interpreted as evidence for interaction with an asymmetric CSM. Its evolution therefore illustrates that a broad-lined SLSN-II classification does not exclude a contribution from CSM interaction at later phases.

However, the comparison with SN\,2020jgv demonstrates an important limitation of this classification. At $+153$\,d, SN\,2020jgv exhibits Balmer profiles as broad as those of the two broad-lined SLSNe-II, whereas its $-23$\,d spectrum shows clear narrow exponential Balmer profiles characteristic of SNe IIn. No spectra of SN\,2022lvm are available between peak and $+174.0$\,d. We therefore cannot exclude the presence of narrow SN Balmer emission at intermediate phases. This illustrates how sparse spectroscopic coverage can obscure the distinction between SLSNe-II and SLSNe-IIn that develop broad profiles at late phases. 

To calculate the observed fraction of SLSN-II in our sample, we restrict the counting to the 2021--2022 interval to avoid overlap with the K22 sample, which was already excluded in the parent sample. Since we identify only one broad-lined event among 36 H-rich SLSNe within the 2021--2022 interval, this corresponds to a fraction of $2.8\%$. This is lower than the fraction of approximately $20\%$ reported by K22 for 2018--2020. However, given the small number of events and the strong dependence of this classification on spectroscopic phase coverage, the difference does not necessarily indicate a temporal change in the intrinsic fraction of broad-lined events among ZTF H-rich SLSNe.

\subsection{Balmer lines} \label{sec:BL}
Balmer lines are ubiquitous features across our sample. Since the line profiles are highly heterogeneous, we measure the Balmer line properties using a customised manual procedure. For any given spectrum with a confident H$\alpha$ identification, we select two regions of local continuum adjacent to the H$\alpha$ line and fit them with a 1st-order polynomial. In cases where the H$\alpha$ wing is blended with a broad but separable non-H$\alpha$ component, we isolate and extract only the narrower H$\alpha$ profile. Otherwise, we include all narrow and broad components associated with H$\alpha$. Then, we subtract the fitted continuum from the observed spectrum to isolate the line profile. To estimate uncertainties, we combine error propagation from the absolute flux calibration (\cref{sec:calibration}), the intrinsic spectral noise (evaluated via a Monte Carlo approach), and the continuum-placement uncertainty (estimated by repeating the manual selection process three times). From this clean, continuum-subtracted H$\alpha$ profile, we integrate the line flux to get the total H$\alpha$ luminosity, the pseudo-equivalent width (pEW), and fit exponential and/or Gaussian profiles to measure the full width at half-maximum (FWHM). We apply the identical procedure to H$\beta$. Finally, for spectra where both H$\alpha$ and H$\beta$ are reliably identified, we also compute the Balmer decrement. The numerical values of all extracted properties are available in the supplementary material.

\subsubsection{Line luminosity, pEW and FWHM}

\begin{figure*}
    \centering
    \includegraphics[width=\linewidth]{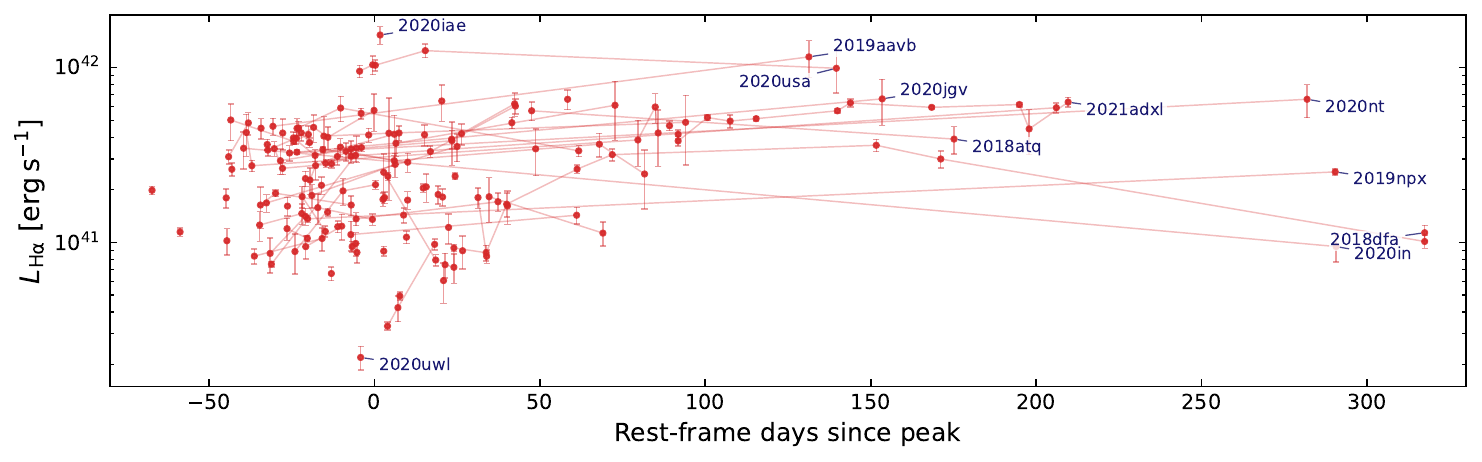}
    \caption{Evolution of the H$\alpha$ luminosity in the full sample. Selected objects are labelled.}
    \label{fig:sample_halpha_luminosity}
\end{figure*}

We present the H$\alpha$ luminosity evolution of the sample in \Cref{fig:sample_halpha_luminosity}, where the individual spectra have been placed on an absolute flux scale according to the procedure in \cref{sec:calibration}. The evolution is highly heterogeneous. The measured luminosities span almost two orders of magnitude, from $2.2\times10^{40}$\,erg\,s$^{-1}$ for SN\,2020uwl to $1.5\times10^{42}$\,erg\,s$^{-1}$ for SN\,2020iae, with a median value of $2.6\times10^{41}$\,erg\,s$^{-1}$. This diversity is expected from the broad range of optical luminosities in the parent sample. Since H$\alpha$ falls within the ZTF $r$ band over much of the redshift range of the sample, the H$\alpha$ luminosity is expected to broadly track the heterogeneous $r$-band luminosity. As shown by \citet{Pessi_2025_SLSNeIISample}, the peak $r$-band absolute magnitudes span from $-19.8$ to $-21.6$\,mag, with a median of $-20.4$\,mag. Physically, H$\alpha$ luminosity can track the strength of ejecta--CSM interaction. The large dispersion in H$\alpha$ luminosity therefore reflects the diversity of the underlying interaction properties across the population.

\begin{figure*}
    \centering
    \includegraphics[width=\linewidth]{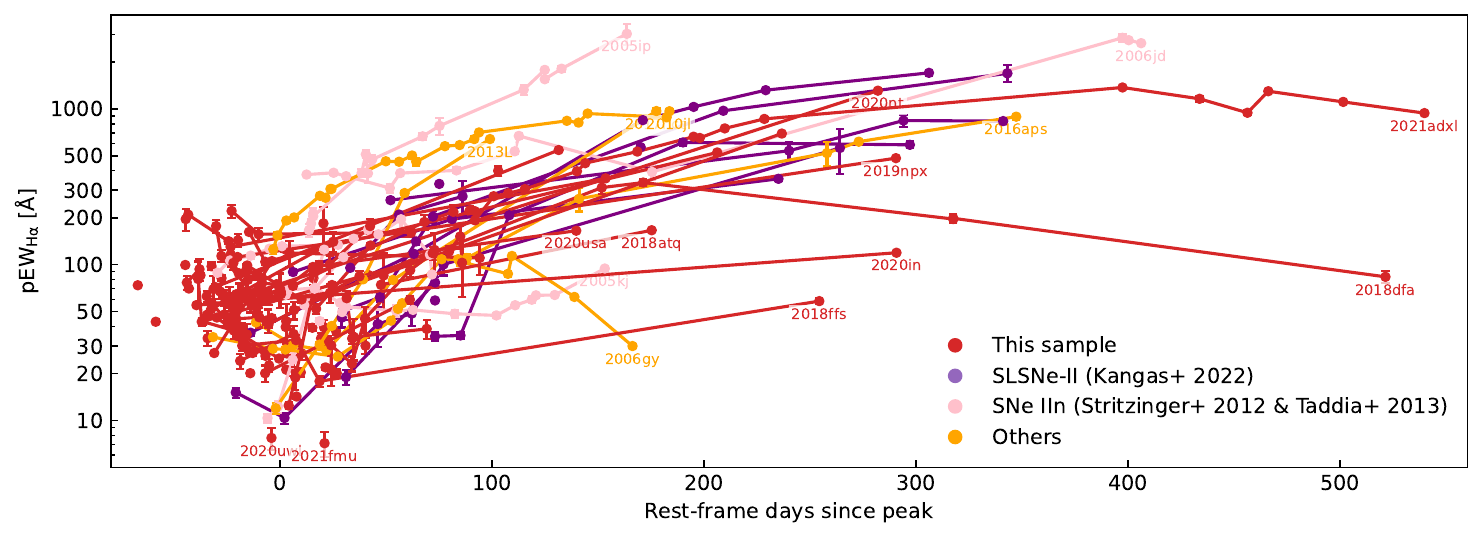}
    \caption{Evolution of the H$\alpha$ pEW, measured relative to the local pseudo-continuum. The full sample is shown in red. Broad-lined SLSNe-II from \citet{Kangas_2022_SLSNeIISample} are shown in purple, SNe IIn from the Carnegie Supernova Project \citep{Stritzinger_2012_Carnegie1,Taddia_2013_Carnegie2} in pink, and selected well-studied SLSNe-IIn in orange.}
    \label{fig:sample_halpha_pEW}
\end{figure*}
\begin{figure*}
    \centering
    \includegraphics[width=\linewidth]{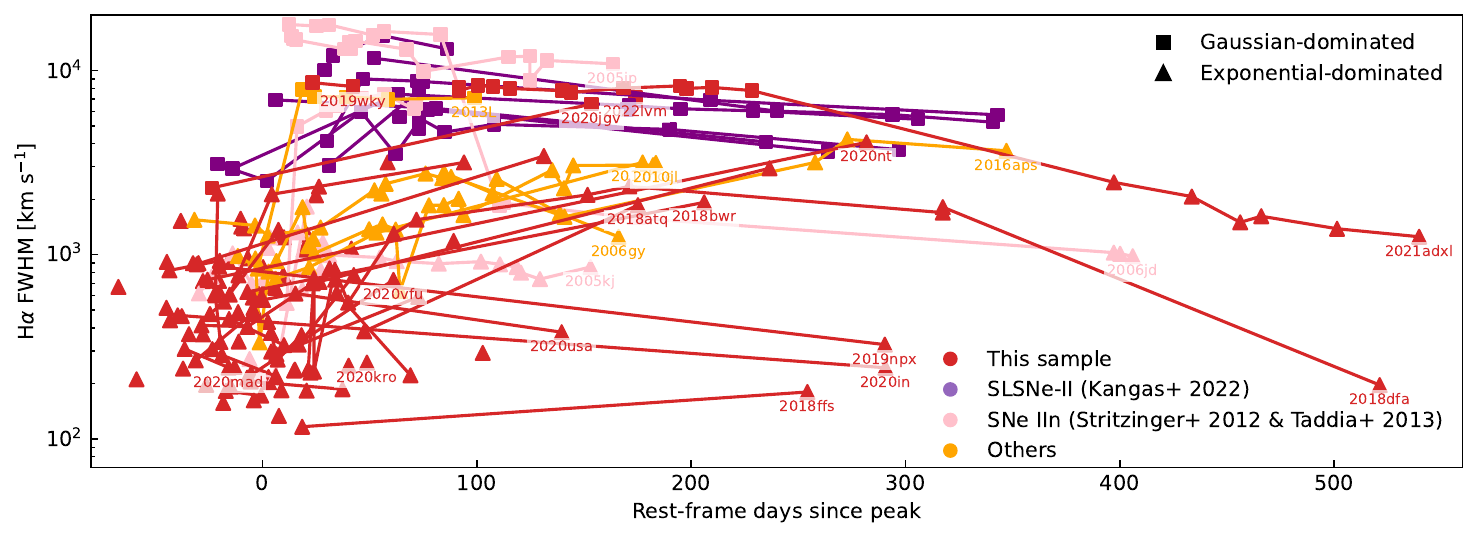}
    \caption{Evolution of the H$\alpha$ FWHM in the full sample and the comparison samples, using the colour scheme of \Cref{fig:sample_halpha_pEW}. For the narrow-lined objects, triangles show the FWHM of the single exponential when that model is preferred, while squares show the Gaussian-component FWHM when the composite exponential-plus-Gaussian model is preferred. The broad-lined SLSNe-II are fitted with a single Gaussian.}
    \label{fig:sample_halpha_FWHM}
\end{figure*}

We also present the H$\alpha$ pEW evolution of the sample in \Cref{fig:sample_halpha_pEW}. The pEW measures the strength of a spectral feature relative to the local pseudo-continuum and is defined as $\mathrm{pEW} =\int_{\lambda_1}^{\lambda_2}(F_{\lambda,\mathrm{H}\alpha}- F_{\lambda,\mathrm{cont}})/{F_{\lambda,\mathrm{cont}}}\,\mathrm{d}\lambda$. The H$\alpha$ pEW generally increases with time. Before and around peak, the pEW is typically of the order of several tens to $\sim200$\,\AA. Beyond 200\,d, the distribution becomes increasingly diverse. In SN\,2021adxl, the pEW approaches $1000$\,\AA, indicating that H$\alpha$ emission becomes increasingly dominant relative to the local continuum as the SN evolves. This is expected if the continuum fades while interaction-powered H$\alpha$ emission remains strong. By contrast, in objects such as SNe\,2018dfa and 2018atq, the pEW decreases to below $100$\,\AA, indicating that H$\alpha$ becomes less dominant relative to the continuum.

Unlike the line luminosity, the pEW is less sensitive to absolute flux-calibration error. Hence, pEW can be calculated over a wider time range compared to H$\alpha$ luminosity. This allows a comparatively homogeneous analysis of literature spectra. We select several comparison samples: 13 broad-lined SLSNe-II from K22; seven SNe IIn from the Carnegie Supernova Project \citep{Stritzinger_2012_Carnegie1,Taddia_2013_Carnegie2}; and four well-studied luminous or superluminous SNe IIn, including SNe\,2006gy \citep{Ofek_2007_2006gy}, 2010jl \citep{Stoll_2011_2010jl, Fransson_2014_2010jl}, 2013L \citep{Taddia_2020_2013L}, and 2017hcc \citep{Moran_2023_2017hcc}. For these comparison objects, we find their public spectra on \textsc{WISeREP} and apply the same manual extraction procedure described in \cref{sec:BL} to measure the H$\alpha$ pEW. The broad-lined SLSN-II comparison sample occupies a relatively narrow space in the pEW evolution plot. Its H$\alpha$ pEW is generally below $100$\,\AA\ around peak and increases to approximately $500$--$1000$\,\AA\ by 300\,d. In contrast, the SNe IIn span a broader range than either the broad-lined SLSNe-II or our sample, despite their smaller sample size. Some objects, including SNe\,2005ip and 2006jd, reach larger pEWs than most of the broad-lined SLSNe-II at late times. These objects exhibit a broad or blue-shoulder component underlying the narrow H$\alpha$ emission, which contributes substantially to the integrated pEW \citep{Taddia_2013_Carnegie2, Taddia_2020_2013L}. By contrast, objects such as SNe\,2005kj and 2006gy maintain comparatively low pEWs. While the H$\alpha$ emission predominantly traces ejecta--CSM interaction, the underlying continuum can contain varying contributions from the photosphere and from interaction luminosity thermalised in optically thick CSM. The broad distribution among SNe IIn and SLSNe-IIn reflects a more heterogeneous source contributing to their optical emission compared to SLSNe-II.

Other than luminosity and pEW, a more line-profile-dependent measurable quantity is the H$\alpha$ FWHM\footnote{We exclude SEDM spectra from the FWHM analysis because their low spectral resolution ($R\sim100$) causes substantial instrumental broadening}. For our sample and the (SL)SN IIn comparison samples, we fit each extracted H$\alpha$ profile with both a single exponential and a composite exponential+Gaussian model, and select the best fit. When the composite model is selected, we adopted the FWHM of the full fitted profile, except four objects with prominent broad or blue-excess components: SNe\,2019wky, 2020jgv, 2021adxl and 2022lvm. For them the FWHM of the Gaussian component is used. This provides a practical measure of the characteristic width of the main resolved component at different phases. At early times, the H$\alpha$ profiles of interacting SNe are shaped by electron scattering in optically thick CSM and can typically be described by a single exponential. At later times, some objects develop a broad component beneath the narrow emission and are better fitted by the composite model. In these cases, the Gaussian component better captures the characteristic width, whereas the exponential component primarily describes the narrow core. For the broad-lined SLSN-II comparison sample, we instead fit each profile with a single Gaussian and adopt its FWHM. We plot the FWHM evolution in \Cref{fig:sample_halpha_FWHM}.

\begin{figure}
    \centering
    \includegraphics[width=\linewidth]{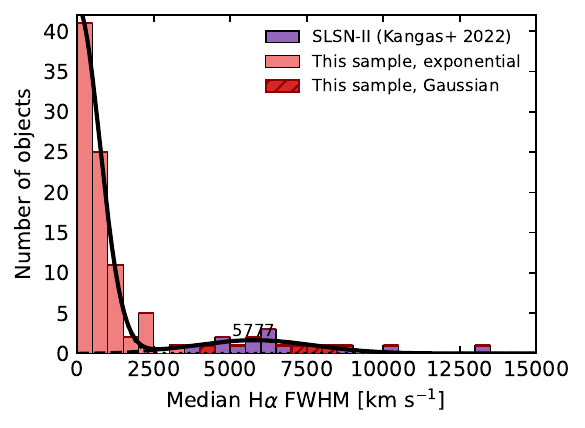}
    \caption{Histogram of the median H$\alpha$ FWHM per object for the full ZTF H-rich SLSN sample until 12 Dec 2022. The stacked colours distinguish the SLSNe-IIn and broad-lined SLSNe-II, while the solid curve shows the fitted two-component Gaussian mixture.}
    \label{fig:FWHM_hist}
\end{figure}

The FWHM evolution also shows a distinction between (SL)SNe IIn and broad-lined SLSNe-II. Most (SL)SNe IIn have FWHM values below $1000\,\kms$ before and around peak, which increase with time but generally remain below $3000\,\kms$. The main exceptions are objects with a blue shoulder. Their FWHM can be at $\sim8000\,\kms$, such as SNe\,2021adxl and 2013L. It can even exceed $10000\,\kms$, as in SN\,2005ip and 2006jd at early times. All broad-lined SLSNe-II develop large FWHM above $5000\,\kms$ after peak. This distinction is more clearly illustrated by the distribution of the median FWHM per object in \Cref{fig:FWHM_hist}. The figure shows the full ZTF H-rich SLSN sample until 12 Dec 2022, which is a combination of our sample with the broad-lined SLSNe-II from \citet{Kangas_2022_SLSNeIISample}. The distribution can be described by a two-component Gaussian mixture: the SLSNe-IIn dominate the lower-width component with the majority below approximately $2500\,\kms$, while the broad-lined SLSNe-II dominate the higher-width component, centred at $5777\,\kms$. This distribution also separates objects with fast bulk motion of material visible from those without.

\subsubsection{Balmer decrement}\label{sec:BD}
\begin{figure*}
    \centering
    \includegraphics[width=\linewidth]{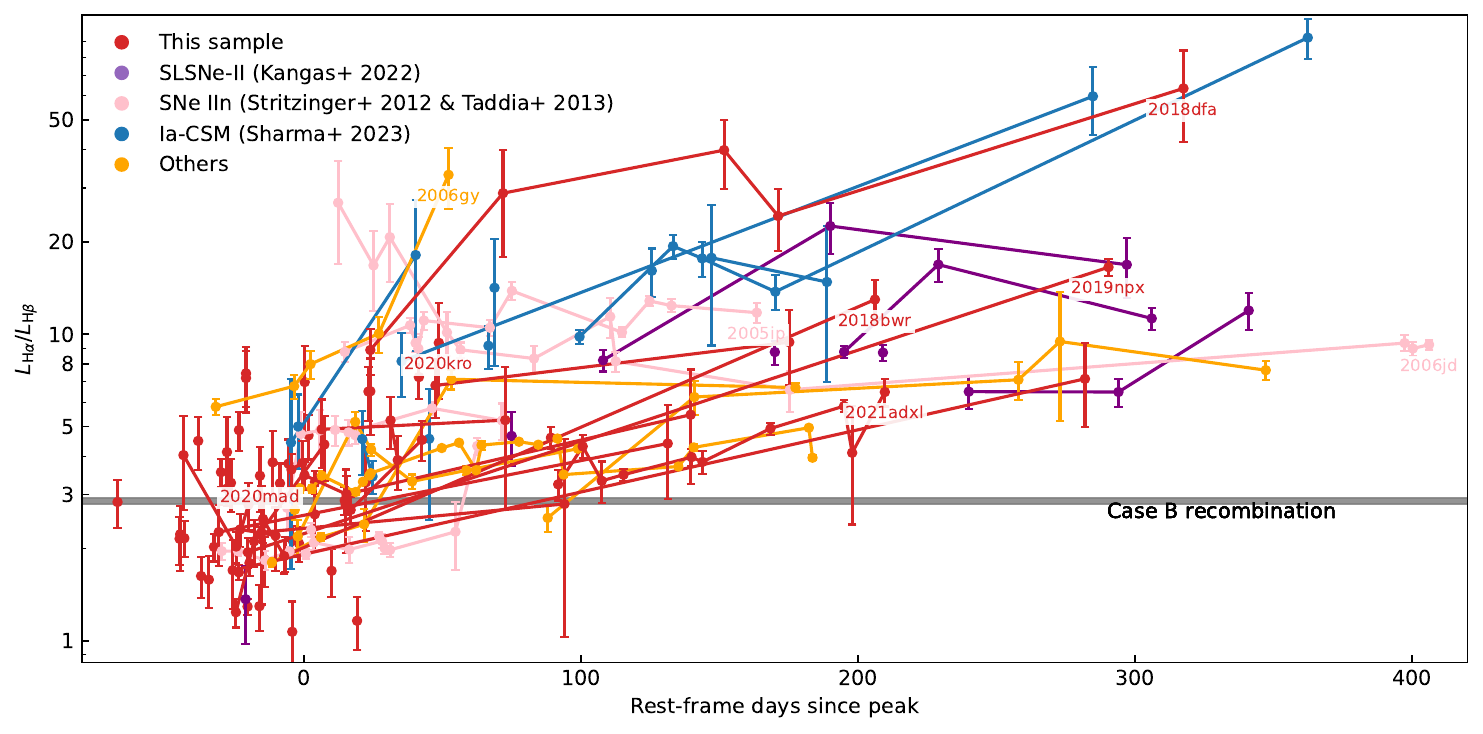}
    \caption{Time evolution of the Balmer decrement for spectra with reliable H$\alpha$ and H$\beta$ profile extractions. Red markers show objects in our sample with multiple spectroscopic epochs, while pink markers show objects with only one epoch. Blue markers show objects from the Type Ia-CSM sample of \citet{Sharma_2023_IaCSM}, and purple markers show objects from the SLSN-II sample of \citet{Kangas_2022_SLSNeIISample}. The Case B recombination value of 2.86 is shown as a grey horizontal band.
}
    \label{fig:BD}
\end{figure*}

The evolution of the Balmer decrement, defined as the total H$\alpha$-to-H$\beta$ flux ratio, is shown in \Cref{fig:BD}. We include both narrow and broad hydrogen components where present. Broad non-hydrogen components underlying the Balmer features are excluded from the measurement, and spectra significantly affected by telluric absorption are also excluded. 

\begin{figure}
    \centering
    \includegraphics[width=\linewidth]{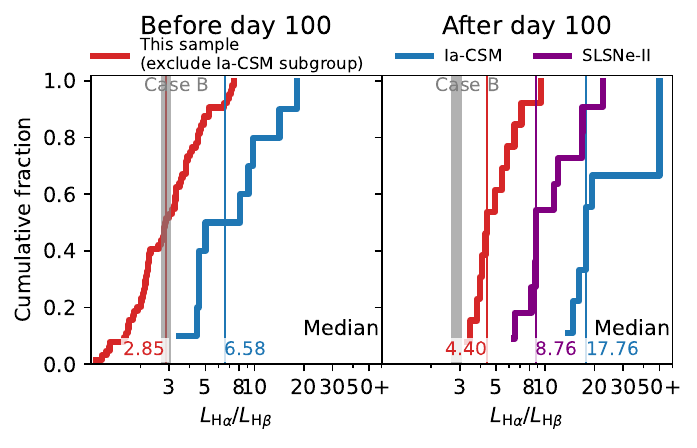}
    \caption{Cumulative distributions of the Balmer decrement for the spectra shown in \Cref{fig:BD}, split into phases before rest-frame day 100 (left) and after day 100 (right). The colour scheme follows \Cref{fig:BD}.}
    \label{fig:CBD}
\end{figure}

Our SLSN-IIn sample shows an overall shift towards larger Balmer decrements with increasing phase, although the dispersion at any given epoch is substantial. In the cumulative distributions shown in \Cref{fig:CBD}, the median increases from 2.85 before $100$\,d to 4.40 after $100$\,d. The early-time median is close to the Case B recombination value of H$\alpha$/H$\beta=2.86$ expected for photoionised gas under typical low-density nebular conditions \citep[e.g.][]{Hummer_1987_recombination, Osterbrock_2006_AGN}. This is broadly consistent with Balmer line emission coming from H recombination in photoionised CSM. However, the early-time distribution shows substantial scatter, with Balmer decrements ranging from $\sim1$ to 9. In particular, a non-negligible fraction of objects have Balmer decrements below the Case B value. Such low values are difficult to explain within the simplest Case B picture. One plausible explanation is optical-depth effects in very dense gas. The dust-free radiative-transfer calculations of \citet{Yan_2026_recombination} show in their figure 2 that Balmer decrements can fall below 2 in two distinct density regimes. For column densities of $N_{\rm H}\sim10^{22}$--$10^{22.5}$\,cm$^{-2}$, these occur at approximately volume density $n_{\rm H}\sim10^{4}$--$10^{7}$\,cm$^{-3}$ and $n_{\rm H}\sim10^{11}$--$10^{12}$\,cm$^{-3}$. The high-density branch approaches the conditions inferred for the opaque inner CSM of SLSNe-IIn 2006gy. \citet{Smith_2010_2006gy} inferred a mass density above $10^{-13}$\,g\,cm$^{-3}$ at $r\sim2\times10^{15}$\,cm. For fully ionised H-rich CSM, this corresponds to $n_{\rm H}\gtrsim4\times10^{10}$\,cm$^{-3}$. This may indicate dense shells or clumps in inner CSM. Low Balmer decrements may therefore constrain combinations of the volume and column densities during the early phases of H-rich SN evolution. We leave a detailed investigation of this behaviour to future work. The late-time increase also indicates a departure from simple Case B recombination. Enhanced Balmer decrements can be produced by collisional excitation, optical-depth effects \citep{Silverman_2013_IaCSM, Sharma_2023_IaCSM}, and differential extinction \citep{Osterbrock_2006_AGN}. These mechanisms cannot be separated using H$\alpha$/H$\beta$ alone. 

To place the Balmer-decrement evolution of our sample in a broader context, we apply the same manual extraction and calculation procedure to the K22 SLSN-II sample, the SN IIn sample from the Carnegie Supernova Project \citep{Stritzinger_2012_Carnegie1,Taddia_2013_Carnegie2} and the four well-studied SNe IIn 2006gy, 2010jl, 2013L, and 2017hcc. In addition, we include the Type Ia-CSM sample of \citet{Sharma_2023_IaCSM} in this analysis. SNe Ia-CSM are thermonuclear SNe interacting with dense H-rich CSM and can therefore exhibit narrow Balmer emission resembling that of SNe IIn. Previous studies have found that SNe Ia-CSM tend to show larger Balmer decrements than SNe IIn \citep{Silverman_2013_IaCSM,Leloudas_2015_IaCSM,Sharma_2023_IaCSM}. We therefore test whether SNe Ia-CSM also exhibit systematically larger Balmer decrements than the SLSNe-IIn in our sample. 

The Balmer-decrement evolutions of all aforementioned samples are compared with our sample in \Cref{fig:BD}. We also compare the cumulative distributions in \Cref{fig:CBD}, but in this cumulative distribution plot we exclude SNe\,2018dfa, 2018bwr, 2019npx, 2020kro, 2020mad, 2020vfu in our sample. We argue that these six objects show Ia-CSM characteristics in \Cref{sec:Ia-CSM} and therefore exclude them. We assess population differences using two-sample Kolmogorov--Smirnov (KS) and Anderson--Darling (AD) tests on one median decrement per object in each phase interval. The available broad-lined SLSNe-II have a late-time median of 8.76, higher than the corresponding value for the SLSN-IIn sample. However, the SLSN-II sample is small and the measurements are sparse, and neither test establishes a significant difference ($p_{\rm KS}=0.36$, $p_{\rm AD}=0.11$). The external SNe Ia-CSM sample has median Balmer decrements of 6.58 before $100$\,d and 17.76 after $100$\,d, compared with 2.85 and 4.40, respectively, for our SLSN-IIn sample. Both tests support a difference between these populations in the earlier interval ($p_{\rm KS}=2.1\times10^{-4}$, $p_{\rm AD}=2.0\times10^{-5}$) and the later interval ($p_{\rm KS}=0.016$, $p_{\rm AD}=0.016$). This difference is consistent with previous comparisons between SNe Ia-CSM and SNe IIn \citep{Silverman_2013_IaCSM,Sharma_2023_IaCSM}, although its physical origin is not uniquely established. Following \citet{Silverman_2013_IaCSM}, \citet{Sharma_2023_IaCSM} note that interaction with thin, dense CSM shells could generate sufficiently high densities and Balmer-line optical depths for collisional excitation and Balmer self-absorption. Thus, one possible interpretation is that the Balmer-line-forming regions of SNe Ia-CSM are, on average, denser and/or more optically thick than those of SLSNe-IIn, although differential extinction and other radiative-transfer effects could also contribute.

\subsection{Trends and subgroups within the sample}
Building on the spectroscopic properties presented above and the light-curve properties reported by \citet{Pessi_2025_SLSNeIISample}, we examine the diversity among our full sample, focusing on the evolution of their Balmer-line profiles and possible associations with their photometric properties. Six objects exhibit additional spectral characteristics resembling those of SNe Ia-CSM and are treated as a possible Ia-CSM/IIn subgroup later in \Cref{sec:Ia-CSM}. The remaining 99 narrow-lined objects show no comparable evidence for a thermonuclear origin. Among these 99 objects, the post-peak H$\alpha$ profiles do not appear to separate cleanly into discrete morphological classes. Instead, they seem to span a range from symmetric exponential profiles to asymmetric, multi-component profiles with blue excess to objects with strong asymmetric blue excess, hereafter referred to as blue shoulder. We use representative objects below to illustrate the two ends of the observed distribution and different evolutionary pathways between them. We then separately examine the six objects with Ia-CSM-like characteristics.

\subsubsection{Objects with persistent, approximately symmetric Balmer lines} \label{sec:symmetric}
\begin{figure*}[tbp]
    \centering
    \includegraphics[width=\linewidth]{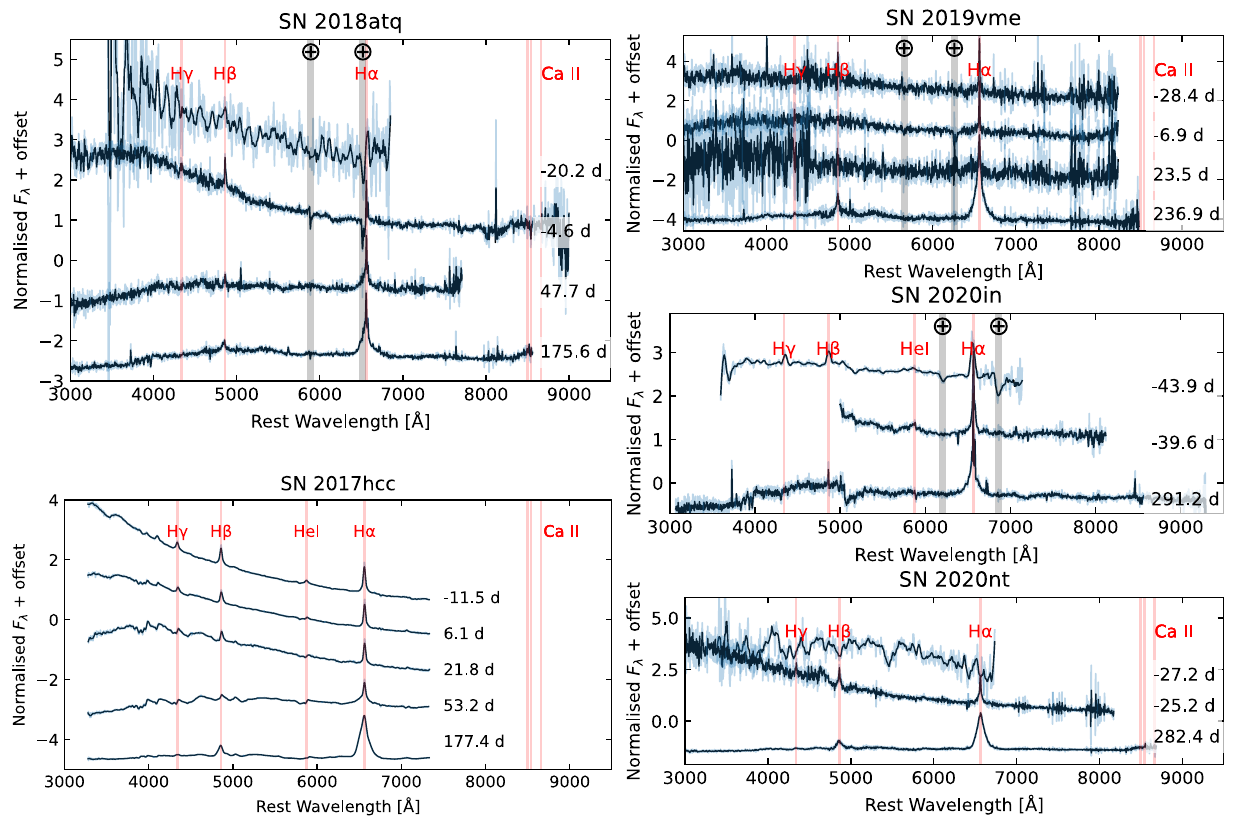}
    \caption{Spectral sequences of SNe\,2018atq, 2020nt, 2019vme, 2020in, and 2017hcc from \citet{Moran_2023_2017hcc}. The phases labelled to the right of each spectrum are relative to the rest-frame ZTF $r$-band peak. Common spectral emission features are labelled in red, and wavelength regions affected by telluric absorption are marked by vertical grey bands.}
    \label{fig:spectral_sequence1}
\end{figure*}

\begin{figure}[tbp]
    \centering
    \includegraphics[width=\columnwidth]{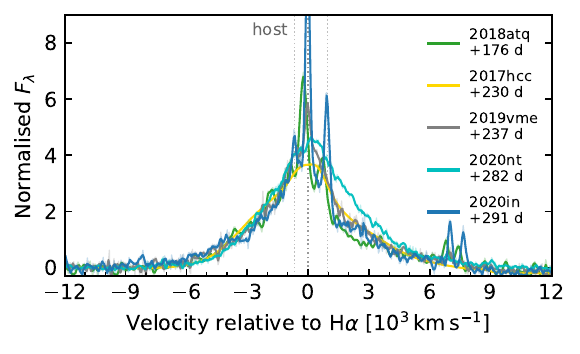}
    \caption{Continuum-subtracted and independently normalised late-time H$\alpha$ profiles of SNe\,2018atq, 2019vme, 2020nt, and 2020in, compared with that of SN\,2017hcc \citep{Moran_2023_2017hcc}. Narrow host-galaxy emission lines are marked by vertical dotted lines.}
    \label{fig:spectral_sequence1_Halpha}
\end{figure}

To illustrate one end of the observed spectroscopic diversity, we select four events that retain approximately symmetric, exponential Balmer lines throughout their observed evolution: SNe\,2018atq, 2020nt, 2019vme, and 2020in. Their spectral evolution is shown in \Cref{fig:spectral_sequence1}. All four objects were observed before peak and initially exhibit blue continua with narrow, exponential Balmer emission, and with Balmer decrement near or below the Case B value. As they evolve, their continua become redder and metal-line structure becomes more prominent, but Balmer lines remain the dominant emission feature. Most importantly, the H$\alpha$ profile develops broader, approximately symmetric and exponential wings without a distinct high-velocity blue excess. The similarity is particularly clear in the late-time H$\alpha$ profiles shown in \Cref{fig:spectral_sequence1_Halpha}. Between approximately $+175$ and $+290$\,d, the profiles remain exponential and approximately symmetric, with characteristic FWHM values of order $3000\,\kms$ and wings extending to substantially higher projected velocities. Their morphology is broadly similar to that of the well-studied SN\,2017hcc at a comparable phase \citep{Moran_2023_2017hcc}. The persistence of strong H$\alpha$ emission beyond $200$\,d indicates that ongoing CSM interaction remains an important contributor to the optical emission long after maximum light. 

\begin{figure}[tbp]
    \centering
    \includegraphics[width=\columnwidth]{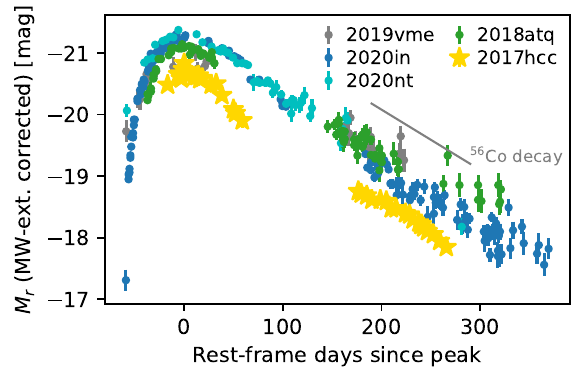}
    \caption{Milky Way-extinction-corrected absolute $r$-band light curves of SNe\,2018atq, 2020nt, 2019vme, and 2020in, compared with that of SN\,2017hcc \citep{Moran_2023_2017hcc}. The grey line indicates the decline rate expected from \Co\ decay.}
    \label{fig:spectral_sequence1_LC}
\end{figure}

These four objects occupy a similar region of light-curve parameter space. Their $r$-band peaks lie in the range $-21.3\lesssim M_r\lesssim-20.9$\,mag, placing them among the brighter objects in the sample, and their light curves are broad and slowly evolving (\Cref{fig:spectral_sequence1_LC}). At phases after peak, their  decline rates are broadly comparable to the \Co\ decay rate. SN\,2018atq shows a flattening after approximately $250$\,d, providing additional evidence for sustained energy input from CSM interaction. SN\,2017hcc is somewhat fainter at peak but shows a comparable late-time H$\alpha$ morphology and similarly slow photometric evolution.

Approximately symmetric, exponential wings can be produced when initially narrow line photons undergo multiple Thomson scatterings in optically thick, slowly moving CSM. There are two possible interpretations of what these profiles imply for the underlying shock and photon-emitting shell. First, the profiles could indicate that the photon-emitting shell has been strongly decelerated by dense CSM. As we later demonstrate in the line-profile models presented in \Cref{sec:modelling}, increasing the photon-emitting shell velocity primarily strengthens the blue excess. The absence of a prominent blue shoulder in these four objects favours relatively low velocities at the observed epochs. Such low velocities could indicate that the shock and the photon-emitting shell have been efficiently decelerated by dense, massive CSM. The resulting efficient conversion of ejecta kinetic energy into radiation could help explain the high optical luminosities and similar light-curve evolution of these objects. Alternatively, the shock can remain embedded behind optically thick, unshocked material. Taking this interpretation, the absence of a blue excess would then provide only a weak constraint on the true photon-emitting shell velocity.

\subsubsection{Objects with strong blue excess/shoulder} \label{sec:shoulder}
\begin{figure*}[!t]
    \centering

    \begin{minipage}{\textwidth}
        \centering
        \includegraphics[
            width=\linewidth
        ]{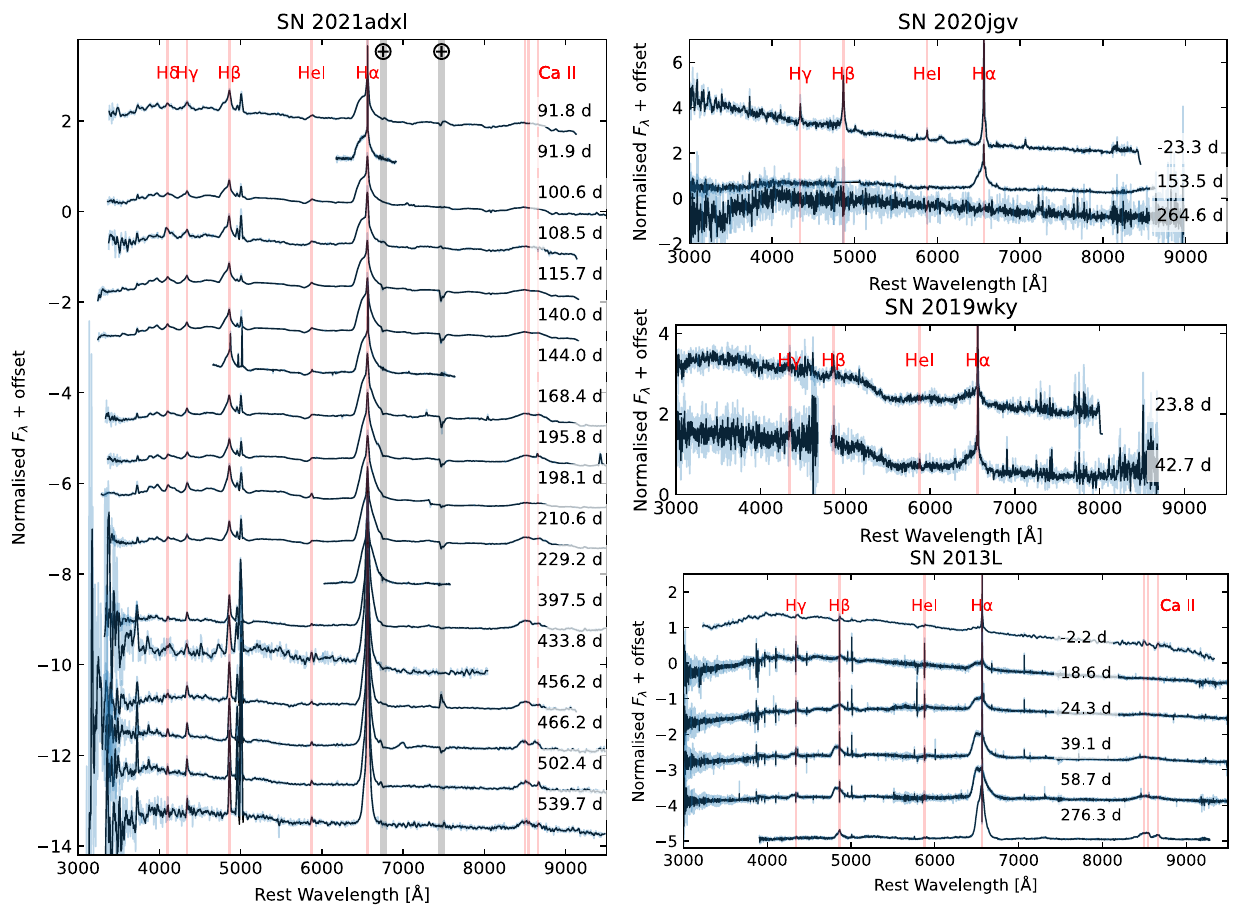}
        \caption{Spectral sequences of SNe\,2021adxl, 2020jgv, 2019wky,
        and 2013L \citep{Taddia_2020_2013L}. The phases labelled to the
        right of each spectrum are relative to the rest-frame $r$-band
        peak. Common spectral features, including H$\alpha$, H$\beta$,
        H$\gamma$, He\,I, and the Ca\,II triplet, are labelled in red.
        Wavelength regions affected by telluric absorption are marked by
        vertical grey bands.}
        \label{fig:spectral_sequence2}
    \end{minipage}

    \vspace{0.8em}

    \begin{minipage}[t]{0.49\textwidth}
        \vspace{0pt}
        \centering
        \includegraphics[
            width=\linewidth
        ]{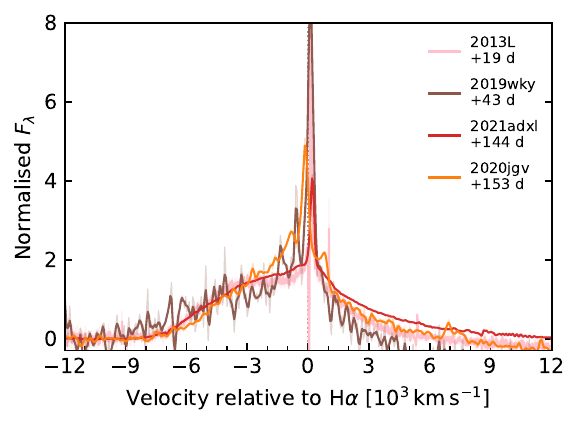}
        \caption{Continuum-subtracted and independently normalised
        H$\alpha$ profiles of SNe\,2019wky, 2020jgv, and 2021adxl,
        compared with that of SN\,2013L
        \citep{Taddia_2020_2013L}.}
        \label{fig:spectral_sequence2_Halpha}
    \end{minipage}
    \hfill
    \begin{minipage}[t]{0.49\textwidth}
        \vspace{0pt}
        \centering
        \includegraphics[
            width=\linewidth
        ]{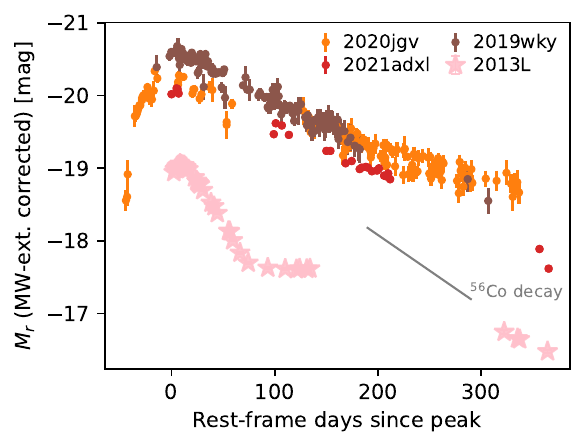}
        \caption{MW-extinction-corrected absolute $r$-band light curves
        of SNe\,2021adxl, 2020jgv, and 2019wky, compared with that of
        SN\,2013L. Data for the objects in our sample are from
        \citet{Pessi_2025_SLSNeIISample}, and those for SN\,2013L are
        from \citet{Taddia_2020_2013L}.}
        \label{fig:spectral_sequence2_LC}
    \end{minipage}
\end{figure*}

At the other end of the observed spectroscopic diversity, several objects develop an excess of emission on the blueshifted side of the narrow H$\alpha$ core. The morphology varies continuously from an extended slanted blue wing to a more clearly separated shoulder-like component. Representative blue excess examples are SNe\,2019wky and 2020jgv, and the blue shoulder object is SN\,2021adxl, whose spectral evolution is shown in \Cref{fig:spectral_sequence2}. We also include the lower-luminosity Type IIn SN\,2013L \citep{Taddia_2020_2013L} as a comparison object with a similar H$\alpha$ blue shoulder morphology.

SN\,2019wky already exhibits a strongly asymmetric H$\alpha$ profile between approximately $+24$ and $+43$\,d, with excess emission extending to velocity offsets approaching $-9000\,\kms$. SN\,2020jgv shows narrow, approximately symmetric Balmer emission at $-23.3$\,d but develops a prominent blue excess extending to approximately $-9000\,\kms$ by $+153.5$\,d. SN\,2021adxl displays a particularly well-observed, multi-component H$\alpha$ profile, with a blue shoulder extending to approximately $-8000\,\kms$ between $+90$ and $+200$\,d. Its blue-side asymmetry subsequently weakens, although the profile remains broader and more complex than a single exponential, and strong Balmer emission persists beyond $+500$\,d. A similar blue shoulder is observed in SN\,2013L, a luminous Type IIn SN that peaked at approximately $M_r=-19$\,mag, whose H$\alpha$ profile also extends to velocity offsets of approximately $-8000\,\kms$ \citep{Taddia_2020_2013L}. The asymmetry in SN\,2013L is already visible at $-2.2$\,d, although the broad component is diluted by the strong continuum at this early phase. The asymmetric profile remains visible at $+276.3$\,d, indicating that, in the available spectra, the blue shoulder persists for longer than in SN\,2021adxl. These observations demonstrate that even among the blue excess/shoulder objects, the H$\alpha$ morphology is phase-dependent and they are not a permanent spectroscopic subgroup.

The three objects from our sample are also fainter than the persistent symmetric-profile objects discussed above, reaching approximately $-20.6\lesssim M_r\lesssim-20.1$\,mag (\Cref{fig:spectral_sequence2_LC}). Their available late-time light curves generally decline more slowly than the \Co\ decay rate, demonstrating that CSM interaction remains an important power source even though their peak luminosities are lower. As we demonstrate in \Cref{sec:modelling}, the blue excess can be interpreted as a kinematic signature of fast shocked material, and increasing the photon-emitting shell velocity strengthens the blue shoulder. The lower peak luminosities and higher inferred  velocities of these objects may indicate a different partition of the explosion energy from that of the brighter, persistent symmetric-profile objects. If the underlying ejecta energies are broadly comparable, interaction with a less massive or less optically thick CSM would decelerate the shock less efficiently, leaving a larger fraction of the explosion energy in kinetic form and converting a smaller fraction into radiation around maximum light. This interpretation is model-dependent because the observed asymmetry can also be affected by optical depth, CSM geometry, and obscuration.  

\subsection{Subgroup with Ia-CSM characteristics} \label{sec:Ia-CSM}
\begin{figure*}
    \centering
    \includegraphics[width=\linewidth]{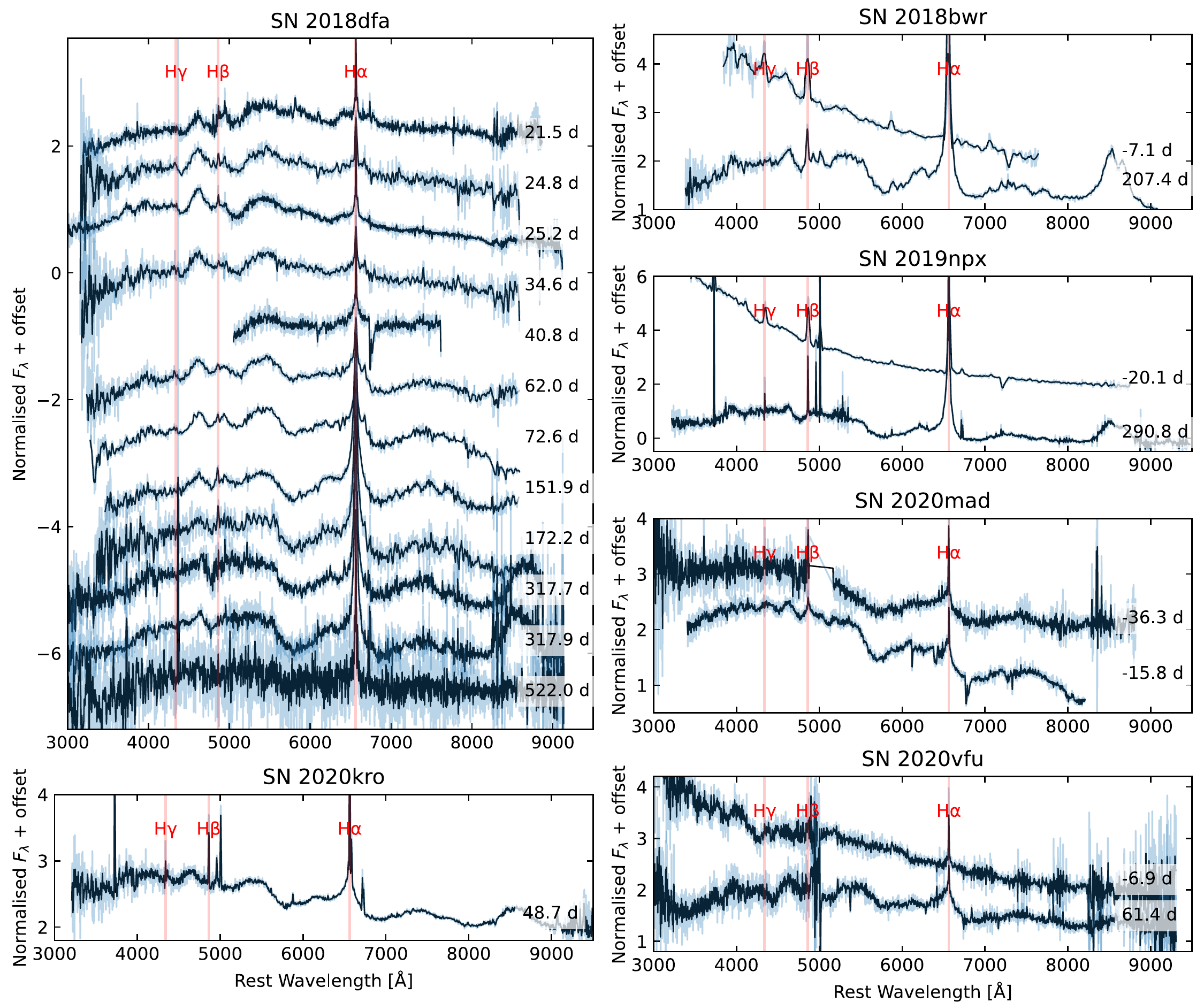}
    \caption{Rest-frame spectral sequences of the Ia-CSM candidate subset in our sample. SEDM spectra are omitted here. Red vertical lines mark Balmer emission lines.}
    \label{fig:IaCSM}
\end{figure*}

\begin{figure*}
    \centering
    \includegraphics[width=\linewidth]{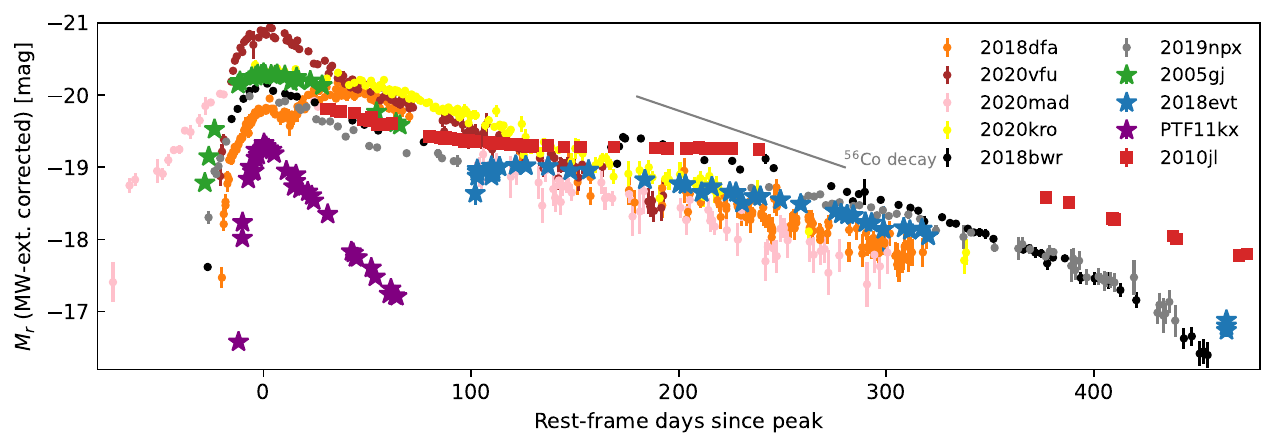}
    \caption{Observed $r$-band light curves of the Ia-CSM candidate subset in our sample compared with those of SNe Ia-CSM 2005gj, PTF11kx, 2018evt, and SLSN-IIn 2010jl. The light curves are corrected for MW extinction. The grey line shows the \Co\,decay slope.}
    \label{fig:IaCSM_photometry}
\end{figure*}

\begin{figure*}
    \centering
    \includegraphics[width=\linewidth]{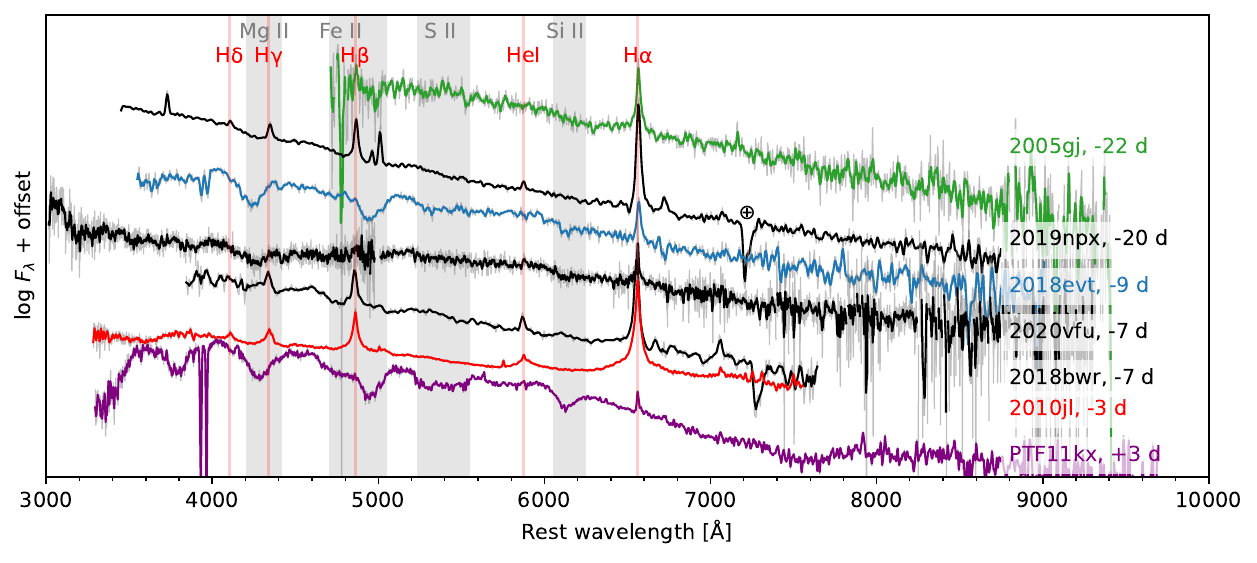}
    \caption{Pre-peak spectra of the Ia-CSM candidate subset in our sample compared with those of SNe Ia-CSM 2005gj, PTF11kx, 2018evt, and SLSN-IIn 2010jl at similar phases. Sample objects are shown in black and literature objects in colour. Vertical red lines mark emission lines, while grey bands mark absorption features. Spectra are arranged in ascending phase order.}
    \label{fig:IaCSM_early}
\end{figure*}

\begin{figure*}
    \centering
    \includegraphics[width=\linewidth]{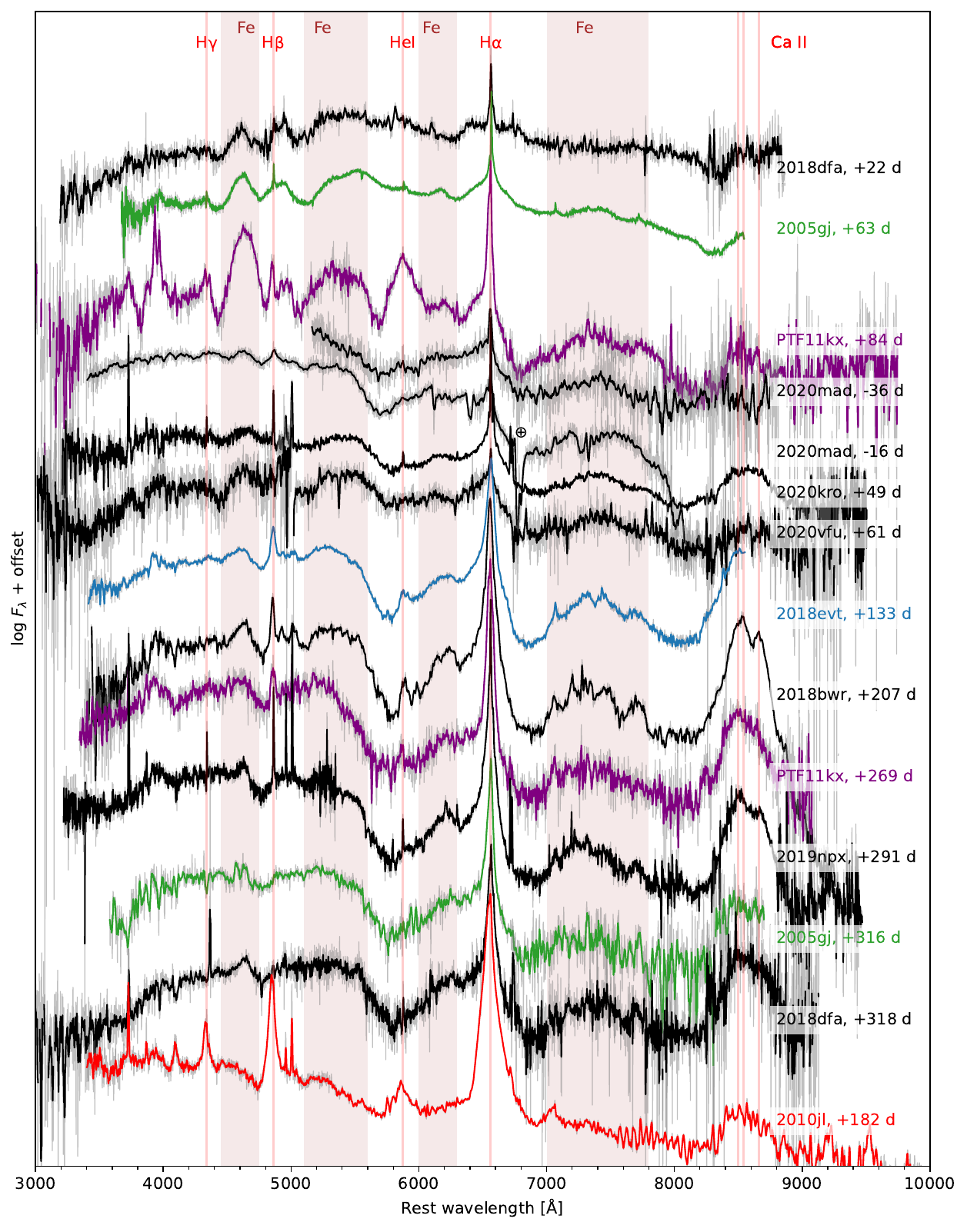}
    \caption{Same comparison as in \Cref{fig:IaCSM_early}, but for post-peak spectra, except for SN\,2020mad, whose pre-peak spectra exhibit late-time characteristics. The spectrum of SN\,2010jl is plotted at the bottom. Vertical brown bands mark Fe forests.}
    \label{fig:IaCSM_late}
\end{figure*}

We identify six objects in our sample that exhibit spectral characteristics resembling those of SNe Ia-CSM at one or more epochs: SNe\,2018dfa, 2018bwr, 2019npx, 2020mad, 2020kro, and 2020vfu. Their spectral sequences are shown in \Cref{fig:IaCSM}. We compare their light curves and spectra with those of the SNe Ia-CSM 2005gj, 2018evt, and PTF11kx (SN\,2011km). We also include SN\,2010jl as a well-studied example of SLSNe-IIn with persistent CSM interaction. The $r$-band light-curve comparison is shown in \Cref{fig:IaCSM_photometry}. The six objects reach peak $r$-band magnitudes of approximately $-20.0$ to $-20.9$\,mag, placing them towards the luminous end of SNe Ia-CSM. The pre- and post-peak spectral comparisons are presented in \Cref{fig:IaCSM_early} and \Cref{fig:IaCSM_late}. 

SN\,2018dfa was mentioned as an ambiguous Ia-CSM candidate in the appendix of \citet{Sharma_2023_IaCSM}. It has a multi-peaked light curve (\Cref{fig:IaCSM_photometry}), with the second peak reaching $-19.94$\,mag in the $r$ band and $-20.14$\,mag in the $i$ band. No spectrum was obtained before the first peak, while 14 spectra were taken at or after the second peak. In the $+21.5$\,d spectrum, H$\alpha$ emission is prominent and consists of a narrow exponential core superposed on a broad, undulating quasi-continuum, which we do not interpret as a broad H$\alpha$ component. H$\beta$ is weak, resulting in a large Balmer decrement of $L_{\mathrm{H}\alpha}/L_{\mathrm{H}\beta}\sim8$ (see \cref{sec:BD}). Shallow absorption features near 6200 and 8200\AA\ may correspond to diluted Si\,II and Ca\,II features observed in SNe Ia and in some SNe Ia-CSM \citep[e.g.][]{Silverman_2013_IaCSM}. We also identify Fe-dominated blends at 4500--4700\AA, including a triangular feature, and at 5000--5500\AA. This spectrum closely resembles that of SN\,2005gj at $+63$\,d and also resembles the $+84$\,d spectrum of PTF11kx, although the features in PTF11kx are more pronounced and it additionally exhibits strong He\,I emission. In subsequent spectra, the H$\alpha$ feature broadens while retaining its exponential shape and gradually blends into the surrounding quasi-continuum. These spectra show no obvious [O\,I] $\lambda\lambda6300,6364$ emission and contain strong Fe-dominated blends with broadly similar relative strengths. From the $+72.6$\,d spectrum onwards, a broad feature develops blueward of H$\alpha$, spanning approximately $\sim6150$--$6300$\AA. We associate this feature with blended Fe\,II emission \citep{Dessart_2022_HeliumStarModel}.

SNe\,2018bwr and 2019npx show similar spectral evolution. Their $r$-band peaks are slightly brighter than $-20.0$\,mag. SN\,2018bwr exhibits a second bump at $\sim200$\,d, while for SN\,2019npx the corresponding phase is not covered. Their early spectra are dominated by very narrow Balmer lines superposed on a blue continuum and are indistinguishable from those of SNe IIn at early phases, such as SN\,2010jl. A similar early-time morphology is also observed in the SN Ia-CSM 2005gj, demonstrating that these early spectra are not diagnostic of the underlying explosion. At later phases, the spectra of both objects develop a morphology resembling that of SN\,2018dfa and the Ia-CSM comparison objects, including strong Fe-dominated blends, weak higher-order Balmer lines, and no obvious [O\,I] emission. 

SNe\,2020kro and 2020vfu reach slightly brighter $r$-band peaks than the other objects in this group. Their latest available spectra, obtained at $+48$--$62$\,d, already exhibit similar Fe-dominated undulations and weak higher-order Balmer lines similar to the late-time morphology described above, although the undulations remain shallower, possibly because ongoing CSM interaction still dilutes the underlying features. SN\,2020mad has the longest rise time in this group, with an estimated $t_{1/e}$ rise time of $50$\,d relative to peak \citep{Pessi_2025_SLSNeIISample}. Although its two spectra were obtained during the rise, the long rise time means that they do not necessarily represent a physically early stage after explosion. Their morphology already resembles the later spectra of the other objects in this subset, but with a less prominent Fe\,II blend at 6000--6300\AA.

We apply the spectral-classification tool \textsc{SNID} \citep{Blondin_2007_SNID} to the spectra included in the late-phase comparison in \Cref{fig:IaCSM_late}. For all six objects, the highest-ranked matches include SNe Ia-CSM 2005gj, 2016iks, and IIn/Ia-CSM events like SNe\,1997cy and 2012ca (see Appendix~\ref{app:C}). For SN\,2020mad, the highest-ranked \textsc{SNID} match is a normal Type Ib SN, but its inferred redshift is inconsistent with the host-galaxy redshift, and the visual agreement is poorer than for the second-ranked match, SN\,1997cy. The six objects exhibit a coherent late-time morphology that more closely resembles the Ia-CSM comparison spectra than that of SN\,2010jl and distinguishes them from most narrow-lined objects in our sample. However, strong CSM interaction can veil the underlying ejecta in both thermonuclear and core-collapse explosions, while Fe-dominated blends, large Balmer decrements, and the absence of obvious [O\,I] emission are not unique to SNe Ia-CSM. The available observations therefore neither establish a thermonuclear origin nor rule out a core-collapse origin. We consequently flag these objects as having Ia-CSM/IIn-like spectral characteristics. We also suggest that, to establish the true nature of a strongly interacting SN, spectroscopic observations should ideally extend into the nebular phase.


\section{Modelling} \label{sec:modelling}
To investigate the physical origin of the observed diversity in this sample, we use Monte Carlo models to study the effect of shock and electron scattering on shaping the observed Balmer line profiles. We then use the characteristic velocity scale inferred from these models to obtain order-of-magnitude estimates of the progenitor mass-loss rates.

\subsection{Line-profile modelling}
\begin{figure}
    \centering
    \includegraphics[width=\linewidth]{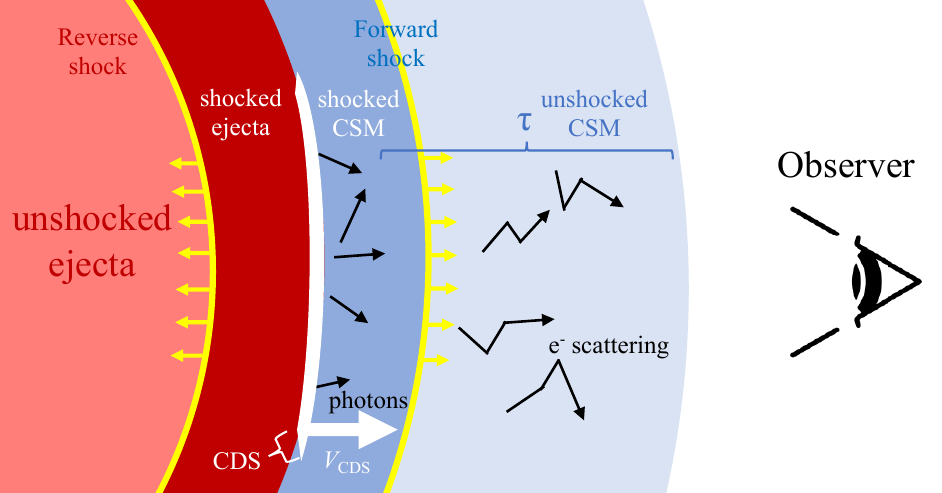}
    \caption{Schematic illustration of the shock structure and electron-scattering line formation in an interacting SN.}
    \label{fig:escatter_schematics}
\end{figure}

The Type IIn-like line profiles observed throughout the sample can be explained by shock physics and electron scattering. \Cref{fig:escatter_schematics} shows a schematic illustration of this process. As the SN ejecta interact with dense CSM, a forward shock propagates into the unshocked CSM, and a reverse shock propagates back into the ejecta in mass coordinate. Between the two shocks lies a thin, dense shell of shocked material, commonly referred to as the cold dense shell (CDS; e.g. \citealt{Chugai_1994_1988Z, Dessart_2009_1994W}) because the shocked gas radiates efficiently and cools to relatively low temperatures. The CDS is optically thick and can obscure emission from material behind it, so photons that can reach the observer predominantly originate from regions beyond the CDS. Then, as these photons travel outward, free electrons in the unshocked, ionised CSM can broaden the line through Thomson scattering.

\textsc{escatter}\footnote{\url{https://github.com/Astro-Sean/escatter}, based on the scattering treatment of \citet{Pozdnyakov_1983_eScattering}} is a 1-D Monte Carlo code that computes electron-scattering line profiles in interacting SNe, with applications to H-rich SLSNe such as SN\,2021adxl \citep{Brennan_2024_2021adxl} and SN\,2021aaev \citep{Hu_2026_2021aaev}. Like other codes developed for similar purposes \citep[e.g.][]{Fransson_2014_2010jl, Taddia_2020_2013L}, \textsc{escatter} is not a radiation-hydrodynamical code and does not follow the time evolution of the interaction. Instead, each model represents a single-epoch snapshot of the line-forming region. In our setup, a packet of $10^5$ photons is emitted at the rest wavelength of H$\alpha$ from region beyond the CDS, with an emission probability proportional to $r^{-4}$. With this emission prescription, the model is describing a scenario where most photons are emitted from a shell right in front of the CDS, as indicated by the straight black arrows in \Cref{fig:escatter_schematics}. This photo-emitting shell has velocity $v_{\mathrm{s}}$ and should be order-of-magnitude similar to the CDS velocity $v_{\mathrm{CDS}}$ and the forward shock velocity $v_{\mathrm{shock}}$, since the CDS and forward shock are spatially close. The CSM is assumed to be stationary\footnote{What matters to the profile is the relative velocity between CSM and photon-emitting shell, so setting $v_{\mathrm{CSM}}=0$ is a valid simplification.} and follows a power-law density profile, $\rho=\rho_0r^{-s}$, between CSM radii $r_{\mathrm{in}}$ and $r_{\mathrm{out}}$. Motivated by the numerical simulations on Type IIn SNe by \citet{Dessart_2015_SimulationIIn}, we set the electron temperature $T_{\mathrm{e}}$ to decline linearly from $2\times10^4$\,K at $r_{\mathrm{in}}$ to $10^4$\,K at $r_{\mathrm{out}}$. We adopt $[r_{\mathrm{in}},r_{\mathrm{out}}]=[1.00,1.01]\times10^{14}$\,cm. In our tests, varying the radial extent of the scattering region produced only minor changes in the resulting line profiles, so we fix these radii throughout the model grid. The remaining free parameters are therefore density exponent $s$, photon-emitting shell velocity $v_{\mathrm{s}}$, and optical depth $\tau$. We do not explicitly model emission originating in the unshocked CSM. A self-consistent prediction of its strength relative to the CDS emission would require calculating the ionising radiation field and the resulting ionisation and excitation of the CSM, which is beyond the scope of this work. Including this component would be a useful future extension of \textsc{escatter}.

\begin{figure}
    \centering
    \includegraphics[width=\linewidth]{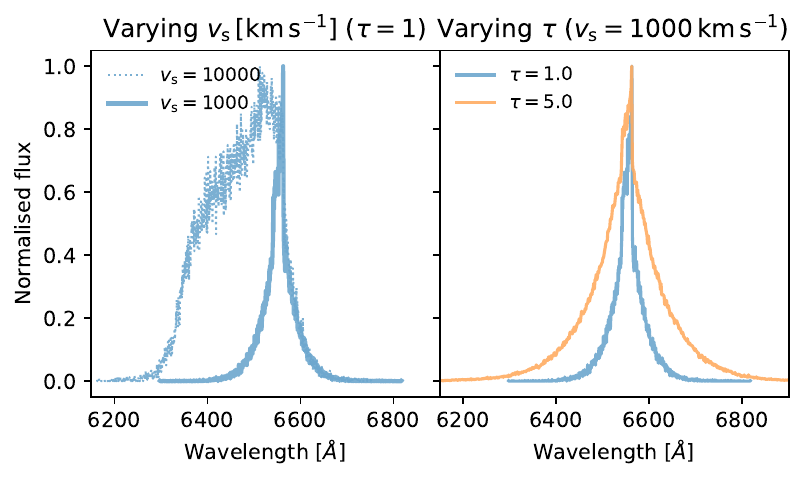}
    \caption{Example H$\alpha$ line profiles generated with \textsc{escatter}. The left panel shows the effect of varying $v_{\rm s}$ at fixed $\tau$. The right panel shows the effect of varying $\tau$ at fixed $v_{\rm s}$. All profiles are normalised to the peak.}
    \label{fig:escatter_comparison}
\end{figure}

We run \textsc{escatter} over a grid with $s=0$ and $2$, $v_{\mathrm{s}}=100$--$10000\,\kms$ in steps of $100\,\kms$, and $\tau=1$--10 in steps of 0.1, producing a grid of H$\alpha$ line-profile models. The resulting profiles are nearly insensitive to the choice of density exponent, with negligible differences between $s=0$ and $s=2$. In contrast, $v_{\mathrm{s}}$ and $\tau$ have distinct effects on the line morphology. \Cref{fig:escatter_comparison} shows model profiles in which one parameter is varied while the other is held fixed. Increasing $v_{\mathrm{s}}$ primarily enhances the blue shoulder of the profile. Increasing $\tau$ instead produces broader exponential wings, with a weaker effect on the overall profile asymmetry. 

We compare the continuum-subtracted H$\alpha$ profiles extracted in \Cref{sec:BL} with the model grid. The fitting is performed over the rest-frame wavelength intervals $6300$--$6535$\,\AA\ and $6575$--$6826$\,\AA, excluding the central region where unresolved host and narrow CSM emission can dominate. We allow an overall wavelength shift of up to $\pm10$\,\AA\ and mask additional contaminated regions where necessary. For each spectrum, we select the model that minimises the residuals and record its values of $v_{\rm s}$ and $\tau$. The fitting result is listed in Appendix~\ref{app:D}. \Cref{fig:escatter_example} shows two example applications to the H$\alpha$ spectral sequence of SNe\,2018atq and 2021adxl. For SN\,2018atq, the H$\alpha$ profiles are affected by host line contamination, and we manually exclude the host line region from the fitting. For SN\,2021adxl, there is an additional narrow component that comes from CSM, and we also manually exclude the region. For each fitted spectrum, we obtain the best-fit values of $\tau$ and $v_{\rm s}$, and use the residuals to assess the quality of the fit. 

\begin{figure}
    \centering
    \includegraphics[width=\linewidth]{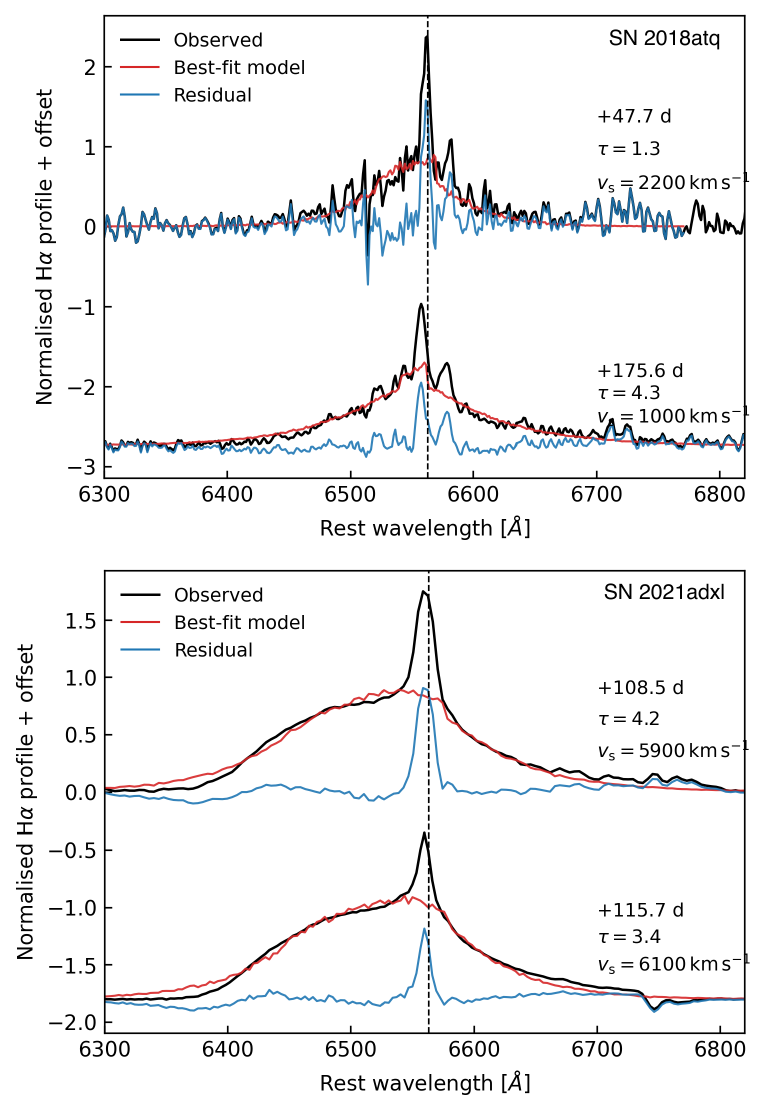}
    \caption{Example \textsc{escatter} fits to selected H$\alpha$ spectral sequence of SNe\,2018atq and 2021adxl. Observed spectra are shown in black, best-fit models in red, and residuals in blue. The residual can come from the host, as is the case in SN\,2018atq, or a narrow CSM component, as is the case in SN\,2021adxl.}
    \label{fig:escatter_example}
\end{figure}

\begin{figure}
    \centering
    \includegraphics[width=\linewidth]{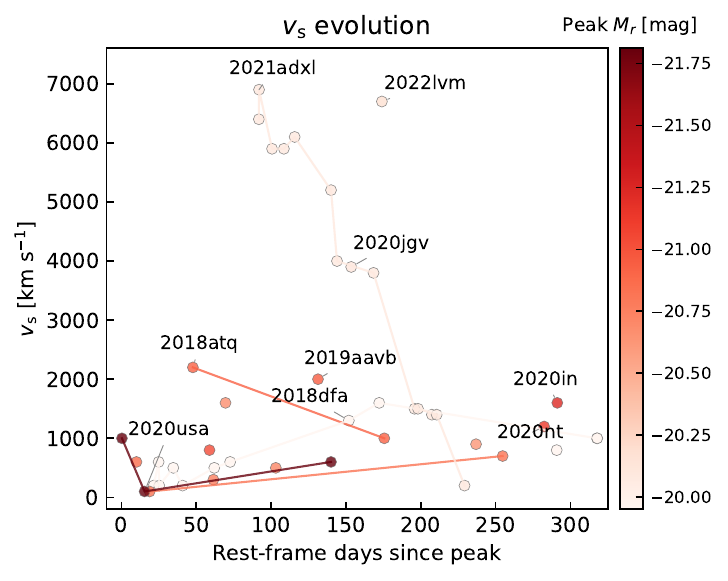}
    \caption{Evolution of the fitted photon-emitting shell-velocity parameter $v_{\rm s}$. Selected objects are labelled, and marker colour indicates the peak ZTF $r$-band absolute magnitude.}
    \label{fig:escatter_vs_evolution}
\end{figure}

We apply the same procedure to the rest of the sample and show the resulting $v_{\rm s}$ evolution in \Cref{fig:escatter_vs_evolution}, with the peak $r$-band absolute magnitude indicated by colour. The inferred $v_{\rm s}$ evolution is highly heterogeneous. SN\,2020usa, one of the most luminous objects in the sample, shows persistent narrow Balmer lines for more than 100\,d and is consistently best fitted by low values of $v_{\rm s}$. Within the adopted model, its H$\alpha$ profile therefore requires little contribution from a rapidly expanding photon-emitting shell. This could indicate substantial shock deceleration by dense CSM, or that emission from faster shocked material remains concealed by optically thick material. Several other luminous objects, including SNe\,2018atq, 2020nt, and 2020in, are also best fitted with $v_{\rm s}\leq2200\,\kms$, although the spectral coverage is limited. In contrast, the fainter SNe\,2020jgv and 2021adxl exhibit prominent blue excesses or blue shoulders and higher fitted velocities at some epochs. SN\,2020jgv has $v_{\rm s}=3900\,\kms$ at $153.5$\,d, while SN\,2021adxl declines from $6400$--$6900\,\kms$ at $92$\,d to $1500\,\kms$ at $196$--$198$\,d and $200\,\kms$ at $229$\,d. The broad-lined SN\,2022lvm also yields a high fitted velocity of $6700\,\kms$ at $174$\,d, although its position should be interpreted separately because its H$\alpha$ profile may be dominated by kinematic broadening rather than the same electron scattering mechanism as the narrow-lined objects. The Ia-CSM/IIn-like SN\,2018dfa shows a non-monotonic evolution, increasing from a few hundred $\kms$ near peak to $1600\,\kms$ at $172$\,d before decreasing to $1000\,\kms$ at $318$\,d.

\begin{figure}
    \centering
    \includegraphics[width=\linewidth]{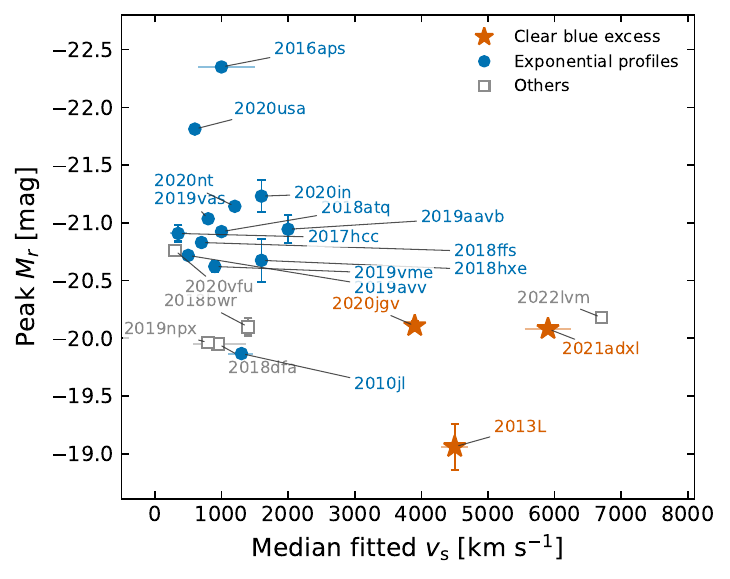}
    \caption{Peak ZTF $r$-band absolute magnitude over median $v_{\mathrm{s}}$ between rest-frame days 50 and 300. Orange stars denote objects with clear blue excess, blue circles denote objects with exponential profiles, and grey squares denote other objects. The comparison sample includes SNe\,2010jl \citep{Fransson_2014_2010jl}, 2013L, 2016aps \citep{Nicholl_2020_2016aps}, and 2017hcc \citep{Moran_2023_2017hcc}. Horizontal bars show the interquartile range across the selected epochs; vertical bars show the adopted magnitude uncertainties.}
    \label{fig:escatter_vs_magnitude}
\end{figure}

We further explore the relations between peak luminosity and the fitted velocity scale and electron-scattering optical depth in the wider context of luminous H-rich SNe. For each object, we compute the day 50--300 median $v_{\mathrm{s}}$. This range avoids early spectra, which are often dominated by narrow exponential profiles with little diversity, and very late spectra, which are difficult to describe with a single electron-scattering model. \Cref{fig:escatter_vs_magnitude} shows the peak absolute magnitude $M_r$ as a function of median $v_{\rm s}$ for 18 objects in our sample, together with the well-studied H-rich SLSNe 2010jl, 2016aps, and 2017hcc from \textsc{WISeREP}. The most luminous objects, SNe\,2020usa and 2016aps, reach $M_r<-21.5$\,mag and are characterised by persistent narrow lines. A group of luminous objects clusters around $M_r\approx-21$\,mag and median $v_{\rm s}<2000\,\kms$, including SN\,2018atq and 2017hcc, whose H$\alpha$ profiles remain relatively narrow and symmetric. At the fainter, higher-velocity end of the distribution, SNe\,2020jgv, 2021adxl and 2013L show prominent blue excess or shoulders and higher inferred velocities. The broad-lined SN\,2022lvm also has a high fitted velocity. A separate group of fainter, low-velocity objects clusters around $M_r\approx-20$\,mag and $v_{\rm s}<2000\,\kms$. They all belong to the objects with Ia-CSM-like spectral characteristics identified in \Cref{sec:Ia-CSM}. Interestingly, under this framework SN\,2010jl also belongs to this region. This suggests that after removing the Fe contamination blueward of H$\alpha$ in the Ia-CSM/IIn subgroup, their H$\alpha$ morphology is comparable to SN\,2010jl. 

Combining the clear-blue-excess and exponential-profile groups (i.e. without the SLSN-II and the Ia-CSM candidates), we find a positive Pearson correlation between $M_r$ and fitted $v_{\rm s}$ ($r=0.63$, $p=0.010$). Thus, brighter objects generally have lower fitted velocity scales, whereas fainter objects have higher values. If these velocities trace shock motion, this trend would be consistent with denser or more massive CSM producing both stronger shock deceleration and more efficient conversion of ejecta kinetic energy into radiation. However, the fitted velocity scale is not necessarily a reliable measure of shock velocity for symmetric, electron-scattering-dominated profiles. The comparison also involves a small, spectroscopically selected subset with heterogeneous phase coverage. The physical interpretation therefore remains tentative.

\begin{figure}
    \centering
    \includegraphics[width=\linewidth]{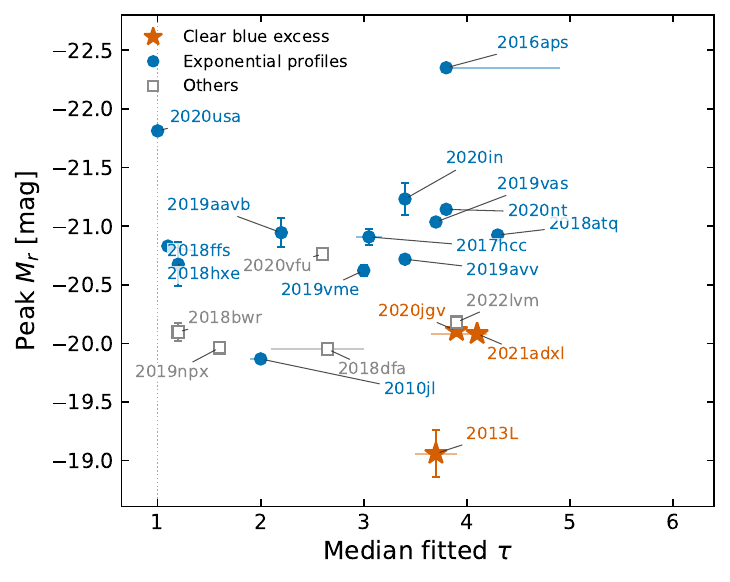}
    \caption{Peak $r$-band absolute magnitude versus median fitted optical depth. Object selections, phase intervals, symbols, and error bars follow \Cref{fig:escatter_vs_magnitude}. The dotted vertical line marks the lower boundary of the model grid at $\tau=1$.}
    \label{fig:escatter_tau_magnitude}
\end{figure}

The corresponding comparison with median fitted $\tau$ is shown in \Cref{fig:escatter_tau_magnitude}. No clear monotonic trend with peak magnitude is apparent, and the clear-blue-excess and exponential-profile groups overlap substantially. The optical-depth comparison therefore does not independently support a simple relation between peak luminosity and CSM density or mass. In particular, $\tau$ probes the electron column through the ionised scattering region and also depends on its geometry and ionisation state.

\subsection{Mass-loss rate estimate} \label{sec:massloss}

We estimate the mass-loss rate following the scaling relation between the shock luminosity and the kinetic power processed by the interaction \citep{Chugai_1994_1988Z}. For a radiative shock propagating through a steady wind, the luminosity can be written as
\begin{equation}
    L = \frac{1}{2}\epsilon \dot{M}\frac{v_{\rm s}^{3}}{u_{\rm w}},
\end{equation}
where $L$ is the radiated luminosity, $\epsilon$ is the conversion efficiency, $\dot{M}$ is the progenitor mass-loss rate, $v_{\rm s}$ is the shock velocity, and $u_{\rm w}$ is the pre-SN wind velocity. Rearranging gives
\begin{equation}
    \dot{M} = \frac{2L}{\epsilon}\frac{u_{\rm w}}{v_{\rm s}^{3}} .
    \label{eq:mass_loss_rate}
\end{equation}

For each usable spectrum, we estimate $L$ by integrating the calibrated and extinction-corrected flux. We fit a blackbody to continuum regions, and use it to extend the spectrum to 3000 and 20000\,\AA\ where coverage is missing. The measured spectrum is retained within its observed wavelength range. For the fitted temperatures, 1.8--9.1\% of the total blackbody luminosity lies beyond 20000\,\AA. Radiation below 3000\,\AA\ is omitted. We therefore use the resulting 3000--20000\,\AA\ pseudo-bolometric luminosity as a proxy for $L$.

Unlike \citet{Pessi_2025_SLSNeIISample}, who used peak pseudo-bolometric luminosities and assumed shock velocities of 2500 and $10000\,\kms$, we estimate the luminosity at each spectroscopic epoch. For SN\,2021adxl, we pair this luminosity with the fitted velocity $v_{\mathrm s}$ from the previous section. For the remaining objects, we adopt $5000\,\kms$, a value close to what is inferred from the median $v_{\mathrm s}$ of SN\,2021adxl and 2013L, since in this case there is no blue shoulder and the fitted velocity does not generally provide a secure shock-velocity measurement. For SN\,2020jgv, which shows clear blue excess, we additionally consider its fitted velocity of $3900\,\kms$ to define an assumed range of 3900--$5000\,\kms$.

\begin{figure}
    \centering
    \includegraphics[width=\linewidth]{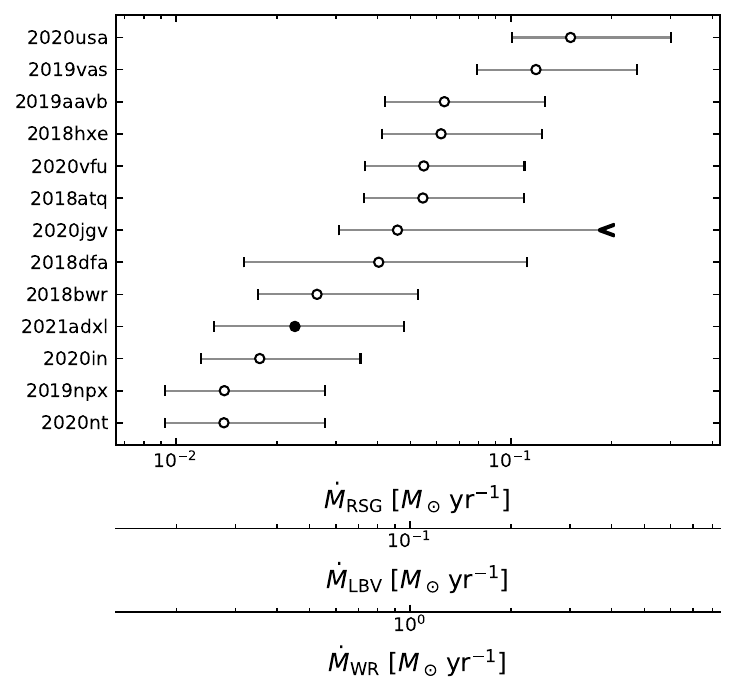}
    \caption{Estimated mass-loss rates inferred from the pseudo-bolometric luminosity and the shock velocity measured from the H$\alpha$ line-profile modelling for 2021adxl or $5000\,\kms$ for the rest. The $<$ symbol for SN\,2020jgv denotes an upper bound under the assumed velocity and efficiency ranges. The main axis assumes $u_{\rm w}=50\,\kms$, while the additional axes show the corresponding values for $u_{\rm w}=100\,\kms$ and $1000\,\kms$, using the linear scaling $\dot{M}\propto u_{\rm w}$.}
    \label{fig:mass_loss}
\end{figure}

We calculate $\dot{M}$ separately for each spectrum and report the per-object median at a fiducial $\epsilon=0.5$, marked as the circles in \Cref{fig:mass_loss}. The error bars show the range obtained by varying $\epsilon$ between 0.25 and 0.75. Since the inferred $\dot{M}$ scales linearly with the assumed wind velocity, we present the results for three representative choices: $u_{\rm w}=50\,\kms$, appropriate for a red-supergiant-like wind \citep[e.g.][]{vanLoon_2005_RSG}; $u_{\rm w}=100\,\kms$, representative of a slower LBV-like wind \citep[e.g.][]{Humphreys_1994_LBV}; and $u_{\rm w}=1000\,\kms$, representative of a WR-like wind \citep[e.g.][]{Crowther_2007_WR}. We estimate mass-loss rates for 13 objects, and they are shown in \Cref{fig:mass_loss}. We report the median value of the per-spectrum mass-loss rates. Other objects have no spectra with both flux-calibration and model-inferred $v_{\rm s}$ and are excluded from this estimate.

For $u_{\rm w}=100\,\kms$ and $\epsilon=0.5$, the reference estimates span approximately $0.028$--$0.30\,M_\odot\,\mathrm{yr}^{-1}$, with $\dot{M}\approx0.045\,M_\odot\,\mathrm{yr}^{-1}$ for SN\,2021adxl. These values suggest substantial pre-explosion mass loss under the adopted interaction model. However, the absolute values depend sensitively on the assumed conversion efficiency $\epsilon$ and the inferred shock velocity, which is highly model-dependent. The calculation further assumes a steady and spherically symmetric wind. Therefore, the inferred mass-loss rates should be regarded as order-of-magnitude estimates rather than precise measurements of the progenitors' mass-loss histories.



\section{Discussion} \label{sec:discussion}

\subsection{Are SLSNe-IIn spectroscopically distinct from lower-luminosity SNe IIn?}
SLSNe-IIn and SNe IIn are both characterised by narrow Balmer emission from slowly moving H-rich CSM. This raises the question whether SLSNe-IIn form a separate population or simply represent the luminous end of SNe IIn. Our sample was selected to have a peak rest-frame absolute magnitude of $M\leq-19.9$\,mag in at least one of the ZTF $gri$ bands, while the SN IIn spectroscopic comparison sample used in this work is small, heterogeneous, and compiled from the literature. We therefore cannot compare the distributions or relative frequencies of their spectral properties across the luminosity boundary. We can instead ask whether the main spectral morphologies found in SLSNe-IIn also occur in lower-luminosity SNe IIn. 

The overall morphology of persistent, narrow and symmetric H$\alpha$ objects in our sample discussed in \Cref{sec:symmetric} has clear counterparts at lower luminosities. Within the Type IIn comparison sample by \citet{Taddia_2013_Carnegie2}, SN\,2005kj shows a narrow, symmetric H$\alpha$ component over epochs from $+6.8$\,d to $+159.5$\,d and peaked at $M_r=-18.5$\,mag. SN\,1994W peaked at $M_{V}=-19.5$\,mag and also retained narrow Balmer cores with electron-scattering wings throughout its observed bright phase, lasting from $+18$\,d to $+121$\,d \citep{Sollerman_1998_1994W, Dessart_2009_1994W}. Multi-component and asymmetric profiles discussed in \Cref{sec:shoulder} are also not unique to SLSNe-IIn. SNe\,2005ip and 2006jd show broad emission underlying the narrow H$\alpha$ component \citep{Stritzinger_2012_Carnegie1,Taddia_2013_Carnegie2}, resembling the prominent blue excess that is developed in SN\,2020jgv. The lower-luminosity SN\,2013L ($M_{r}=-19.0$\,mag) developed a prominent blue shoulder similar to those observed in SN\,2021adxl \citep{Taddia_2020_2013L}. The measured Balmer-line properties in \Cref{sec:BL} show a similar qualitative overlap. The H$\alpha$ pEW, FWHM, and Balmer-decrement ranges of this sample overlap with the selected SN IIn comparison objects. The comparison SNe IIn even span a wider range of late-time H$\alpha$ pEW than our sample, despite their smaller sample size. We therefore do not identify a Balmer line morphology that occurs only in SLSNe-IIn. However, the small and heterogeneous comparison sample does not allow us to determine whether these morphologies occur at the same relative rates in SNe IIn and SLSNe-IIn.

High-ionisation emission lines of He\,II $\lambda4686$, N\,III $\lambda\lambda4634,4641$, and C\,III $\lambda4649$, discussed in \Cref{sec:early}, occur in both ordinarily bright SNe IIn and SLSNe-IIn, but can persist substantially longer in the superluminous events. \citet{Bruch_2023_Flash} found that flash features in their ZTF H-rich SN sample typically last approximately 5\,d. The rarer events with features lasting longer than 10\,d tend to be brighter and rise more slowly. \citet{Hu_2026_2021aaev} reported high-ionisation emission lasting approximately 24\,d in both the SLSN-IIn SN\,2021aaev and the broad-lined SLSN-II SN,2023gpw. Our study adds two further superluminous events with such high-ionisation features, SNe\,2020uwl and 2021fmu. High-ionisation emission in SN\,2021fmu persists to approximately 18.2\,d after detection, while that in SN\,2020uwl is present at least 3.3\,d after detection, with no later spectra of sufficient quality to constrain its disappearance. SN\,2021fmu provides additional evidence that high-ionisation features can persist for several weeks in SLSNe-IIn. Recent works on SN\,2019vxm also suggest that this is a SLSN-IIn flasher peaking at $M_{R}=-20.3$\,mag with flash duration longer than several weeks \citep[e.g.][]{Rameshan_2026_2019vxm, Cai_2026_2019vxm}. Together, these events suggest that SLSNe-IIn can sustain high-ionisation emission for longer than the few-day durations typical of lower-luminosity Type II(n) flashers. This may indicate that their dense CSM extends to larger radii.

It is worth noting that in terms of light-curve comparisons, the study by \citet{Hiramatsu_2024_IIn} finds no transition in SN IIn properties at the conventional $M\simeq-20$\,mag boundary and argues that the SLSN-IIn designation should not be treated as a separate physical class. \citet{Hiramatsu_2024_IIn} argues for a bimodality in the luminosity-timescale space, that there is a `luminous-slow' and a `faint-fast' light-curve morphology group whose median peak magnitudes are $\sim-20.1$\,mag and $-18.2$\,mag respectively in optical bands, and that what are commonly referred to as `SLSNe-IIn' belong to the `luminous-slow' group. Our results show that our SLSN-IIn sample, which falls onto the brighter end of their `luminous-slow' group, is itself spectroscopically diverse. While we do see two distinct ends of Balmer line morphology in our sample, i.e. narrow-symmetric versus multi-component-asymmetric, the photometric bimodality found by \citet{Hiramatsu_2024_IIn} does not correspond to the same spectroscopic bimodality separated by peak luminosity, as similar spectroscopic morphology has appeared in our sample and lower-luminosity Type IIn objects.

Overall, we do not identify a Balmer line morphology unique to SLSNe-IIn selected by the $-19.9$\,mag cut in the ZTF sample when compared with lower-luminosity SNe IIn in the literature. A possible distinction lies in the persistence of high-ionisation features, which can last substantially longer in many SLSNe-IIn than in ordinary H-rich flashers, including lower-luminosity SNe IIn. However, the small number of events prevents us from establishing a systematic difference between SLSNe-IIn and lower-luminosity SNe IIn in terms of high-ionisation features. We therefore suggest future work to treat `SLSN-IIn' as a useful operational label for exceptionally luminous, narrow-lined H-rich SNe in the search for exceptional explosion and powering mechanisms, but the current spectroscopic evidence does not support the luminosity threshold as a fundamental physical division.

\subsection{Mass loss and progenitors of SLSNe-IIn}
In our sample, we observe the persistence of narrow Balmer lines for tens to hundreds of days, which require dense H-rich CSM capable of sustaining prolonged interaction. The estimates in \Cref{sec:massloss} provide reference mass-loss rates for 13 objects with suitable calibrated spectra. For the adopted CSM velocity of $100\kms$ and conversion efficiency of $\epsilon=0.5$, these estimates span approximately $0.028$--$0.30\,M_{\odot}\,\mathrm{yr}^{-1}$. Most estimates assume a shock velocity of $5000\,\mathrm{km\,s^{-1}}$. Adopting a CSM velocity of $50\kms$ reduces them by a factor of two, but the order of magnitude is unchanged. Line-driven winds of OB stars typically have mass-loss rates of order $10^{-7}$--$10^{-5}\,M_{\odot}\,\mathrm{yr}^{-1}$ \citep[e.g.][]{Muijres_2012_MassLoss}, and dusty red-supergiant winds commonly reach $10^{-6}$--$10^{-4}\,M_{\odot}\,\mathrm{yr}^{-1}$ \citep[e.g.][]{vanLoon_2005_RSG}. Our estimates substantially exceed these rates, suggesting that ordinary steady stellar winds are insufficient under the adopted interaction model.

There is evidence that enhanced mass loss is a common feature of Type II SN progenitors shortly before core-collapse \citep[e.g.][]{Bruch_2021_MassLoss, Bruch_2023_Flash, Hinds_2025_SNeII}. In discussing the parent sample of our work, \citet{Pessi_2025_SLSNeIISample} also pointed towards elevated mass loss shortly before explosion as a possible explanation for the inferred CSM densities. An LBV-like giant eruption is one possible channel for producing such material. The nineteenth-century Great Eruption of $\eta$ Carinae expelled an estimated $12$--$20\,M_{\odot}$. \citet{Smith_2006A_MassLoss} reported a lower limit of $0.5\,M_{\odot}\,\mathrm{yr}^{-1}$ averaged over approximately 20 years of eruption, or even several $M_{\odot}\,\mathrm{yr}^{-1}$ for the model where most of the ejection occurred within $5$\,yr. Hence, LBV eruptive episodes can produce much greater mass loss than ordinary stellar winds. Pulsational pair instability is another possible channel that can produce massive shell ejections in stars if the progenitors are very massive stars ($\sim70$--$140\,M_{\odot}$) with large helium cores \citep[$\sim30$--$64\,M_{\odot}$;][]{Woosley_2007_PPI, Woosley_2017_PPI}. Direct evidence for rapidly expanding CSM shells has been found in some SLSNe-I. \citet{Lunnan_2018_iPTF16eh,Gkini_2025_EruptiveMassloss,Gkini_2026_EruptiveMassloss} identified SLSNe-I showing Mg\,II absorption components tracing CSM shells with velocities of $\sim3000$--$5000\,\kms$ expelled likely decades before explosion, although they do not uniquely establish a pulsational pair-instability origin. Due to limitations in the UV coverage in our sample, it is still unclear whether SLSNe-IIn can exhibit such features.

Binary interaction is another channel to reach enhanced mass loss \citep{Langer_2012_Binary,Ercolino_2026_Binary}. Case B Roche-lobe overflow begins after core hydrogen exhaustion but before core helium ignition, whereas Case C begins after core helium exhaustion, when the donor undergoes renewed expansion \citep{Ercolino_2026_Binary}. In the calculations of \citet{Tsai_2026_Binary}, Case B transfer briefly reaches $\sim10\,M_{\odot}\,\mathrm{yr}^{-1}$, but occurs at $\sim10^{6}$\,yr before core collapse, too early to explain the interaction observed in our sample. Their Case C model undergoes mass transfer during all of the final $\sim10^{4}$\,yr before collapse, reaching a peak rate of approximately $10^{-2}$--$10^{-3}\,M_{\odot}\,\mathrm{yr}^{-1}$ and sustaining approximately $10^{-3}$--$10^{-4}\,M_{\odot}\,\mathrm{yr}^{-1}$ until collapse. This Case C model is more consistent with the lower end of estimated mass-loss rates and timing in our sample, but cannot explain the higher mass loss estimated for objects reaching $10^{-1}\,M_{\odot}\,\mathrm{yr}^{-1}$.

\subsection{A hidden Ia-CSM population in SLSNe-IIn}
Six of the 105 narrow-lined objects in our sample (\Cref{sec:Ia-CSM}), approximately $6\%$, exhibit Ia-CSM-like spectral characteristics. This is despite the filtering done in \citet{Pessi_2025_SLSNeIISample} using SLSN-II labels in Fritz and GROWTH marshal. This implies that the original classifications in interacting SNe do not necessarily distinguish thermonuclear explosions from core-collapse events, especially when this classification relies on a single early-time spectrum. Although the six objects' thermonuclear origin remains unconfirmed, their presence suggests that samples of luminous SNe IIn (above $-19$\,mag) and SLSNe-IIn without extensive spectroscopic follow-up may contain a hidden SN Ia-CSM contamination. Such objects could affect population-level conclusions about the mass-loss processes and progenitors of H-rich interacting SNe.

The six Ia-CSM/IIn objects identified in \Cref{sec:Ia-CSM} demonstrate the challenges of inferring the underlying explosion mechanism for H-rich interacting SNe. Strong interaction features can overwhelm the observed spectrum and lead to the formation of a photosphere in the CSM which hides SN ejecta features beneath. From our sample and literature, if the interaction feature kicks in at early times (e.g. SNe\,2018bwr, 2019npx, 2005gj), the spectrum is typically indistinguishable from a Type IIn spectrum, with exceptions such as SN\,2020mad where ejecta features had already appeared before peak. This also implies that, for SNe brighter than approximately $-19.0$\,mag, a Type IIn classification based solely on pre-peak spectra does not securely establish a core-collapse origin. If strong interaction develops at later phases, SN Ia ejecta features may remain visible. Hence, we encourage late-time spectroscopic follow-up of SNe classified as Type IIn and brighter than approximately $-19.0$\,mag to help identify and characterise a potentially hidden Ia-CSM population.


\section{Conclusions} \label{sec:conclusion}
In this work, we presented and analysed 304 optical spectra of 106 H-rich SLSNe observed by ZTF. The objects were selected from the parent photometric sample by \citet{Pessi_2025_SLSNeIISample} to reach a peak rest-frame absolute magnitude of $M\leq-19.9$\,mag in at least one of the ZTF $gri$ bands, excluding 13 events with only broad Balmer emission lines in \citet{Kangas_2022_SLSNeIISample}. We examined their spectroscopic evolution, measured their Balmer-line properties, and modelled their H$\alpha$ profiles to investigate the conditions of CSM interaction. Our main findings are:
\begin{enumerate}
  \item Objects with narrow Balmer emission overwhelmingly dominate our sample. 105 out of 106 objects show narrow Balmer lines and can be classified as SLSNe-IIn. Only one object shows exclusively broad Balmer emission lines with FWHM$\geq5000\kms$ and can be classified as SLSN-II.
  \item Six narrow-lined objects show additional Ia-CSM-like spectral characteristics. These Type Ia-CSM candidates comprise $\sim6\%$ of our SLSN-IIn sample and suggest a possible hidden Ia-CSM population within samples selected as luminous SNe IIn.
  \item Three objects show high-ionisation features of He\,II $\lambda4686$, N\,III $\lambda\lambda4634,4641$, and/or C\,III $\lambda4649$. SN\,2021aaev has an estimated high-ionisation emission duration of approximately 24\,d, while tentative features in SN\,2021fmu are detected 18.2\,d after discovery. SN\,2020uwl shows such feature 3.3\,d after discovery, but its disappearance is unconstrained.
  \item Our SLSN-IIn sample shows substantial diversity in Balmer-line morphology post peak. Of the 45 objects with spectra of sufficient quality between peak and $+200$\,d, $64\%$ show narrow, approximately symmetric H$\alpha$ profiles without a clear blue excess, $22\%$ show a very mild blue excess, $11\%$ show a more extended, slanted blue excess, and one object (SN\,2021adxl) shows a broad blue shoulder.
  \item Beyond rest-frame $+200$\,d, most objects retain prominent H$\alpha$ emission. None shows significant [O\,I] $\lambda\lambda6300,6364$ emission.
  \item The H$\alpha$ pEW and FWHM evolution of our SLSN-IIn sample overlaps with the lower-luminosity SN IIn comparison sample, but is more diverse than in the SLSN-II comparison sample.
  \item Excluding the six SN Ia-CSM candidates, the median Balmer decrement in our SLSN-IIn sample increases from 2.85 before $+100$\,d to 4.40 after. Both medians are lower than those of the comparison Type Ia-CSM sample. The early median value is close to the Case B recombination value, but a fraction of objects has Balmer decrement significantly lower than the Case B recombination value.
  \item Balmer-line modelling yields higher photon-emitting shell velocities for objects with blue excesses and lower velocities for those with symmetric profiles. The former tend to be fainter, peaking near $M_r\simeq-20$\,mag, while the latter are brighter, near $M_r\simeq-21$\,mag. Assuming comparable explosion energies, this trend may reflect differences in shock deceleration and the conversion of kinetic energy into radiation.
  \item Our sample objects have estimated mass-loss rates spanning $\sim0.028$--$0.3\,M_{\odot}\,\mathrm{yr}^{-1}$. Although sensitive to the assumed shock velocities, these estimates suggest enhanced pre-explosion mass loss beyond ordinary stellar winds, and would require very extreme mass loss processes, such as extreme cases in binary mass transfer, LBV-like eruptions or pulsational pair instability explosions.
  \item We identify no Balmer-line morphology unique to SLSNe-IIn defined by the $-19.9$\,mag cut compared with lower-luminosity SNe IIn. However, SLSN-IIn showing high-ionisation features tend to have longer durations than lower-luminosity SN IIn flashers.
\end{enumerate}

Our study presented the largest spectroscopic sample of H-rich SLSNe to date, revealing substantial diversity within the dominant SLSN-IIn population and in the inferred properties of the CSM interaction powering them. Nevertheless, our study is limited by the small SN IIn comparison sample and uneven spectroscopic coverage at very early and late phases. Future studies should extend spectroscopic comparisons across the wider SN IIn population, without imposing a fixed magnitude cut. Two observational priorities are to search for flash-ionisation features shortly after discovery and to obtain late-time spectra to reveal the underlying ejecta. The latter can also help identify Type Ia-CSM events hidden among objects initially classified as SNe IIn. Our work provides a starting point for connecting the spectroscopic diversity of H-rich SLSNe to their CSM and progenitor systems.

\begin{acknowledgements}
We thank Jacco H. Terwel for very constructive discussions on SNe Ia-CSM. This project is funded by the European Union (ERC, project number 101042299, TransPIre). Views and opinions expressed are however those of the author(s) only and do not necessarily reflect those of the European Union or the European Research Council Executive Agency. Neither the European Union nor the granting authority can be held responsible for them. Based on observations obtained with the Samuel Oschin Telescope 48-inch and the 60-inch Telescope at the Palomar Observatory as part of the Zwicky Transient Facility project. ZTF is supported by the National Science Foundation under Grants No. AST-1440341 and AST-2034437 and a collaboration including current partners Caltech, IPAC, the Weizmann Institute of Science, the Oskar Klein Center at Stockholm University, the University of Maryland, Deutsches Elektronen-Synchrotron and Humboldt University, the TANGO Consortium of Taiwan, the University of Wisconsin at Milwaukee, Trinity College Dublin, Lawrence Livermore National Laboratories, IN2P3, University of Warwick, Ruhr University Bochum, Northwestern University and former partners the University of Washington, Los Alamos National Laboratories, and Lawrence Berkeley National Laboratories. Operations are conducted by COO, IPAC, and UW. SED Machine is based upon work supported by the National Science Foundation under Grant No. 1106171. The Gordon and Betty Moore Foundation, through both the Data-Driven Investigator Program and a dedicated grant, provided critical funding for SkyPortal. Based on observations collected at the European Organisation for Astronomical Research in the Southern Hemisphere, Chile, as part of ePESSTO+ (the advanced Public ESO Spectroscopic Survey for Transient Objects Survey – PI: Inserra). ePESSTO+ observations were obtained under ESO program IDs 106.216C, 108.220C and 111.24PR. T. E. M. B. is funded by Horizon Europe ERC grant no. 101125877. P.C. acknowledges financial support from the Secretary of Universities and Research (Government of Catalonia) and by the Horizon 2020 Research and Innovation Programme of the European Union under the Marie Sk\l{}odowska-Curie and the Beatriu de Pin\'os 2024 BP 00125 programme, and from the Centro Superior de Investigaciones Cient\'ificas (CSIC) under the Spanish program Unidad de Excelencia María de Maeztu CEX2020-001058-M, financed by MCIN/AEI/10.13039/501100011033, and by the MaX-CSIC Excellence Award MaX4-SOMMA-ICE. M. N. is supported by the Science and Technology Facilities Council (grant number UKRI4107) and by the European Research Council (grant agreement No.~948381). This publication has emanated from research conducted with the financial support of Taighde \'Eireann--Research Ireland under Grant No.~24/FPP/12959. T.-W. Chen acknowledges financial support from the Yushan Fellow Program by the Ministry of Education, Taiwan (MOE-111-YSFMS-0008-001-P1), and the National Science and Technology Council, Taiwan (NSTC grant 114-2112-M-008-021-MY3). I.A. is supported by the National Science Foundation award AST 2505775, NASA grant 24-ADAP24- 0159, Scialog award LSST2025-112b. GL was supported by a research grant (VIL60862) from VILLUM FONDEN. 

\end{acknowledgements}
\nolinenumbers
\clearpage
\onecolumn
\appendix
\section{Logs of observations} \label{app:A}
Observation log for the ZTF hydrogen-rich SLSN sample. There are 304 spectra of 106 objects. The full table is available in the supplementary material.
\begin{table}[ht]
\caption{First 10 rows of the spectroscopic observation log of the sample.}
\label{tab:spectral_log}
\centering
\begin{tabular}{llrlrll}
\hline
IAU name & ZTF name & MJD & Telescope/Inst. & Exp. time [s] & Grism/grating & Sources \\
\hline
2018atq & ZTF18aahmhxu & 58230.4 & P60/SEDM & 3300 & - & Fritz marshall \\
2018atq & ZTF18aahmhxu & 58237.0 & LT/SPRAT & 300 & - & Fritz marshall \\
2018atq & ZTF18aahmhxu & 58255.2 & P200/DBSP & 600 & blue+red & Fritz marshall \\
2018atq & ZTF18aahmhxu & 58316.2 & P200/DBSP & 900 & blue+red & \citet{Dahiwale_2020_2020-06-02} \\
2018atq & ZTF18aahmhxu & 58465.5 & P200/DBSP & 1200 & blue+red & Fritz marshall \\
2018bwr & ZTF18aavskep & 58271.2 & P60/SEDM & 1600 & - & \citet{Fremling_2018_2018-06-05} \\
2018bwr & ZTF18aavskep & 58279.0 & LT/SPRAT & 300 & - & Fritz marshall \\
2018bwr & ZTF18aavskep & 58503.3 & NOT/ALFOSC & 1200 & Gr\#4 & Fritz marshall \\
2018ddq & ZTF18aavjcpf & 58286.3 & Keck I/LRIS & 325 & blue+red & Fritz marshall \\
2018ddq & ZTF18aavjcpf & 58316.2 & P200/DBSP & 600 & blue+red & \citet{Fremling_2018_2018-07-26} \\
\hline
\end{tabular}
\end{table}

\section{Calculated properties of spectra} \label{app:B}
\begin{table}[ht]
\caption{First 10 rows of the sample spectroscopic properties. Phases are measured in the rest frame relative to $r$-band maximum, or the $g/i$-band maximum if necessary. The full table is available in the supplementary material. Unavailable measurements are marked with --.}
\label{tab:spectral_properties}
\centering
\setlength{\tabcolsep}{3pt}
\begin{tabular}{lrrrrrrrr}
\hline
IAU name & MJD & Phase & SNR
& $L_{\mathrm{H}\alpha}$
& $L_{\mathrm{H}\beta}$
& pEW$_{\mathrm{H}\alpha}$
& pEW$_{\mathrm{H}\beta}$
& FWHM$_{\mathrm{H}\alpha}$ \\
& & [d] &
& $10^{41}$ [erg s$^{-1}$]
& $10^{41}$ [erg s$^{-1}$]
& [\AA]
& [\AA]
& [km s$^{-1}$] \\
\hline
2018atq & 58230.4 & -25.8 & 19.7 & -- & -- & -- & -- & -- \\
2018atq & 58237.0 & -20.2 & 3.3 & -- & -- & -- & -- & -- \\
2018atq & 58255.2 & -4.6 & 18.9 & $1.19\pm0.11$ & -- & $14.2\pm0.6$ & -- & -- \\
2018atq & 58316.2 & 47.7 & 18.7 & $5.66\pm0.69$ & $0.83\pm0.15$ & $74.3\pm2.5$ & $9.8\pm1.4$ & 383 \\
2018atq & 58465.5 & 175.6 & 22.7 & $3.90\pm0.70$ & $0.41\pm0.09$ & $166.2\pm3.0$ & $14.2\pm1.7$ & 1857 \\
2018bwr & 58271.2 & -14.5 & 50.8 & $2.84\pm0.15$ & $1.14\pm0.17$ & $162.1\pm3.5$ & $27.8\pm4.5$ & -- \\
2018bwr & 58279.0 & -7.1 & 24.4 & $3.12\pm0.27$ & $1.65\pm0.14$ & $104.3\pm4.3$ & $27.3\pm1.2$ & 627 \\
2018bwr & 58503.3 & 207.4 & 15.3 & $5.90\pm0.39$ & $0.45\pm0.07$ & $524.8\pm16.6$ & $23.3\pm3.9$ & 1922 \\
2018ddq & 58286.3 & -15.3 & 19.5 & $3.40\pm0.13$ & $1.53\pm0.06$ & $98.8\pm2.7$ & $24.0\pm0.7$ & 447 \\
2018ddq & 58316.2 & 10.2 & 9.6 & $2.88\pm0.36$ & -- & $59.8\pm2.9$ & -- & 323 \\
\hline
\end{tabular}
\end{table}
\section{Ia-CSM Classification} \label{app:C}

\begin{figure*}
    \centering
    \includegraphics[width=0.48\linewidth]{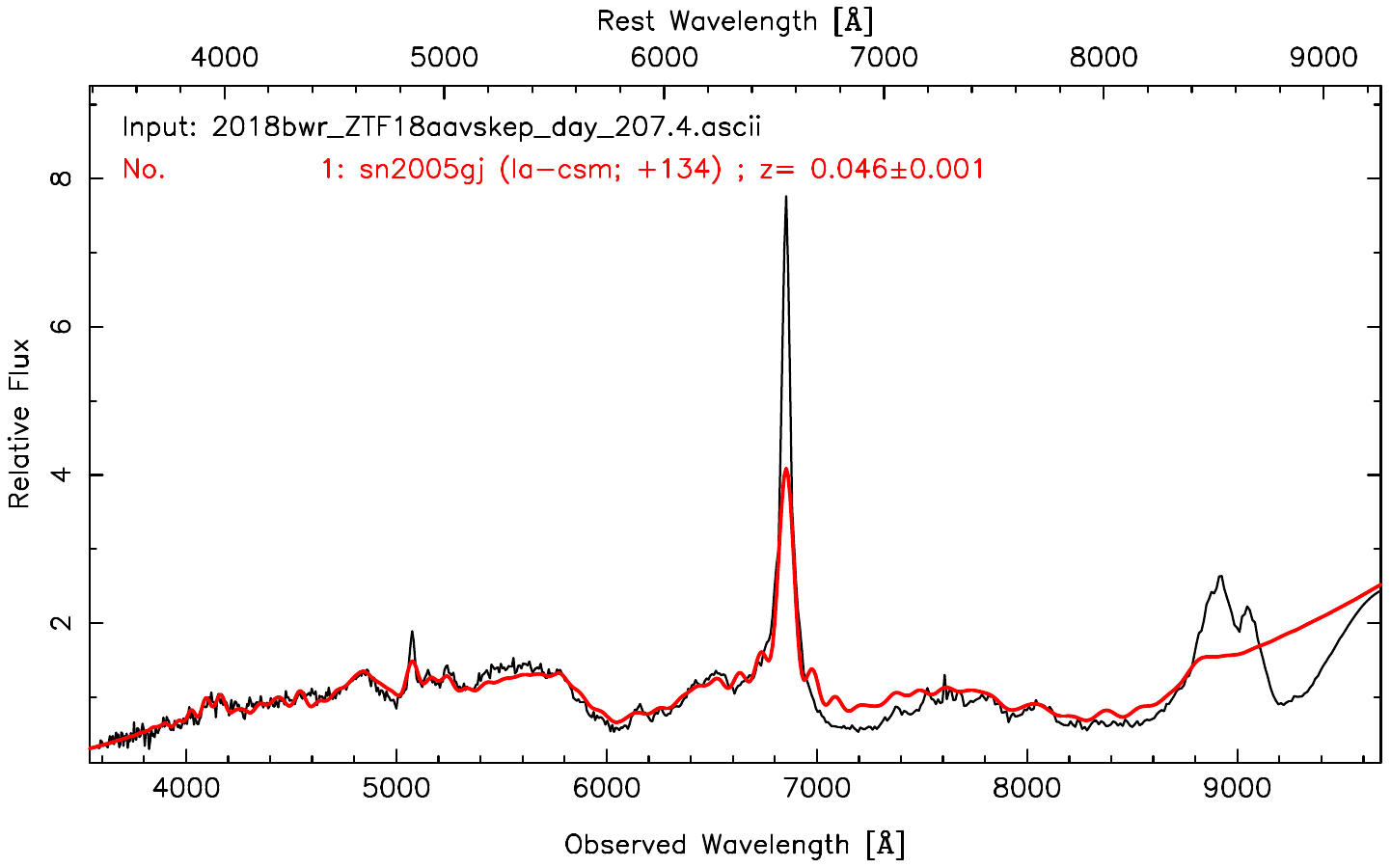}
    \includegraphics[width=0.48\linewidth]{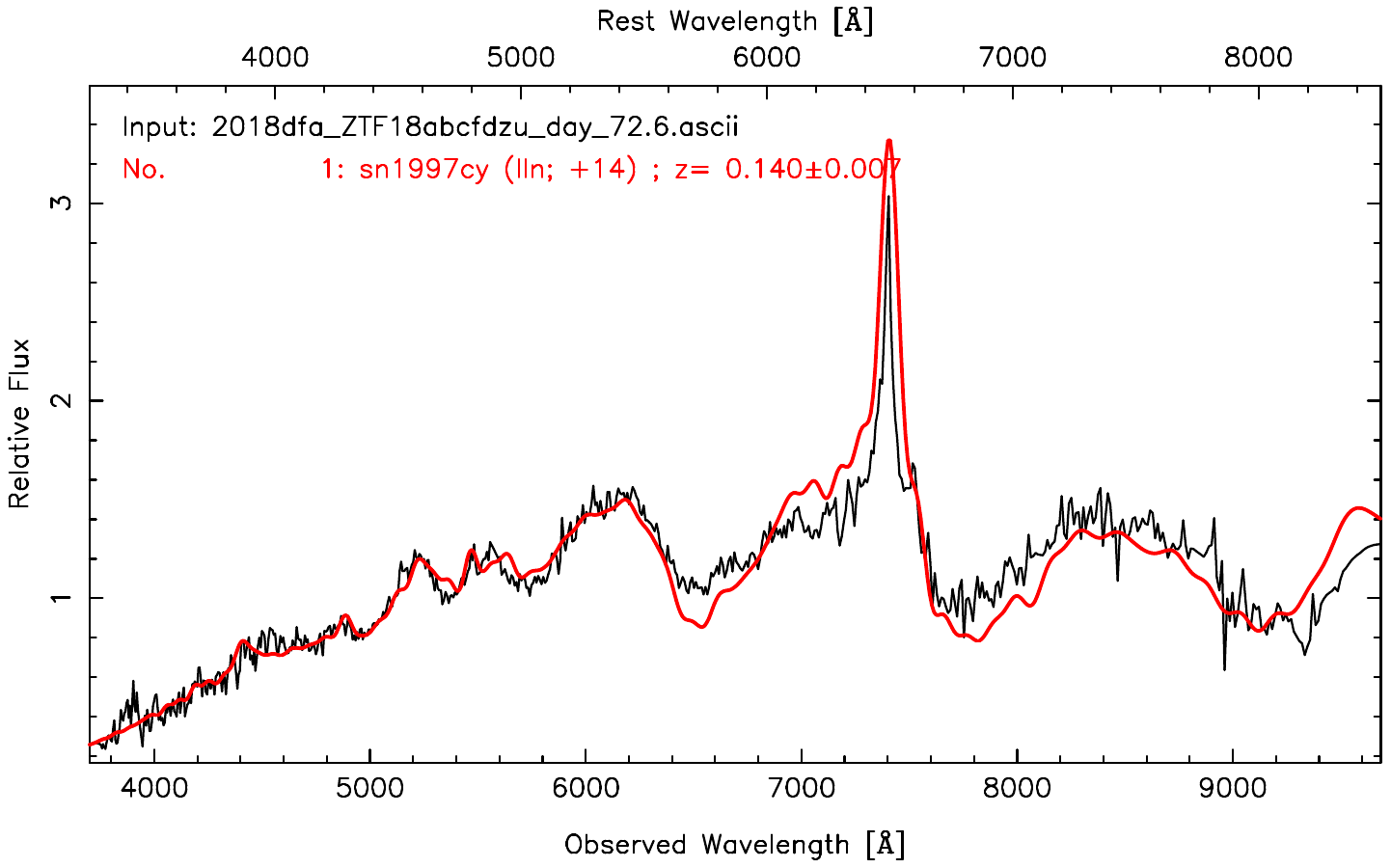}\\
    \includegraphics[width=0.48\linewidth]{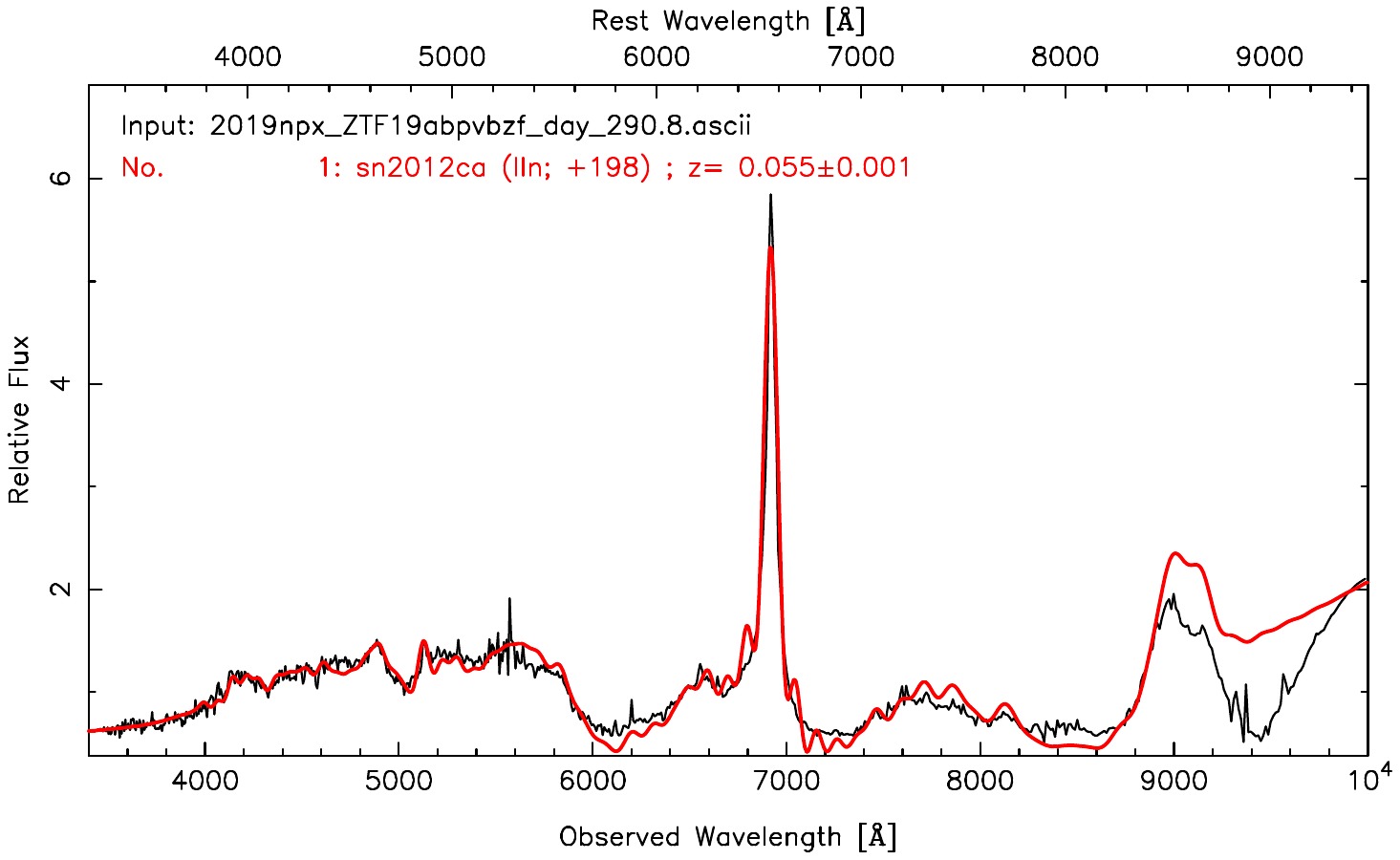}
    \includegraphics[width=0.48\linewidth]{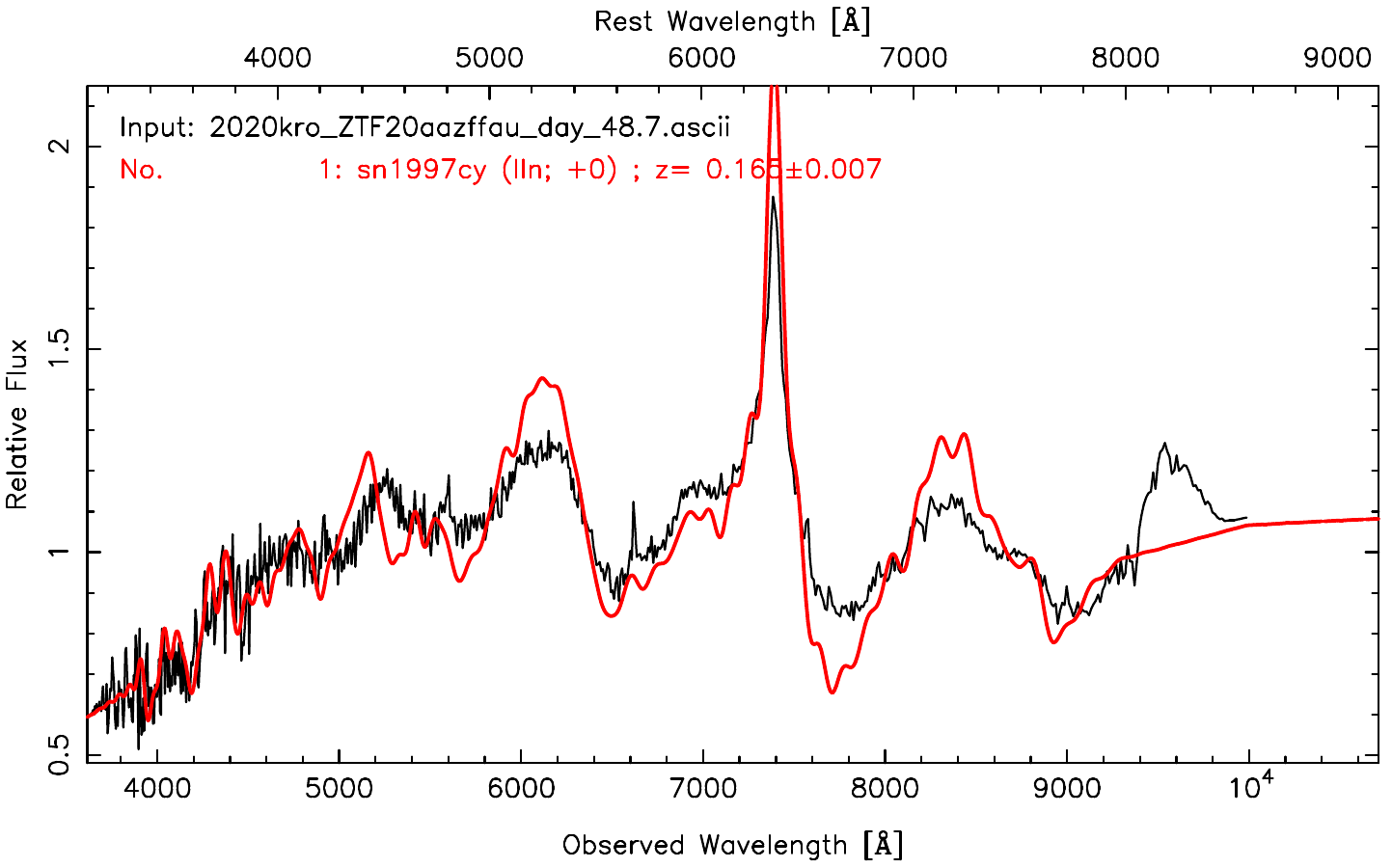}\\
    \includegraphics[width=0.48\linewidth]{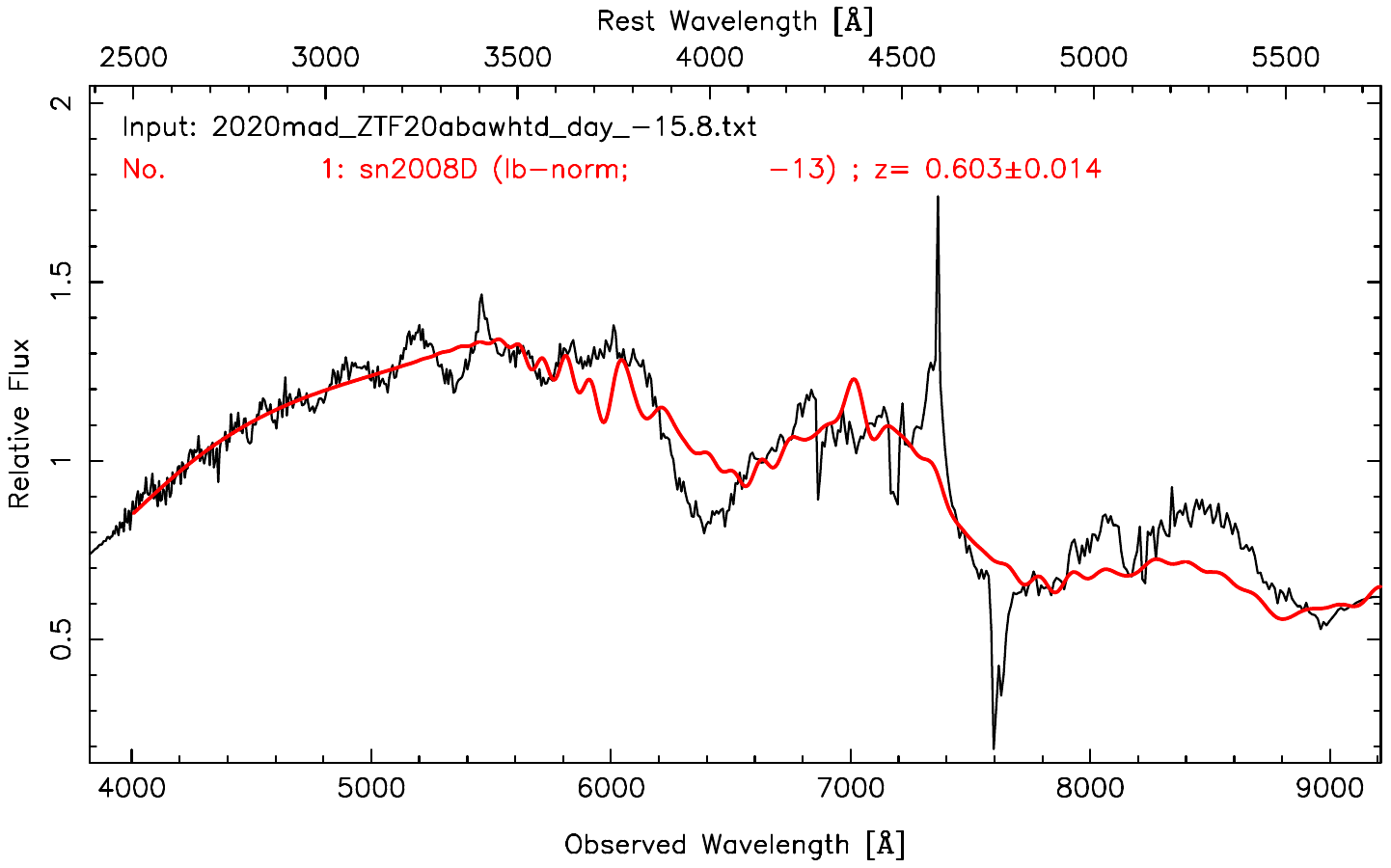}
    \includegraphics[width=0.48\linewidth]{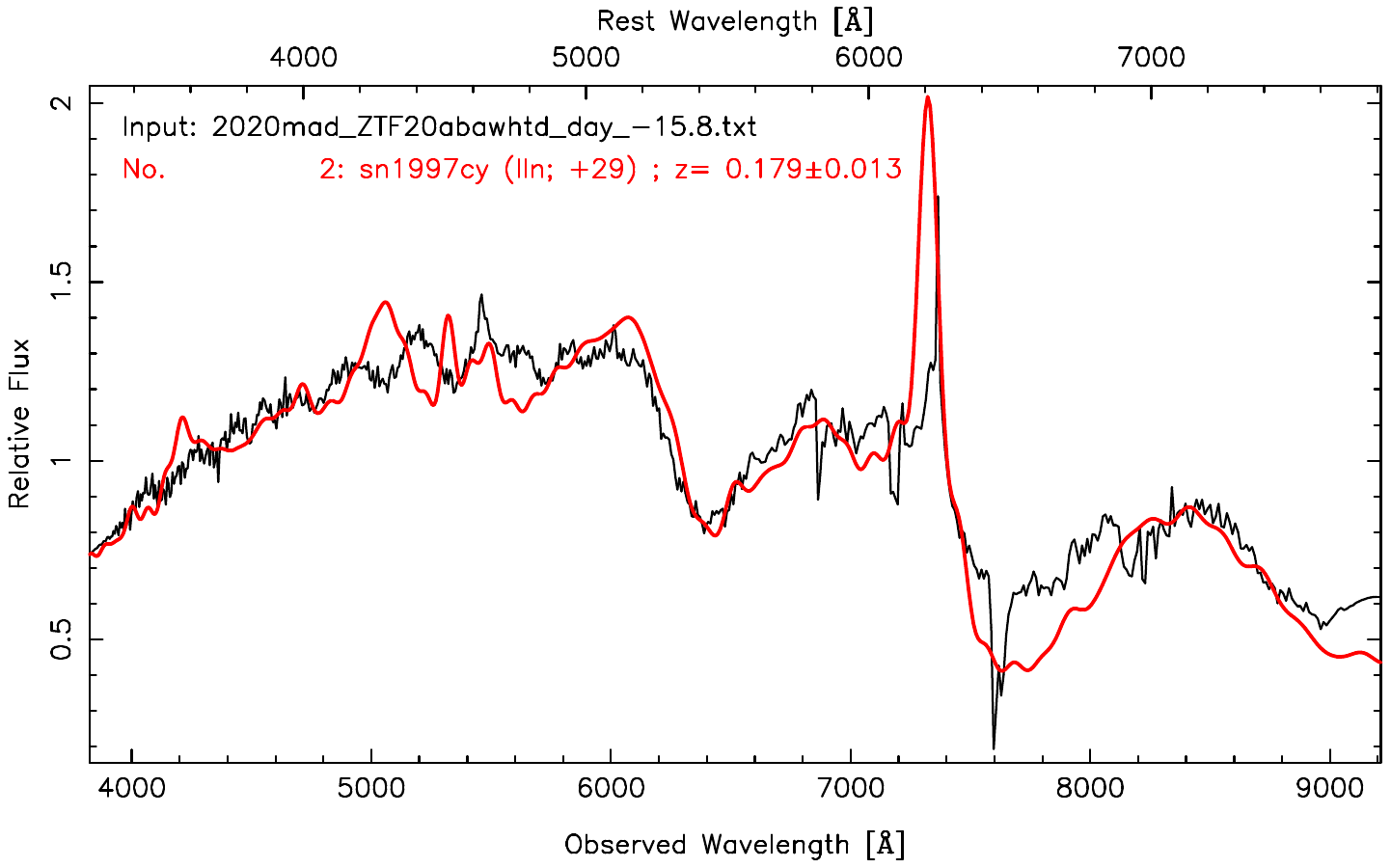}\\
    \includegraphics[width=0.48\linewidth]{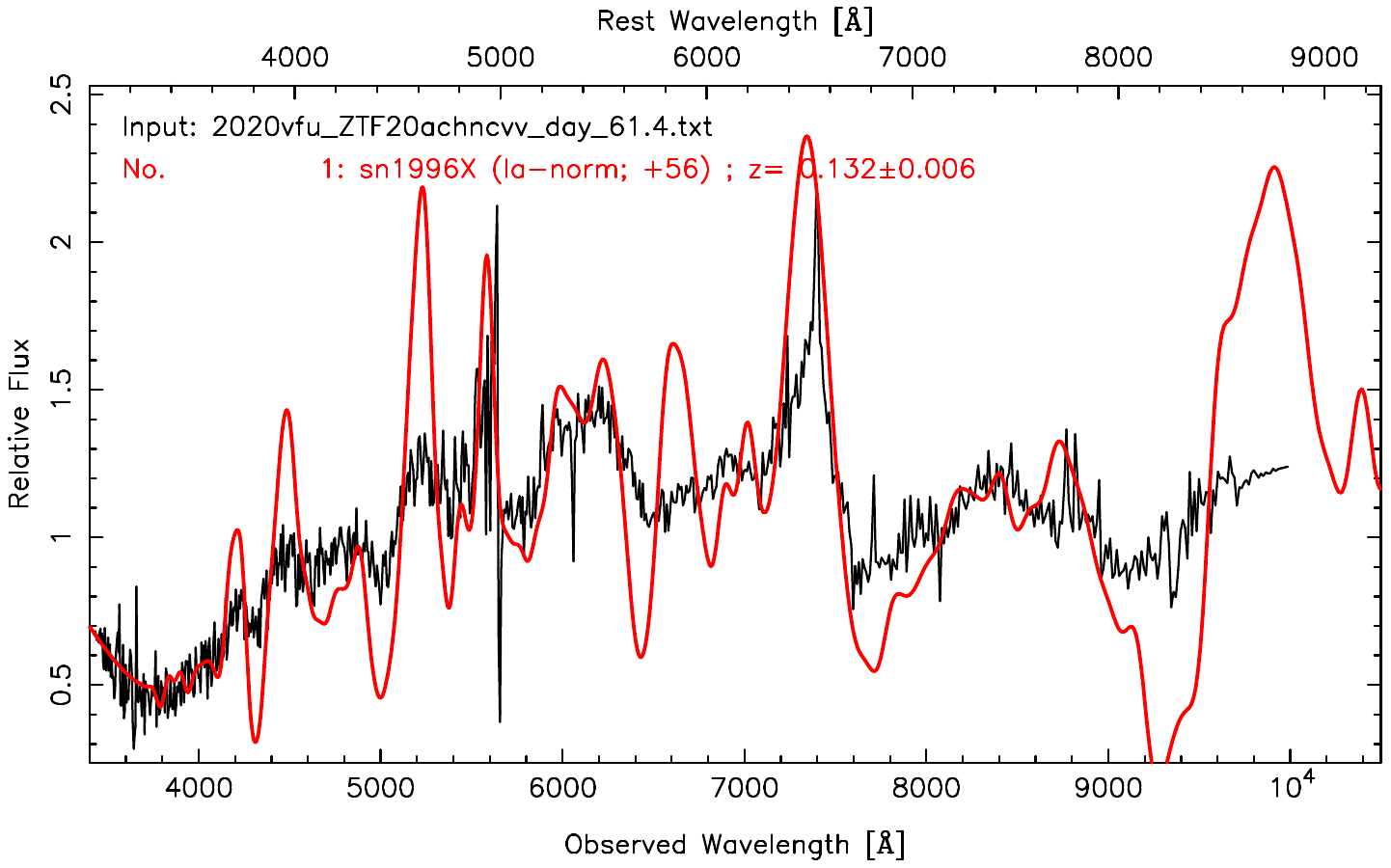}
    \includegraphics[width=0.48\linewidth]{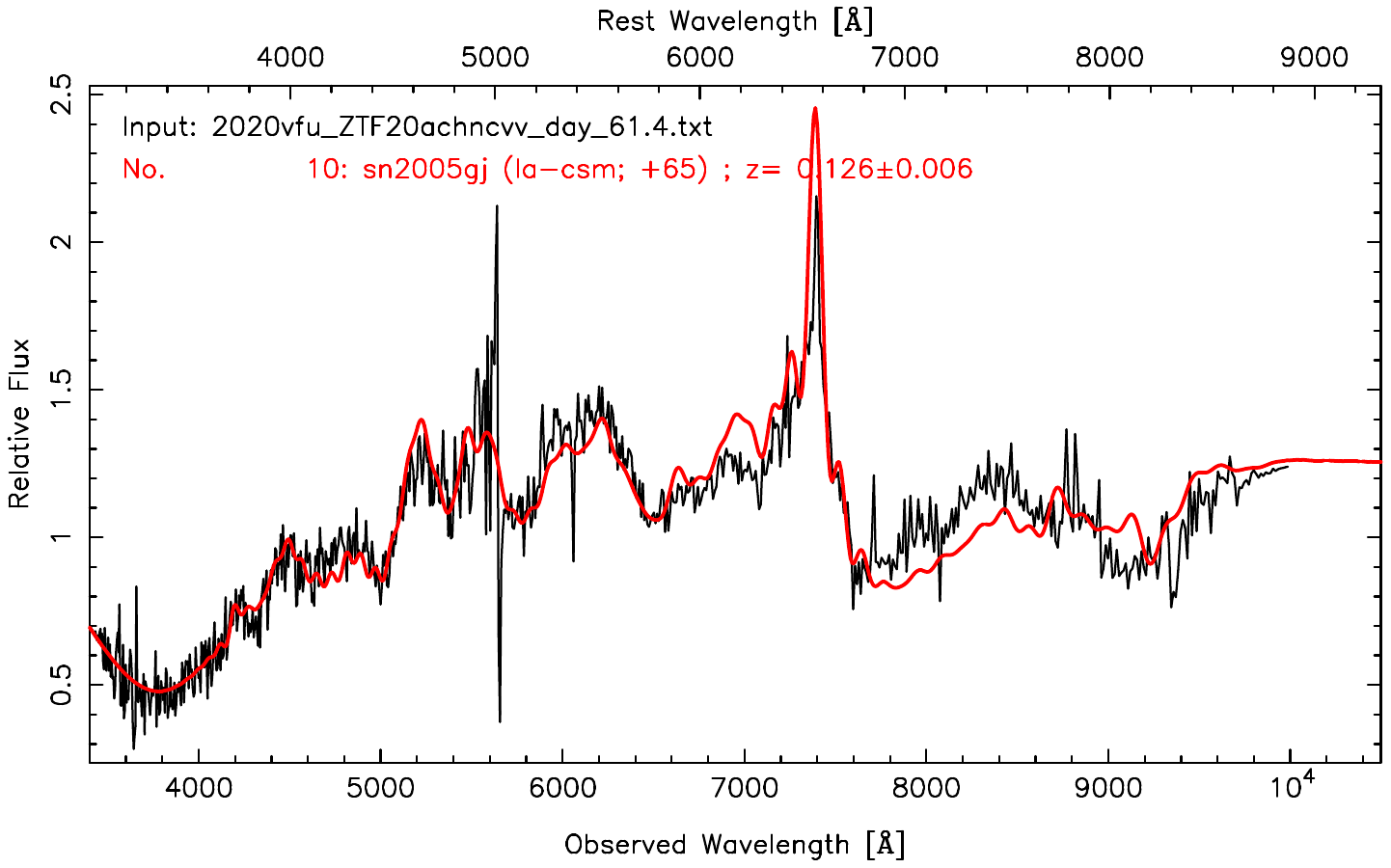}\\
    \caption{SNID template comparisons for SNe\,2018bwr, 2018dfa, 2019npx, 2020kro, 2020mad, and 2020vfu, ordered from left to right and top to bottom. Observed spectra are shown in black and template spectra in red.}
    \label{fig:snid_comparisons_1}
\end{figure*}
\clearpage
\section{Electron-scattering model parameters} \label{app:D}

\begin{table}[ht]
\caption{Adopted \textsc{escatter} parameters for the 52 fitted spectra of 18 objects in the sample. Phases are measured in the rest frame relative to the adopted photometric peak, as defined in the text. The optical depth $\tau$ is dimensionless, and $v_{\rm s}$ is the best-fit photon-emitting shell velocity.}
\label{tab:escatter_parameters}
\centering
\small
\setlength{\tabcolsep}{6pt}
\begin{tabular}{lrrrr}
\hline
IAU name & MJD & Phase & $\tau$ & $v_{\rm s}$ \\
 & & [d] & & [km s$^{-1}$] \\
\hline
2018atq  & 58316.2 & 47.7   & 1.3 & 2200 \\
2018atq  & 58465.5 & 175.6  & 4.3 & 1000 \\
2018bwr  & 58279.0 & -7.1   & 1.0 & 500  \\
2018bwr  & 58503.3 & 207.4  & 1.2 & 1400 \\
2018dfa  & 58334.2 & 21.5   & 1.0 & 200  \\
2018dfa  & 58337.9 & 24.8   & 1.0 & 600  \\
2018dfa  & 58338.3 & 25.2   & 1.0 & 200  \\
2018dfa  & 58348.9 & 34.6   & 1.1 & 500  \\
2018dfa  & 58355.9 & 40.8   & 1.0 & 200  \\
2018dfa  & 58379.8 & 62.0   & 3.0 & 500  \\
2018dfa  & 58391.8 & 72.6   & 3.0 & 600  \\
2018dfa  & 58481.2 & 151.9  & 2.3 & 1300 \\
2018dfa  & 58504.2 & 172.2  & 1.5 & 1600 \\
2018dfa  & 58668.3 & 317.7  & 1.2 & 1000 \\
2018dfa  & 58668.5 & 317.9  & 1.0 & 1000 \\
2018ffs  & 58397.3 & 18.8   & 1.1 & 100  \\
2018ffs  & 58666.5 & 254.7  & 1.1 & 700  \\
2018hxe  & 58509.5 & 69.5   & 1.2 & 1600 \\
2019aavb & 58904.2 & -30.5  & 1.0 & 400  \\
2019aavb & 59107.0 & 131.3  & 2.2 & 2000 \\
2019avv  & 58666.4 & 103.1  & 3.4 & 500  \\
2019npx  & 58718.1 & -20.1  & 1.0 & 300  \\
2019npx  & 59046.4 & 290.8  & 1.6 & 800  \\
2019vas  & 58910.5 & 58.8   & 3.7 & 800  \\
2019vme  & 59134.4 & 236.9  & 3.0 & 900  \\
2020in   & 58868.0 & -39.6  & 1.0 & 600  \\
2020in   & 59234.6 & 291.2  & 3.4 & 1600 \\
2020jgv  & 58904.2 & -23.3  & 1.0 & 1600 \\
2020jgv  & 59107.0 & 153.5  & 3.9 & 3900 \\
2020nt   & 58891.5 & -25.2  & 1.0 & 600  \\
2020nt   & 59256.6 & 282.4  & 3.8 & 1200 \\
2020usa  & 59165.1 & -4.4   & 1.0 & 600  \\
2020usa  & 59170.0 & -0.5   & 1.1 & 100  \\
2020usa  & 59171.1 & 0.3    & 1.0 & 1000 \\
2020usa  & 59190.1 & 15.4   & 1.1 & 100  \\
2020usa  & 59347.5 & 140.0  & 1.0 & 600  \\
2020vfu  & 59144.3 & -6.9   & 1.0 & 100  \\
2020vfu  & 59221.3 & 61.4   & 2.6 & 300  \\
2021adxl & 59615.0 & 91.8   & 3.9 & 6400 \\
2021adxl & 59615.1 & 91.9   & 2.7 & 6900 \\
2021adxl & 59624.0 & 100.6  & 4.2 & 5900 \\
2021adxl & 59632.0 & 108.5  & 4.2 & 5900 \\
2021adxl & 59639.3 & 115.7  & 3.4 & 6100 \\
2021adxl & 59664.1 & 140.0  & 4.1 & 5200 \\
2021adxl & 59668.1 & 144.0  & 5.0 & 4000 \\
2021adxl & 59693.0 & 168.4  & 5.0 & 3800 \\
2021adxl & 59720.9 & 195.8  & 5.0 & 1500 \\
2021adxl & 59723.2 & 198.1  & 5.0 & 1500 \\
2021adxl & 59735.9 & 210.6  & 5.0 & 1400 \\
2021adxl & 59754.9 & 229.2  & 5.4 & 200  \\
2022lvm  & 59969.0 & 174.0  & 3.9 & 6700 \\
2022opm  & 59818.2 & 10.0   & 1.0 & 600  \\
\hline
\end{tabular}
\end{table}
\clearpage

\twocolumn

\nocite{Angus_2020_2020-10-13}
\nocite{Ayala_2022_2022-08-21}
\nocite{Blanchard_2020_2020-12-09}
\nocite{Brennan_2019_2019-08-23}
\nocite{Brink_2020_2020-08-05}
\nocite{Brinnel_2019_2019-03-20}
\nocite{Carini_2021_2021-10-14}
\nocite{Chu_2021_2021-11-22}
\nocite{Dahiwale_2019_2019-10-04}
\nocite{Dahiwale_2020_2020-01-02}
\nocite{Dahiwale_2020_2020-02-22}
\nocite{Dahiwale_2020_2020-03-02}
\nocite{Dahiwale_2020_2020-04-29}
\nocite{Dahiwale_2021_2021-01-21}
\nocite{Dahiwale_2021_2021-02-05}
\nocite{Dahiwale_2021_2021-05-12}
\nocite{Dahiwale_2021_2021-05-28}
\nocite{Dugas_2019_2019-07-30}
\nocite{Fremling_2018_2018-08-04}
\nocite{Fremling_2018_2018-09-18}
\nocite{Fremling_2018_2018-11-12}
\nocite{Fremling_2018_2018-12-05}
\nocite{Fremling_2018_2018-12-07}
\nocite{Fremling_2019_2019-03-05}
\nocite{Fremling_2019_2019-03-26}
\nocite{Fremling_2019_2019-04-17}
\nocite{Fremling_2019_2019-04-27}
\nocite{Fremling_2019_2019-07-12}
\nocite{Gomez_2020_2020-10-16}
\nocite{Gomez_2021_2021-11-08}
\nocite{Gomez_2021_2021-11-09}
\nocite{Gomez_2022_2022-03-10}
\nocite{Graham_2020_2020-01-20}
\nocite{Graham_2021_2021-09-14}
\nocite{Grayling_2022_2022-09-21}
\nocite{Grayling_2022_2022-09-22}
\nocite{Gromadzki_2018_2018-09-16}
\nocite{Gutierrez_2022_2022-08-27}
\nocite{Harvey_2021_2021-09-16}
\nocite{Ihanec_2020_2020-11-22}
\nocite{Ihanec_2021_2021-10-28}
\nocite{Liang_2019_2019-07-03}
\nocite{Nagao_2025_2021adxl}
\nocite{Pellegrino_2022_2022-06-15}
\nocite{Perez-Fournon_2022_2022-08-01}
\nocite{Perley_2019_2019-07-10}
\nocite{Perley_2021_2021-04-05}
\nocite{Perley_2022_2022-03-07}
\nocite{Perley_2022_2022-05-29}
\nocite{Perley_2022_2022-06-21}
\nocite{Perley_2022_2022-08-25}
\nocite{Pessi_2020_2020-12-08}
\nocite{Pessi_2021_2021-04-03}
\nocite{Pessi_2021_2021-04-04}
\nocite{Pessi_2024_2024-09-11a}
\nocite{Pessi_2024_2024-09-11b}
\nocite{Pessi_2024_2024-09-11c}
\nocite{Pessi_2024_2024-09-11d}
\nocite{Pessi_2024_2024-09-11e}
\nocite{Pessi_2024_2024-09-11f}
\nocite{Pessi_2024_2024-09-11g}
\nocite{Pessi_2024_2024-09-11h}
\nocite{Pessi_2024_2024-09-11i}
\nocite{Pessi_2024_2024-09-11j}
\nocite{Prentice_2021_2021-01-28}
\nocite{Prentice_2021_2021-03-23}
\nocite{Prentice_2021_2021-03-24}
\nocite{Pursiainen_2021_2021-05-22}
\nocite{Schulze_2021_2021-05-11}
\nocite{Schulze_2022_2022-05-12}
\nocite{Schulze_2022_2022-11-08}
\nocite{Siebert_2021_2021-02-12}
\nocite{Smith_2022_2022-03-02}
\nocite{Sollerman_2022_2022-10-19}
\nocite{Soraisam_2020_2020-06-25}
\nocite{Tucker_2019_2019-02-04}
\nocite{Wiseman_2019_2019-08-20}
\nocite{Yan_2020_2020-06-09}
\nocite{Yan_2020_2020-08-30}
\nocite{Yan_2020_2020-12-02}

\bibliographystyle{aa}
\bibliography{ref}

\end{document}